\documentclass[authoryear,review]{elsarticle}
\usepackage{appendix}
\usepackage{setspace}
\usepackage[font=small,  margin=2cm]{caption}
\usepackage[tbtags]{amsmath}
\usepackage{amsthm,amssymb,amsfonts}
\usepackage{natbib}
\usepackage{multirow}
\usepackage{lscape}
\journal{Journal of Econometrics \ Templates}
\usepackage{subfigure}
\usepackage{makecell}
\usepackage{exscale}
\usepackage{booktabs}
\usepackage{threeparttable}
\usepackage{array}
\usepackage{fullpage}
\usepackage{url}
\usepackage{algorithm}
\usepackage{algpseudocode}
\usepackage{bm}
\usepackage{smile}
\usepackage{mathtools}
\usepackage{wrapfig}
\usepackage{lipsum}
\usepackage{mathrsfs}
\usepackage{relsize}
\usepackage{dsfont}
\usepackage{multirow}
\usepackage{apalike}
\usepackage{chngcntr}
\usepackage{graphicx}
\usepackage{algorithmicx}
\usepackage{algpseudocode}
\usepackage[usenames,dvipsnames,svgnames,table]{xcolor}
\usepackage[colorlinks=true,
linkcolor=blue,
urlcolor=blue,
citecolor=blue]{hyperref}

\newcommand{\beq}{\begin{equation}}
\newcommand{\eeq}{\end{equation}}
\renewcommand{\l}{\left}
\renewcommand{\r}{\right}

\numberwithin{equation}{section}
\numberwithin{theorem}{section}
\numberwithin{corollary}{section}
\counterwithout{asmp}{section}
\numberwithin{definition}{section}
\renewcommand{\baselinestretch}{1.6}
\begin{document}
	
	\begin{frontmatter}
		\title{ Huber Principal Component Analysis for Large-dimensional Factor Models}

		\author[myfirstaddress]{Yong He\corref{cor1}}
		\address[myfirstaddress]{Institute for Financial Studies, Shandong University, Jinan, 250100, China}
			
		\ead{heyong@sdu.edu.cn}

			\author[mysecondaddress]{Lingxiao Li}
		\address[mysecondaddress]{School of Mathematics, Shandong University, Jinan, 250100, China}

	\author[mythirdaddress]{Dong Liu}
		\address[mythirdaddress]{Shanghai University of Finance and Economics, Shanghai, 200433, China}
		
			\author[myfourthaddress]{Wen-Xin Zhou\corref{cor1}}
		\address[myfourthaddress]{Department of Mathematics, University of California, San Diego, La Jolla, CA, 92093, USA}
			\cortext[cor1]{Corresponding author. All authors contributed equally to this work.}
		\ead{wez243@ucsd.edu}
		
		\begin{abstract}
Factor models have been widely used in economics and finance.  However,  the heavy-tailed nature of macroeconomic and financial data is often neglected in the existing literature. To address this issue and achieve robustness, we propose an approach to estimate factor loadings and scores by minimizing the Huber loss function, which is motivated by the equivalence of conventional Principal Component Analysis (PCA) and the constrained least squares method in the factor model.  We provide two algorithms that use different penalty forms. The first algorithm, which we refer to as Huber PCA,  minimizes the $\ell_2$-norm-type Huber loss and performs PCA on the weighted sample covariance matrix. The second algorithm involves an element-wise type Huber loss minimization, which can be solved by an iterative Huber regression algorithm. Our study examines the theoretical minimizer of the element-wise Huber loss function and demonstrates that it has the same convergence rate as conventional PCA when the idiosyncratic errors have bounded second moments.
We also derive their asymptotic distributions under mild conditions. Moreover, we suggest a consistent model selection criterion that relies on rank minimization to estimate the number of factors robustly. We showcase the benefits of Huber PCA through extensive numerical experiments and a real financial portfolio selection example. An R package named ``HDRFA" \footnote{\url{%
https://cran.r-project.org/web/packages/HDRFA/index.html}} has been developed to implement the proposed robust factor analysis.
		\end{abstract}

	\end{frontmatter}

		\section{Introduction}
 The large dimensionality of contemporary data is undeniably one of the primary challenges of modern statistics, as it is pervasive in most domains related to data science. Time series analysis is no exception to this trend, and the study of large-dimensional time series or equivalently, large cross-sections of univariate time series, also referred to as panels, is now one of the most active topics in both theoretical and applied econometrics.
Thus far, the most effective tools for analyzing and predicting large-dimensional time series are the large-dimensional (approximate) factor models \citep{Chamberlain1983Arbitrage}. Factor models, with various forms, essentially decompose the observations, say, a large cross-section of time series with complex interrelations, into two mutually orthogonal (all leads, all lags) components: the common component, driven by a small number of factors or common shocks, and an idiosyncratic component. The definitions of ``common" and ``idiosyncratic" may vary, and assumptions are made regarding these components.
Inference of large-dimensional Approximate Factor Models (AFM) has been extensively studied in the literature, and  can be categorized into two main methods: the principal component analysis (PCA) method and the maximum likelihood estimation (MLE) method. PCA-based methods are straightforward to implement and provide consistent estimators for the factor scores and loadings when both the cross-section $N$ and time dimension $T$ tend to infinity; see for example \cite{Bai2002Determining}, \cite{stock2002forecast}, \cite{Bai2003Inferential}, \cite{onatski2009testing}, \cite{ahn2013eigenvalue}, \cite{fan2013large}, \cite{Trapani2018A}, \cite{yu2019robust}, \cite{barigozzi2020consistent} and \cite{barigozzi2022estimation}.
 Furthermore, the PCA method has been shown to be equivalent to a constrained least squares method, while MLE-based methods are more efficient but computationally more demanding; see, for example, \cite{Bai2012Statistical}, \cite{Bai2014Theory}, \cite{Bai2016Maximum} and \cite{barigozzi2019quasi}.

In  economics and finance, it is well recognized that the collected data often have heavy tails \citep{Fama1963Mandelbrot,Cont2001Empirical}, making conventional PCA or MLE-based methods unsuitable \citep{barigozzi2022inference}. Despite the vast literature on AFM, very few studies have explored robust factor analysis. \cite{He2020large} proposed a Robust Two-Step (RTS) method under the joint elliptical distribution assumption of the factors and idiosyncratic errors.
\cite{Chen2021Quantile} proposed the Quantile Factor Model (QFM) for extracting quantile-dependent factors, and the corresponding estimation procedure at quantile level $\tau = 0.5$ can be considered as a form of robust factor analysis, denoted by QFA (Quantile Factor Analysis) in this paper. \cite{he2020quantile} provided a theoretical analysis of the iterative estimators, whereas \cite{Chen2021Quantile} focused on the theoretical minimizers. Neither of these methods requires moment conditions on the idiosyncratic errors. Compared to QFM, the mean factor model is more suitable for practical financial problems such as portfolio selection, as the Mean-Variance (MV) framework introduced by \cite{Markowitz1952Portofolio} forms the foundation of modern portfolio theory.


A natural question is how to perform robust factor analysis under the mean factor model without making the restrictive elliptical assumption. In this work, we aim to tackle this important question by employing Huber's loss  \citep{Huber1964Robust} as a tool, inspired by the link between the PCA method and the constrained least squares method. It is widely recognized that least squares based methods are susceptible to the effects of outliers, which is why PCA may perform poorly under factor models with heavy-tailed idiosyncratic errors.
One approach to making estimators less susceptible to heavy-tailedness is to substitute the $\ell_2$ loss with a robust alternative. The $\ell_1$ loss achieves robustness against outliers at the cost of asymptotic efficiency under light-tailed distributions. The minimizer of the $\ell_1$-type loss in QFM lacks an analytical closed-form, necessitating the use of an iterative algorithm to locate the stationary points of the optimization problem. This algorithm is sensitive to the starting point and may stuck at a local minimizer.
To achieve a tradeoff among robustness, statistical and numerical efficiencies, we suggest using the Huber loss function as a robust alternative to the $\ell_2$ loss for fitting high-dimensional factor models, and propose two algorithms for different usages of the Huber loss. One is based on minimizing the $\ell_2$-norm-type Huber loss in \eqref{equ:Huberloss}, which is equivalent to performing PCA on the weighted sample covariance matrix and is therefore referred to as ``Huber PCA" (HPCA). The other is based on minimizing the element-wise-type Huber loss in \eqref{equ:elementHuberloss}, which can be solved by an iterative Huber regression algorithm, {hence the name  ``IHR" .}


\begin{figure}[!h]
	\centerline{\includegraphics[width=17cm,height=8cm]{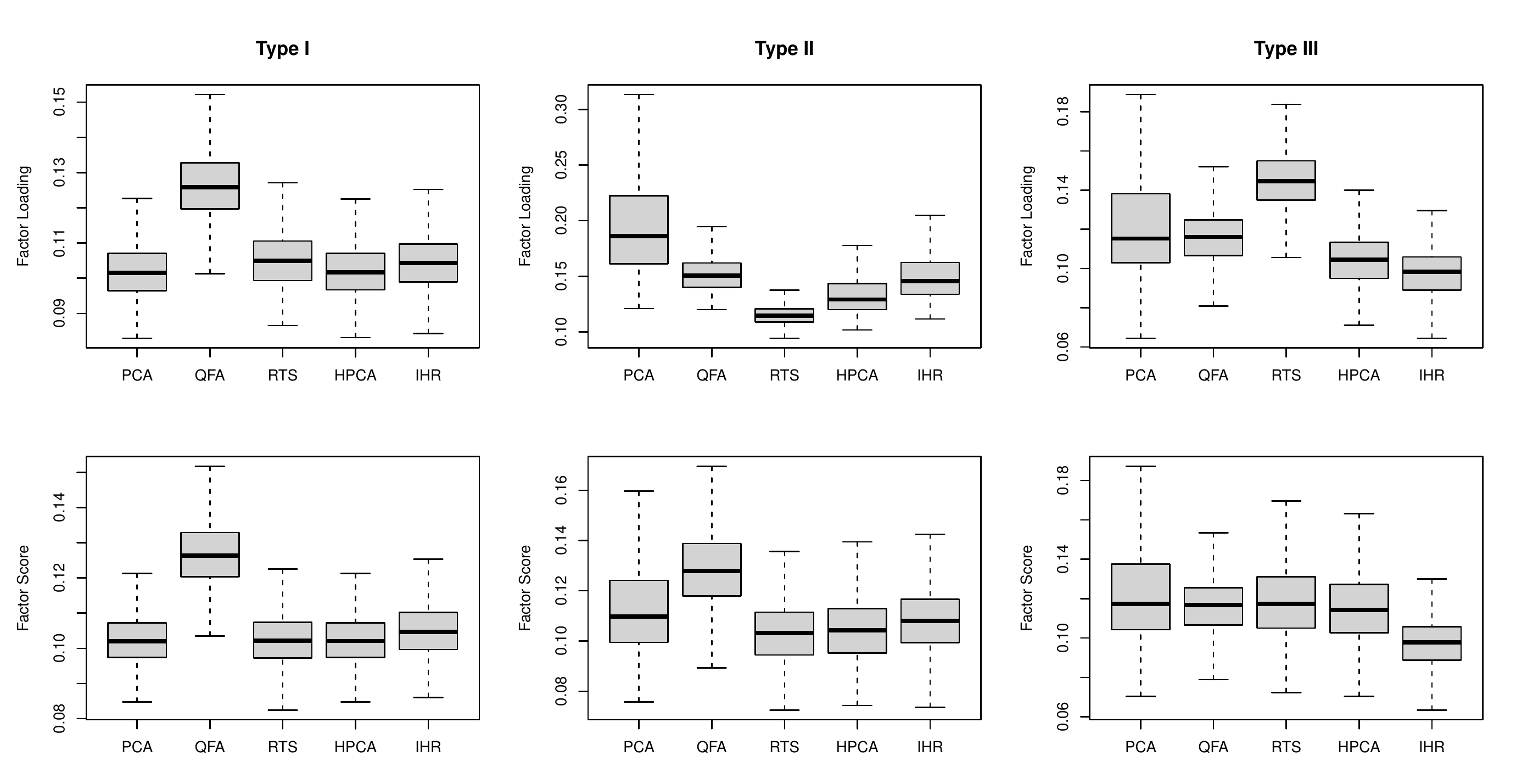}}
	\caption{Boxplots of the estimation errors of the estimated factor loadings and scores by four methods under $T=N=100$. Type \uppercase\expandafter{\romannumeral1}: $(\bbf_t^\top,\bv_t^\top)^\top$ are from multivariate standard Gaussian distribution. Type \uppercase\expandafter{\romannumeral2}: $(\bbf_t^\top,\bv_t^\top)^\top$ are from multivariate standard $t_3$ distribution. Type \uppercase\expandafter{\romannumeral3}: $\bbf_t$ are from multivariate skewed $t_3$ distribution and $\bv_t$ are from $\alpha$-Stable distribution.}
	\label{fig:bp}
\end{figure}

Let us use a synthetic data experiment to demonstrate the sensitivity of PCA and robust methods such as {HPCA/IHR}, QFA, and RTS to the tail properties of idiosyncratic errors. Figure~\ref{fig:bp} depicts boxplots of estimation errors for the factor loading and factor score spaces based on 500 replications. QFA performs the worst in the light-tailed Gaussian settings, due to efficiency loss, but outperforms the other methods as the distribution tails become heavier. This numerical example shows the {HPCA/IHR} is robust to heavy-tailed idiosyncratic errors and performs comparably to RTS when the elliptical assumption holds.
{More importantly, both HPCA and IHR  perform well under non-elliptical distributions,} and thus exhibit high degree of robustness in more general contexts.


The contributions of this work can be summarized in the following aspects: first, our work provides a much-needed addition to the limited literature on robust factor analysis for large-dimensional time series. Secondly, the proposed HPCA and IHR methods are computationally efficient and IHR is proven to achieve the same convergence rates as conventional PCA under a more relaxed second-moment condition on the idiosyncratic errors than the  fourth or even higher moment conditions typically imposed in the literature. We also derived the asymptotic distributions of the theoretical minimizers corresponding to IHR under mild conditions.  Thirdly, we introduce a rank minimization estimator for determining the number of factors, which also complements the scarce literature on robust determination of the factor number and is of independent interest. Finally, we have developed an R package, ``HDRFA", which implements related robust factor analysis methods found in the literature and is available on CRAN. \footnote{\url{%
https://cran.r-project.org/web/packages/HDRFA/index.html}}  Although ``Huber PCA" and ``IHR" have almost identical empirical performances, theoretical analysis of the former is more complicated, and we defer it to future research.

The rest of the paper proceeds as follows. In Section 2, we introduce the HPCA and IHR methods for robust factor analysis and provide estimators of the factor loadings and scores. Section 3 establishes the consistency and the convergence rates of the estimated factor loadings and scores via IHR. Their asymptotic distributions are also derived.  We conduct extensive numerical studies in Section 4. In Section 5, we demonstrate the effectiveness of our proposed methods through real financial data analysis. We discuss possible future research directions in Section 6 and conclude the article. Detailed proofs of the main theorems and technical lemmas are provided in the Appendix.

To end this section, we introduce the following notations that will be used throughout this paper. For any vector $\bmu=(\mu_1,\ldots,\mu_p)^\top \in \RR^p$, let $\|\bmu\|_2=(\sum_{i=1}^p\mu_i^2)^{1/2}$, $\|\bmu\|_\infty=\max_i|\mu_i|$. For a real number $a$, denote  $[a]$ as the largest integer smaller than or equal to $a$. Let $I(\cdot)$ be the indicator function. Let ${\rm diag}(a_1,\ldots,a_p)$ be a $p\times p$ diagonal matrix, whose diagonal entries are $a_1\ldots,a_p$.   For a matrix $\Ab$, let $\mathrm{A}_{ij}$ (or $\mathrm{A}_{i,j}$) be the $ij$ entry of $\Ab$, $\Ab^\top$ the transpose of $\Ab$, ${\rm Tr}(\Ab)$ the trace of $\Ab$, $\text{rank}(\Ab)$ the rank of $\Ab$ and $\text{diag}(\Ab)$ a vector composed of the diagonal elements of $\Ab$. Denote $\lambda_j(\Ab)$ as the $j$-th largest eigenvalue of a nonnegative definitive matrix $\Ab$, and let $\|\Ab\|$ be the spectral norm of matrix $\Ab$ and $\|\Ab\|_F$ be the Frobenius norm of $\Ab$.
{Let $\operatorname{sgn}(a) = 1$ if $a \geq 0$ and $\operatorname{sgn}(a) = -1$ if $a < 0$. For a square matrix $\Ab$, define $\operatorname{sgn}(\Ab)$ as a diagonal matrix with $i$th diagonal elements $\operatorname{sgn}(\mathrm{A}_{ii})$.} For two series of random variables, $X_n$ and $Y_n$, $X_n\asymp Y_n$ means $X_n=O_p(Y_n)$ and $Y_n=O_p(X_n)$. For two random variables (vectors) $\bX$ and $\bY$, $\bX\stackrel{d}{=}\bY$ means the distributions of $\bX$ and $\bY$ are the same. The constants $c, C_1, C_2$ in different lines can be nonidentical.

\section{Methodology}

This section introduces our robust factor analysis methods using the Huber loss. In Section 2.1, we introduce the Huber Principal Component Analysis (HPCA) method and provide a detailed algorithm for it. In Section 2.2, we present the iterative Huber regression algorithm, which is tailored to the element-wise Huber loss.

	\subsection{Huber Principal Component Analysis (HPCA)}\label{sec:hpca}
In this section, we propose the Huber Principal Component Analysis (HPCA) method for robust factor analysis. We first introduce the factor model setup  for a large panel dataset $\{Y_{it}\}_{i\leq N,t\leq T}$.
Factor models decompose the observation $\{Y_{it}\}$ into the sum of two  mutually orthogonal parts: the common component $c_{it}$ and the idiosyncratic component $\epsilon_{it}$, i.e.,
\begin{equation}\label{EFM}
  Y_{it}=c_{it}+\epsilon_{it}=\bl_i^\top \bbf_t+\epsilon_{it}, \hspace{0.5 em} 1\leq  i\leq N, \ \  1\leq  t\leq T, \hspace{0.5em} \text{or in vector form,} \hspace{0.5em} \bY_t=\Lb \bbf_t+\bepsilon_t,
\end{equation}
where $c_{it}=\bl_i^\top \bbf_t$, $\bY_t=(Y_{1t},\ldots,Y_{Nt})^\top$, $\bbf_t\in \RR^r$ are the unobserved factors, $\Lb=(\bl_1,\ldots,\bl_N)^\top$ is the factor loading matrix, and $\bepsilon_t=(\epsilon_{1t},\ldots,\epsilon_{Nt})^\top$ represents the idiosyncratic errors.   For the large-dimensional approximate factor model introduced in  \cite{Chamberlain1983Arbitrage}, the idiosyncratic errors $\bepsilon_t$ are assumed to be cross-sectionally weakly dependent.

It has been shown that for factor models, the principal component analysis (PCA) is equivalent to the least squares \citep{fan2013large}. It's well-known that statistical procedures based on least squares often
behave poorly in the presence of heavy-tailed data. The observed data are often heavy-tailed in areas such as finance and macroeconomics, which motivates us to replace the $\ell_2$ loss function with the Huber loss function \citep{Huber1964Robust}, i.e., we consider the following optimization problem:

\begin{equation}\label{equ:Huberloss}
	\begin{array}{ccc}
		 \min_{\{\mathbf{L},\mathbf{F}_t\}}L_H(\mathbf{L},\bbf_t)=\dfrac{1}{T}\sum\limits_{t=1}^{T}H_{\tau}\Big({\Vert \bY_t - \mathbf{L} \bbf_t  \Vert _2  }\Big),\vspace{2ex} \\
		\text{s.t.}~ \dfrac{1}{N} \mathbf{L}^\top \mathbf{L} = \mathbf{I}_{r},
	\end{array}
\end{equation}
	where the Huber loss $H_{\tau}(x)$ is defined as
	$$ H_{\tau}(x)=
	\begin{cases}
	     \dfrac{1}{2}x^2, & \text{$\vert x \vert \leq \tau$ }, \\
		\tau \vert x \vert - \dfrac{\tau ^2}{2}, & \text{$\vert x \vert > \tau$}.
	\end{cases} $$

	For some fixed time point $t$, the Huber loss $H_{\tau}\Big({\Vert \bY_t - \mathbf{L} \bbf_t  \Vert _2  }\Big)$ can be further expressed as
\begin{equation}\label{equ:huber1}
	H_{\tau}\Big({\Vert \bY_t - \mathbf{L} \bbf_t  \Vert _2  }\Big)=
	\begin{cases}
		 \dfrac{1}{2}\left(\bY_t^\top\bY_t-2\bY_t^\top\Lb\bbf_t+N\bbf_t^\top\bbf_t\right), & {\Vert \mathbf{Y}_t - \mathbf{L} \bbf_t \Vert ^2 _2} \leq \tau^2, \vspace{2ex}  \vspace{0.5em} \\
	\tau\sqrt{\left(\bY_t^\top\bY_t-2\bY_t^\top\Lb\bbf_t+N\bbf_t^\top\bbf_t\right)}-\dfrac{1}{2}\tau^2, & \Vert \mathbf{Y}_t - \mathbf{L} \bbf_t \Vert ^2 _2 > \tau^2.
	\end{cases}
\end{equation}
	For each time point $t$, take ${\partial H_\tau}/{\partial \bbf_t} = 0$, we obtain
	$ \bbf_t=\mathbf{L}^\top \bY_t /N$. Thus by substituting $ \bbf_t=\mathbf{L}^\top \bY_t /N$ in (\ref{equ:huber1}), we further have

			$$
	L_H(\mathbf{L})=
	\begin{cases}
		 \dfrac{1}{2}\left(\bY_t^\top\bY_t-\frac{1}{N}\bY_t^\top\Lb\Lb^\top\bY_t\right), & {\Vert \mathbf{Y}_t - \mathbf{L} \bbf_t \Vert ^2 _2} \leq \tau^2, \vspace{ 0.5em}   \\
	\tau\sqrt{\left(\bY_t^\top\bY_t-\frac{1}{N}\bY_t^\top\Lb\Lb^\top\bY_t\right)}-\frac{1}{2}\tau^2, & \Vert \mathbf{Y}_t - \mathbf{L} \bbf_t \Vert ^2 _2 > \tau^2.
	\end{cases}
		$$
	For the case that $ {\Vert \mathbf{Y}_t - \mathbf{L} \bbf_t \Vert ^2 _2} \leq \tau^2$, we have
	\[
	\dfrac{\partial H_\tau}{\partial \mathbf{L}}=-\dfrac{1}{N}\bY_t\bY_t^\top\Lb,
	\]
	For the case that ${\Vert \mathbf{Y}_t - \mathbf{L} \bbf_t \Vert ^2 _2} >\tau^2$, we have
	\[
	\dfrac{\partial H_\tau}{\partial \mathbf{L}}=-\frac{\tau}{N}\frac{\bY_t\bY_t^\top\Lb}{\sqrt{\left(\bY_t\bY_t^\top-\frac{1}{N}\bY_t^\top\Lb\Lb^\top\bY_t\right)}}.
	\]
	Then the Lagrangian function is introduced as follows:
	 \[\min_{\Lb}\mathcal{L}=L_H(\mathbf{L}) + \text{Tr}\left[ \mathbf{\Theta} (\dfrac{1}{N} \mathbf{L}^\top \mathbf{L} - \mathbf{I}_{r}) \right],
	\]
	where the Lagrangian multipliers $\mathbf{\Theta}$ is a  symmetric matrix. According to the KKT condition, we have
	
\begin{equation}\label{equ:KKT}
\dfrac{\partial \mathcal{L}}{\partial \mathbf{L}} =-\frac{1}{T}\sum_{t=1}^Tw_t\bY_t\bY_t^\top\Lb+\frac{2}{N}\Lb\bTheta =0,
\end{equation}
	where the weights $w_t$ are
	$$
	w_t=\left\{
	\begin{array}{ccc}
		\dfrac{1}{N}, &   {\Vert \mathbf{Y}_t - \mathbf{L} \bbf_t \Vert ^2 _2} \leq \tau^2 \vspace{2ex}\\
		\dfrac{\tau}{N} \dfrac{1}{\sqrt{\left(\bY_t^\top\bY_t-\frac{1}{N}\bY_t^\top\Lb\Lb^\top\bY_t\right)}}, & {\Vert \mathbf{Y}_t - \mathbf{L} \bbf_t \Vert ^2 _2} > \tau^2.
	\end{array}
    \right.
    $$
	By reorganizing the notations, we further have
$\mathbf{L} \mathbf{\Theta} =\bSigma \mathbf{L}$,
where $\bSigma=\sum\limits_{t=1}^{T} w_t^\prime \bY_t \bY_t^\top/T$,	and $w_t^\prime=Nw_t/2$. We denote the first $r$ eigenvectors of $\bSigma$ as $\{\bxi_1,\ldots,\bxi_r\}$ and the corresponding eigenvalues as $\{\theta_1,\ldots,\theta_{r}\}$.
In other words,  we deduce that the leading eigenvectors of $\bSigma$,  $\Lb=(\bxi_1,\ldots,\bxi_r)$ and the corresponding $\bTheta=\text{diag}(\theta_1,\ldots,\theta_{r})$ satisfy the KKT condition in (\ref{equ:KKT}).
Interestingly, minimizing the Huber loss would lead to doing PCA on a weighted sample covariance $\bSigma$ and thus we call our method Huber Principal Component Analysis (HPCA). Note that the weights $w_t^\prime$  depend on the unknown loading matrix $\Lb$ and  the factors $\bbf_t$. In practice, we can first get initial estimators $\mathbf{\widehat{L}}^{(0)}$ and $\{\widehat\bbf_t^{(0)},1\leq t\leq T\}$ from the conventional PCA or the RTS algorithm by \cite{He2020large}. As for the tuning parameter $\tau$, we suggest setting $\tau$ so that half of the observations $\{\bY_t, t=1,\ldots,T\}$ are winsorized, which is justified by extensive simulation studies later. We then compute the weights $\{w_t^{\prime(0)}\}, t=1,\ldots,T$, with the parameter $\tau$  set as the median of $\big\{\Vert \bY_t -  \mathbf{\widehat{L}}^{(0)}\widehat\bbf_t^{(0)}\Vert_2 ,t=1\ldots,T\big\}$. We further define the weighted sample covariance matrix: $$\widehat\bSigma=\sum\limits_{t=1}^{T} w_t^{\prime(0)} \bY_t \bY_t^\top/T,$$	and denotes its leading $r$ eigenvectors as $\{\widehat\bxi_1,\ldots,\widehat\bxi_r\}$. The HPCA estimator for the loading matrix is defined as $\widehat{\Lb}=\sqrt{N}(\widehat\bxi_1,\ldots,\widehat\bxi_r)$ and the HPCA estimators for the factor scores are $\widehat\bbf_t=\widehat\Lb^\top \bY_t /N.$
The detailed algorithm for the HPCA is summarized in Algorithm \ref{alg:HPCA}. An iterative algorithm can be derived as long as we treat the HPCA estimators $\widehat{\Lb}$ and $\widehat\bbf_t$ as new initial estimators and implement the steps 2-4 in Algorithm \ref{alg:HPCA} recursively. Empirical studies show that the iterative algorithm hardly bring any further refinement.

\begin{algorithm}[!h]
		\caption{Huber Principal Component Analysis (HPCA)}\label{alg:HPCA}
		\hspace*{0.02in} {\bf Input:} Data matrices $\{\bY_t, 1\leq t\leq T\}$, the  factor number $r$ \\
		\hspace*{0.02in} {\bf Output:} HPCA estimators for factor loading matrix and factor scores, denoted as  $\mathbf{\widehat{L}}$ and $\{\widehat\bbf_t,1\leq t\leq T\}$.\\
		\begin{algorithmic}[1]
			\State Obtain the initial estimators $\mathbf{\widehat{L}}^{(0)}$ and $\{\widehat\bbf_t^{(0)},1\leq t\leq T\}$ from the {RTS algorithm  by \cite{He2020large}};\\
			Compute the weights $\{w_t^{\prime(0)}\}, t=1,\ldots,T$, with the parameter $\tau$  set as the median of $\big\{\Vert \bY_t -  \mathbf{\widehat{L}}^{(0)}\widehat\bbf_t^{(0)}\Vert_2 ,t=1\ldots,T\big\}$; \\
			 Calculate $\widehat\bSigma=\sum\limits_{t=1}^{T} w_t^{\prime(0)} \bY_t \bY_t^\top/T$,	and denotes its leading $r$ eigenvectors as $\{\widehat\bxi_1,\ldots,\widehat\bxi_r\}$.\\
 Calculate the HPCA estimators $\widehat{\Lb}=\sqrt{N}(\widehat\bxi_1,\ldots,\widehat\bxi_r)$ and $\widehat\bbf_t=\widehat\Lb^\top \bY_t /N.$

		\end{algorithmic}
	\end{algorithm}
\subsection{Iterative Huber Regression Algorithm}\label{sec:ihuberreg}
For the  factor model in (\ref{EFM}), corresponding to the HPCA method, the Huber loss function is applied to the vectors $\{\bY_t-\Lb\bbf_t\}$  and thereby achieving robustness.
Motivated by the quantile factor model \citep{he2020quantile,Chen2021Quantile}, one may also  apply the Huber loss function to each element of $\{\bY_t-\Lb\bbf_t\}$ rather than to the whole vector, i.e., one may consider the following optimization problem:
\beq \label{equ:elementHuberloss}
\begin{gathered}
\min_{\{\Lb,\bbf_t\}}L_{EH}\l(\Lb,\bbf_t\r)=\frac{1}{TN}\sum_{t=1}^T\sum_{i=1}^NH_\tau\l(Y_{it}-\bl_{i}^\top\bbf_t\r),\\
\text{s.t. }\frac{1}{N}\Lb^\top\Lb=\Ib_r.
\end{gathered}
\eeq
There is no explicit solution to this optimization problem. An Iterative Huber Regression algorithm  is proposed to solve the optimization problem, corresponding to the Iterative Quantile Regression (IQR) algorithm in  \cite{he2020quantile} and \cite{Chen2021Quantile}.

For brevity, let $\Lb=\l(\bl_1,\cdots,\bl_N\r)^\top$ and $\Fb=\l(\bbf_1,\cdots,\bbf_T\r)^\top$. Define
$$
L_{i,T}(\bl,\Fb)=\frac{1}{T}\sum_{t=1}^TH_\tau(Y_{it}-\bl^\top\bbf_t) \text{ and } L_{t,N}(\Lb,\bbf)=\frac{1}{N}\sum_{i=1}^NH_\tau(Y_{it}-\bl_i^\top\bbf).
$$
The global minimum of $L_{EH}$ is difficult to locate because this optimization problem is non-convex over both $(\Lb,\Fb)$. But when $\Lb$ is given, $L_{t,N}(\Lb,\bbf)$ is convex with respect to $\bbf$ for each $t$, and similarly, $L_{i,T}(\bl,\Fb)$ is convex with respect to $\bl$ with given $\Fb$ for each $i$. In practice, we first get a normalized initial estimate $\widehat\Fb^{(0)}$ of $\Fb$, for example, from the traditional PCA. Then, $\bl_i^{(0)}$ is estimated as the minimum point of $L_{i,T}(\bl,\widehat\Fb^{(0)})$ for $i=1,\cdots, N$  and then we normalize $\l(\widehat\bl_1^{(0)},\cdots,\widehat\bl_N^{(0)}\r)^\top$ to obtain the initial estimate $\widehat\Lb^{(0)}$ of $\Lb$. For some positive integer $s>0$, the estimate $\widehat\Lb^{(s)}$ is the normalization of $\l(\widehat\bl_1^{(s)},\cdots,\widehat\bl_N^{(s)}\r)^\top$, where  $\widehat\bl_i^{(s)}=\argmin_{\bl}L_{i,T}(\bl,\Fb^{(s-1)})$ and $\widehat\Fb^{(s)}$ is the normalization of $\l(\widehat\bbf_1^{(s)},\cdots,\widehat\bbf_T^{(s)}\r)^\top$, where  $\widehat\bbf_t^{(s)}=\argmin_{\bbf}L_{t,N}(\Lb^{(s-1)},\bbf)$. This process is repeated until the maximum number of iterations is reached or {$\widehat\Fb^{(s)}\widehat\Lb^{(s)\top}$ is close enough to $\widehat\Fb^{(s-1)}\widehat\Lb^{(s-1)\top}$.}
The detailed procedure is summarized in the following Algorithm \ref{alg:IHR}.

\begin{algorithm}[!h]
	\caption{Iterative Huber Regression Algorithm (IHR)}\label{alg:IHR}
	\hspace*{0.02in} {\bf Input:} Data matrices $\{\bY_t, 1\leq t\leq T\}$, the  factor number $r$,  \\
	\hspace*{0.02in} {\bf Output:} Estimators for factor loading matrix and factor scores, denoted as  $\mathbf{\widehat{L}}$ and $\widehat\Fb$.\\
	\begin{algorithmic}[1]
		\State Obtain the normalized initial estimators $\widehat\Fb^{(0)}$ by traditional PCA, compute $\{\widehat\bl_i^{(0)}=\argmin_{\bl}L_{i,T}(\bl,\widehat\Fb^{(0)}),i\leq 1\leq N\}$ and normalize $\l(\widehat\bl_1^{(0)},\cdots,\widehat\bl_N^{(0)}\r)^\top$ as $\widehat\Lb^{(0)}$, $s=0$; \\
		Let $s=s+1$, compute $\{\widehat\bl_i^{(s)}=\argmin_{\bl}L_{i,T}(\bl,\widehat\Fb^{(s-1)}),i\leq 1\leq N\}$ and $\{\widehat\bbf_t^{(s)}=\argmin_{\bbf}L_{t,N}(\widehat\Lb^{(s-1)},\bbf),t\leq 1\leq T\}$, normalize $\l(\widehat\bl_1^{(s)},\cdots,\widehat\bl_N^{(s)}\r)^\top$ as $\widehat\Lb^{(s)}$ and $\l(\widehat\bbf_1^{(s)},\cdots,\widehat\bbf_T^{(s)}\r)^\top$ as $\widehat\Fb^{(s)}$;\\
		Repeat step 2 until the stopping criterion is met, at last let $\mathbf{\widehat{L}}=\widehat\Lb^{(s)}$ and $\mathbf{\widehat{F}}=\widehat\Fb^{(s)}$.
	\end{algorithmic}
\end{algorithm}

{There are many algorithms to do huber regression and for ease of implementation, we resort to R function \texttt{rlm} in the R package \texttt{MASS} which conduct huber  regression by iterative re-weighted least square algorithm with given initial weights, see for example \cite{ven2002} and \cite{huber2011robust} for more details.}

\section{Theoretical Properties}\label{tp}
In this section, we establish the asymptotic properties for the theoretical minimizers of  the element-wise Huber loss function $L_{EH}$, and propose a rank-minimization method to estimate the factor number and derive its consistency.
In Section \ref{sec:asmp}, we first give some mild technical assumptions to derive the theoretical properties. In Section \ref{sec:hp}, we establish the theoretical properties for the estimators of factor loadings and factor scores.
In Section \ref{sec:fn}, we propose the rank-minimization method to estimate the factor number and derive the consistency of the estimator for the factor number.
\subsection{Technical Assumptions} \label{sec:asmp}
We first introduce some notations.
For optimization problem (\ref{equ:elementHuberloss}),{
let $\theta = (\bl_{1}^{\top}, \ldots, \bl_{N}^{\top},\bbf_{1}^{\top}, \ldots, \bbf_{T}^{\top})^{\top}$, and $\theta_0 = (\bl_{01}^{\top}, \ldots, \bl_{0N}^{\top},\bbf_{01}^{\top}, \ldots, \bbf_{0T}^{\top})^{\top}$ be the true parameters. Also
inherit the notations in Section \ref{sec:ihuberreg} that $\Lb = (\bl_{1}, \ldots, \bl_{N} )^{\top}$, $\Lb_0 = (\bl_{01}, \ldots, \bl_{0N} )^{\top}$, $\Fb=(\bbf_{1}, \ldots, \bbf_{T} )^{\top}$ and $\Fb_0=(\bbf_{01}, \ldots, \bbf_{0T} )^{\top}$. To make the factor model identifiable, we propose the following conditions:
\begin{equation}\label{equ:identify}
	\begin{aligned}
		&(i)\ \  \Lb^\top\Lb/N=\Ib_r;\\
		&(ii)\ \ \frac{1}{T}\sum_{t=1}^T\bbf_t\bbf_t^{\top} \text{ is a $r\times r$ positive diagonal matrix with non-increasing diagonal elements. }
	\end{aligned}
\end{equation}
Let $\cL_r,~\cF_r$ be subsets of $\RR^r$ and define
$$
\Theta=\left\{\theta: \bl_i \in \cL_{r}, ~\bbf_t \in \mathcal{F}_r \text { for all } i,~t, ~\{\bl_i\} \text{ and } \{\bbf_t\} \text{ satisfies the identifiability condition (\ref{equ:identify}) }\right\}.
$$
}
We need the following assumptions for further theoretical analysis.

\begin{asmp}\label{asmp:1}
	{\it	
		$\mathcal{L}_{r}$ and $\mathcal{F}_r$ are compact sets
		and $\theta_0 \in \Theta$. The factor vector satisfies
		$$
		\frac{1}{T} \sum_{t=1}^T \bbf_{0t} \bbf_{0t}^{\top} = \boldsymbol{\Sigma}_{0T},
		$$
		where $\boldsymbol{\Sigma}_{0T}$ is a $r \times r$ positive definite diagonal matrix with bounded diagonal elements $\sigma_{0T,1} \geq\ldots\geq\sigma_{0T,r}$, and $\sigma_{0T,j}\rightarrow\sigma_{0j}$ as $T\rightarrow \infty$ for $j=1,\ldots,r$ with $\infty>\sigma_{01}>\ldots>\sigma_{0r}>0$.
	}
\end{asmp}

\begin{asmp}\label{asmp:2}
	{\it	
		Given $\left\{\bbf_{0 t}, t=1, \ldots, T\right\},\left\{\epsilon_{it}\right\}$ are independent across $i$ and $t$.
	}
\end{asmp}

\begin{asmp}\label{asmp:3}
	{\it	
		The conditional distribution functions of $\epsilon_{it} \text{ given } \left\{\bbf_{0 t}\right\}$ have a common support covering an open neighborhood of the origin,  $\EE\left(\left(\epsilon_{it}\right)^{2} \mid\left\{\bbf_{0 t}\right\}\right) < \infty$ for any $i,~t$, and the conditional density function of $\epsilon_{it} \text{ given } \left\{\bbf_{0 t}\right\}$ (written as $f_{it}$) satisfies:\\
		(i) continuous;\\
		(ii) symmetric about the origin;\\
		(iii) for any compact set $\cC\subseteq\RR$ and any $u\in\cC$, there exists $0<\underline f<\infty$(depending on $\cC$), such that $\underline f<f_{it}(u)$ for all $i,~t$.
	}
\end{asmp}

Assumption \ref{asmp:1} is the standard strong factor assumption and is common in the related literature. We assume that $\mathbf{\Sigma}_{0T}$ are diagonal matrices with different diagonal elements for further identifiability. Assumption \ref{asmp:2} assumes that the idiosyncratic errors are independent and identically distributed given the factors $\{\bbf_{0t}\}$, but  may not be {\it i.i.d} unconditionally. Assumption \ref{asmp:3} exerts some conditions on the conditional distribution of the idiosyncratic errors $\{\epsilon_{it}\}$ given the factors $\{\bbf_{0t}\}$. The assumption of the symmetry of its conditional density function ensures that $\EE\left(H_\tau^{(1)}(\epsilon_{it})\mid\left\{\bbf_{0 t}\right\}\right)=0$, where $H_\tau^{(1)}(\cdot)$ is the derivative function of $H_\tau(\cdot)$. In addition, we assume that $\{\epsilon_{it}\}$ have bounded second moments, which relaxes the sub-Gaussian condition or the fourth (or even eighth) moment conditions for the traditional PCA in the literature \citep{Bai2003Inferential}.

Further denote $$\bPhi_i=\lim_{T\rightarrow\infty}\sum_{t=1}^T\EE H_\tau^{(2)}(Y_{it}-\bl_{0i}^\top\bbf_{0t})\bbf_{0t}\bbf_{0t}^{\top}/T \text{ and } \bPsi_t=\lim_{N\rightarrow\infty}\sum_{i=1}^N\EE H_\tau^{(2)}(Y_{it}-\bl_{0i}^\top\bbf_{0t})\bl_{0i}\bl_{0i}^{\top}/N, $$
where $H_\tau^{(2)}(u)$ stands for $(\partial/\partial u)^2H_\tau(u)$. We also need the following assumption hold  to derive the asymptotic distributions for the estimates of factor loadings and scores.
\begin{asmp}\label{asmp:4}
	{\it	
		(1) $\bPhi_i>0$ and $\bPsi_t>0$ for all $i,~t$;
		(2) For any compact set $\cC\subseteq\RR$ and any $u\in\cC$, there exists $0<\overline f<\infty$(depending on $\cC$), such that $\underline f<f_{it}(u)<\overline f$ for all $i,~t$;
		(3) $N \asymp T$ as $T,~N\rightarrow\infty$.
	}
\end{asmp}

\subsection{Asymptotic properties of the estimators for factor loadings and scores }\label{sec:hp}

In this section, we establish the convergence rate of the theoretical minimizers of  (\ref{equ:elementHuberloss}). In detail, we present the asymptotic properties of the theoretical minimizers $\widehat{\theta}$, defined as
$$
\widehat{\theta}=\left(\widehat\bl_{1}^{\top}, \ldots, \widehat\bl_{N}^{\top},\widehat{\bbf}_{1}, \ldots, \widehat{\bbf}_{T} \right)^\top=\underset{\theta \in \Theta}{\arg \min } \frac{1}{TN} \sum_{t=1}^T\sum_{i=1}^NH_{\tau}\l(Y_{it}-\bl_i^\top\bbf_t\r).
$$

The following theorem presents the convergence rate of the theoretical minimizers $\widehat{\theta}$ of the element-wise Huber loss function in (\ref{equ:elementHuberloss}).
\begin{theorem}\label{th:1}
	Let $\widehat{\mathbf{S}}=\operatorname{sgn}\left(\frac{1}{T} \sum_{t=1}^T\left(\widehat{\bbf}_{t} \bbf_{0t}^{\top}\right)\right)$ and $\tau$ is a fixed positive constant. Then, under Assumptions \ref{asmp:1}-\ref{asmp:3}, we have
	$$
	\frac{1}{N}\left\|\widehat{\Lb}-\Lb_0 \widehat{\mathbf{S}}\right\|_F^2=O_p\left(\frac{1}{L}\right), \text { and } \frac{1}{T}\left\|\widehat{\Fb}-\Fb_0\widehat{\mathbf{S}}\right\|_F^2=O_p\left(\frac{1}{L}\right) \text {, }
	$$
	where $L=\min \left\{N,T\right\}$.
\end{theorem}	
The existence of $\widehat{\mathbf{S}}$ is due to the sign indeterminacy of the factors and loadings estimation, where $\widehat{\mathbf{S}}$ is a diagonal matrix with diagonal elements $\{1, -1\}$. Assume we only require $\Lb_0^\top\Lb_0/T=\bSigma_L$ and $\Fb_0^\top\Fb_0/N=\bSigma_F$  for identifiability, then by letting $\Hb=\bSigma_L^{-1/2}\bGamma$ with $\bGamma$ being the matrix of the eigenvectors of $\bSigma_L^{1/2}\bSigma_F\bSigma_L^{1/2}$, it holds that
$$
\frac{1}{N}\left\|\widehat{\Lb}-\Lb_0\Hb \widehat{\mathbf{S}}\right\|_F^2=O_p\left(\frac{1}{L}\right), \text { and } \frac{1}{T}\left\|\widehat{\Fb}-\Fb_0\l(\Hb^\top\r)^{-1}\widehat{\mathbf{S}}\right\|_F^2=O_p\left(\frac{1}{L}\right) \text {, }
$$
as $\Lb_0\Hb$ and $\Fb_0\l(\Hb^\top\r)^{-1}$ satisfies the identifiability condition in (\ref{equ:identify}). The convergence rate derived here is the same as  that of the traditional PCA estimators by \cite{Bai2003Inferential} and that of the Quantile factor model by \cite{Chen2021Quantile}. However, \cite{Bai2003Inferential} assumes the finite eighth-moment condition on the idiosyncratic errors, \cite{Chen2021Quantile} does not require any moment condition, while our results rely on  the finite second-moment condition on the idiosyncratic errors.

The next theorem presents the asymptotic distributions of the element-wise Huber loss estimates of the factor loadings and scores:
\begin{theorem}\label{th:3}
	Let $\widehat{\mathbf{S}}=\operatorname{sgn}\left(\frac{1}{T} \sum_{t=1}^T\left(\widehat{\bbf}_{t} \bbf_{0t}^{\top}\right)\right)$. Then under Assumptions \ref{asmp:1}-\ref{asmp:4}, we have that
	$$\sqrt{T}\left(\widehat{\bl}_i-\widehat{\Sbb} \bl_{0 i}\right) \stackrel{d}{\rightarrow} \mathcal{N}\left(\mathbf0, \bPhi_i^{-1}\bSigma_{L,i}\bPhi_i^{-1} \right),
	$$ and
	$$\quad \sqrt{N}\left(\widehat{\bbf}_t-\widehat{\Sbb} \bbf_{0 t}\right) \stackrel{d}{\rightarrow} \mathcal{N}\left(\mathbf0, \bPsi_t^{-1}\bSigma_{F,t}\bPsi_t^{-1}\right),
	$$
	for each $i$ and $t$, where $$
	\begin{gathered}\bSigma_{L,i}=\lim_{T\rightarrow\infty}\sum_{t=1}^T\l(\int_{-\infty}^{\infty}\min\{u^2,\tau^2\}f_{it}(u)du\r) \bbf_{0t}\bbf_{0t}^\top/T, \\
		\text{ and }\bSigma_{F,t}=\lim_{N\rightarrow\infty}\sum_{i=1}^N\l(\int_{-\infty}^{\infty}\min\{u^2,\tau^2\}f_{it}(u)du\r) \bl_{0i}\bl_{0i}^\top/N.
	\end{gathered}$$
\end{theorem}	

Similarly, assume we only require  $\Lb_0^\top\Lb_0/T=\bSigma_L$ and $\Fb_0^\top\Fb_0/N=\bSigma_F$ for identifiability, it holds that
$$
\begin{aligned}
	& \sqrt{T}\left(\widehat{\bl}_i-\widehat{\Sbb}\Hb^\top \bl_{0 i}\right) \stackrel{d}{\rightarrow} \mathcal{N}\left(\mathbf0, \Hb^{\top}\bPhi_i^{-1}\bSigma_{L,i}\bPhi_i^{-1}\Hb \right),  \\
	& \sqrt{N}\left(\widehat{\bbf}_t-\widehat{\Sbb}\Hb^{-1} \bbf_{0 t}\right) \stackrel{d}{\rightarrow} \mathcal{N}\left(\mathbf0, \Hb^{-1}\bPsi_t^{-1}\bSigma_{F,t}\bPsi_t^{-1} \l(\Hb^{-1}\r)^\top\right),
\end{aligned}
$$
where $\Hb=\bSigma_L^{-1/2}\bGamma$ with $\bGamma$  being the matrix composed of the eigenvectors of $\bSigma_L^{1/2}\bSigma_F\bSigma_L^{1/2}$. To our knowledge, this is the first time that inference for factor loadings and scores are derived under the relaxed finite second moment condition.


\subsection{Rank minimization for the factor number and its Consistency }\label{sec:fn}
In this section, we propose a rank minimization method to estimate the factor number. In detail,
let $k$ be a positive integer that is larger than $r$, i.e.,  $k>r$, $\mathcal{L}_{k}$ and $\mathcal{F}_k$ are compact sets in $\RR^k$. Assume that \[
\left(\bl_{0i}^\top~\mathbf{0}_{1\times (k-r)}^\top\right)^\top \in \cL_k \ \text{and} \ \left(\bbf_{0t}^\top ~\mathbf{0}_{1\times (k-r)}^\top\right)^\top \in \cF_k \ \text{for all} \ i \text{ and} \ t.
\]
Let $\theta^k=\left(\bl_1^{k\top},\ldots,\bl_N^{k\top},\bbf_1^{k\top},\ldots,\bbf_T^{k\top}\right)^\top$,  $\Lb^k=(\bl_1^k,\ldots,\bl_N^k)^\top$ and $\Fb^k=(\bbf_1^k,\ldots,\bbf_T^k)^\top$.

We also assume the following identifiability condition hold:
\beq\label{equ:identify_k}
\begin{aligned}
	&(i)~(\Lb^k)^\top\Lb^k/N=\Ib_k ;\\
	&(ii)~\frac{1}{T}\sum_{t=1}^T\bbf^k_t\bbf_t^{k\top}= \bSigma^k \text{ is a $k \times k$ positive diagonal matrix with non-increasing diagonal elements.}
\end{aligned}
\eeq
Assume that $\widehat\theta^k$ is the theoretical minimizer in the parameter space $\Theta^k$, i.e.,
 $$\widehat\theta^k=\argmin_{\theta^k\in \Theta^k}\frac{1}{TN}\sum_{t=1}^T\sum_{i=1}^NH_{\tau}(Y_{it}-\bl_i^{k\top}\bbf_t^k),$$
where $\Theta^k=\left\{\theta^k: \bl_i^k \in \cL_{k}, ~\bbf_t^k \in \mathcal{F}_k \text { for all } i,~t, ~\{\bl_i^k\} \text{ and } \{\bbf_t^k\} \text{ satisfy the identifiability condition (\ref{equ:identify_k}) }\right\}$, and
further denote $\sum_{t=1}^T \widehat{\bbf}_{t}^k (\widehat{\bbf}_{t}^k)^{\top}/T= \widehat{\boldsymbol{\Sigma}}^k=\diag(\widehat\sigma^k_{T,1},\ldots,\widehat\sigma^k_{T,k})$.

Then the rank minimization estimator of the  factor number $r$ is defined as $$\widehat{r}=\sum_{j=1}^kI\{\widehat\sigma^k_{T,j}>P\},$$
where $P\rightarrow 0$ as $T,N\rightarrow\infty$. In other words, the estimator for the factor number is the number of  the diagonal elements of $\sum_{t=1}^T \widehat{\bbf}_{t}^k (\widehat{\bbf}_{t}^k)^{\top}/T$ which are greater than  the threshold $P$.

The following theorem establishes the consistency for the rank minimization estimator.
\begin{theorem}\label{th:2}
	Under Assumptions \ref{asmp:1}-\ref{asmp:3} ,
	$\PP\left(\widehat{r}=r\right) \rightarrow 1$ as $N, T \rightarrow \infty$ if $k>r, P \rightarrow 0$, and $P L \rightarrow \infty$, where $L=\min \left\{N,T\right\}$.
\end{theorem}	

Theorem \ref{th:2} indicates that if we choose an integer $k$ greater than the true factor number $r$, we can get a consistent estimate. In fact, for the first $r$ diagonal elements of $\sum_{t=1}^T \widehat{\bbf}_{t}^k (\widehat{\bbf}_{t}^k)^{\top}/T$, we have that $\widehat\sigma^k_{T,j}\rightarrow\sigma_{0j}$, where $\sigma_{0j}$ is bounded. For $j>r$, it holds that $\widehat\sigma_{T,j}^k=O_p(1/L)$, which indicates that the first $r$ diagonal elements of $\sum_{t=1}^T \widehat{\bbf}_{t}^k (\widehat{\bbf}_{t}^k)^{\top}/T$ are well separated.

From the proof of the theorem, it can also be seen that if we overestimate the factor number $r$,  Theorem \ref{th:1} still holds for the first $r$ columns of the corresponding estimators $\{\widehat\Lb^k, \widehat\Fb^k\}$. But when we underestimate the number of factors, the estimate would not be consistent. This is also consistent with the findings in related literature.
As for the choice of threshold parameter $P$, by the Theorem \ref{th:2},   $P$ needs to satisfy that
$$
P\rightarrow 0 \text{ and } PL\rightarrow \infty \text{ as } T,N\rightarrow \infty.
$$
In practice, we set $P=L^{-1/3}$ suggested by \cite{Chen2021Quantile} and find that this works well in the following simulation studies.

\section{Simulation Study}\label{ss}
In this section, we conduct thorough  simulation studies to compare the HPCA and {IHR} with the Robust Two-Step (RTS) estimator by \cite{He2020large}, the  Quantile Factor Analysis (QFA) with $\tau=0.5$ by \cite{Chen2021Quantile} and the conventional PCA method.  The initialization of the factors for the QFA algorithm is randomly drawn from $\cN(0,1)$. As a byproduct, we make an \texttt{R} package called \texttt{HDRFA} to implement all the factor analysis methods mentioned above. All the simulation results reported hereafter are based on $500$ replications. In Section \ref{sec:rsf}, we compare different methods in terms of estimating the loading/factor spaces and the recovery of common components.

\subsection{Estimation of the Loading/Factor Spaces and Common Components}\label{sec:rsf}
We use similar data-generating mechanisms as in \cite{ahn2013eigenvalue} and \cite{He2020large}.  In detail, we generate the synthetic dataset from the following model:
\[
	 Y_{it}=\sum\limits_{j=1}^{r}L_{ij}f_{jt}+\sqrt{\theta}u_{it},\quad u_{it}=\sqrt{\frac{1-\rho^2}{1+2J\beta^2}}e_{it},  \]
\[
e_{it}=\rho e_{i,t-1}+(1-\beta)v_{it}+\sum_{l={\rm max}\{i-J,1\}}^{{\rm min}\{i+J,p\}}\beta v_{lt}, \ \ t=1,\ldots,T, \ \ i=1,\ldots,N,
\]
	where  $\bbf_t=(f_{1t},\ldots,f_{rt})^\top$ and $\bv_t=(v_{1t},\ldots,v_{Nt})^\top$ are  generated from the scenarios as described below. We let $L_{ij}$ be independently drawn from the standard normal distribution $\cN(0,1)$. The parameter $\theta$ controls the SNR (signal-to-noise ratio), $\rho$ controls the serial correlations of idiosyncratic errors, and $\beta$ and $J$ control the cross-sectional correlations.

We consider the following data-generating scenarios in the simulation studies.

\vspace{0.5em}

\textbf {Scenario A} Set $r=3,\theta=1,\rho=\beta=J=0$, $(N,T)=\big\{(100,100),(200,200)\big\}$, $(\bbf_t^\top,\bv_t^\top)^\top$ are generated in the following ways:

Case \uppercase\expandafter{\romannumeral1}: $(\bbf_t^\top,\bv_t^\top)^\top$ are \emph{i.i.d.} random samples from multivariate Gaussian distributions $\mathcal{N}(\zero,\Ib_{p+m})$;

Case \uppercase\expandafter{\romannumeral2}: $(\bbf_t^\top,\bv_t^\top)^\top$ are random samples from  multivariate centralized $t_3$ distributions $t_{3}(\zero,\Ib_{p+m})$;

Case \uppercase\expandafter{\romannumeral3}: $\bbf_t$ are \emph{i.i.d.} random samples from multivariate Gaussian distributions $\mathcal{N}(\zero,\Ib_{m})$ while $\bv_t$ are \emph{i.i.d.} random samples from multivariate centralized $t_3$ distributions $t_{3}(\zero,\Ib_{p})$;

Case \uppercase\expandafter{\romannumeral4}: The elements of $(\bbf_t^\top,\bv_t^\top)^\top$ are \emph{i.i.d.} random samples from symmetric $\alpha$-Stable distribution $S_{\alpha}(\kappa, \gamma, \delta)$ with skewness parameter $\kappa=0$, scale parameter $\gamma=1$ and location parameter $\delta=0$, $\alpha=1.9$;

Case \uppercase\expandafter{\romannumeral5}: $\bbf_t$ are \emph{i.i.d.} random samples from multivariate skewed $t_3$ distribution $\cS\cT_{N+r}(\xi=\bm{0},\bOmega=\Ib,\alpha=20,\nu=3)$ by \texttt{R} package \texttt{fMultivar} while the elements of $\bv_t$ are \emph{i.i.d.} random samples from symmetric $\alpha$-Stable distribution $S_{1.9}(0,1,0)$.

\textbf {Scenario B} Set $r=3,\theta=1,\rho=0.5,\beta=0.2, J={\rm max}\{10, p/20\}$, $(N,T)=\big\{(100,100),(200,200)\big\}$ and $(\bbf_t^\top,\bv_t^\top)$ are generated in the same ways as in Scenario A.

In Scenario A, the first two cases  satisfy the moment assumption on idiosyncratic errors for  all robust methods as $(\bbf_t^\top,\bv_t^\top)^\top$ are jointly from an elliptical distribution with finite second moments.
For the remaining three cases,  the assumption of elliptical distribution is not satisfied for the RTS method. Furthermore, for $\alpha$ stable distribution, when $\alpha<2$, its variance is infinite \citep{weron2005computer}, thereby we can illustrate how sensitive  different methods are to the finite second-moment condition. In Scenario A, there are no serial and cross-sectional correlations in the idiosyncratic errors. In Scenario B, both serially and cross-sectionally correlated errors exist.

We consider the following metrics used in \cite{He2020large} to evaluate the empirical performance: the \textbf{ME}dian of the normalized estimation \textbf{E}rrors for \textbf{C}ommon \textbf{C}omponents in terms of the matrix Frobenius norm, denoted as MEE-CC; the \textbf{AV}erage estimation \textbf{E}rror for the \textbf{F}actor \textbf{L}oading matrices, denoted as AVE-FL; and the \textbf{AV}erage estimation \textbf{E}rror for the \textbf{F}actor \textbf{S}core matrices, denoted as AVE-FS. In detail,
\[
\text{MEE-CC}=\text{median}\left\{\|\hat\Lb_m\hat{\Fb}_m^\top-\Lb\Fb^\top\|_F^2/\|\Lb\Fb^\top\|_F^2,m=1,\ldots,M\right\},
\]
\[
\text{AVE-FL}=\frac{1}{M}\sum_{m=1}^M \cD(\hat \Lb_m,\Lb), \ \ \text{AVE-FS}=\frac{1}{M}\sum_{m=1}^M \cD(\hat \Fb_m,\Fb),
\]
where $M$ is the repetition number set as 500, $\hat \Lb_m$ and $\hat \Fb_m$ are the estimators for the loadings and factor scores at the $m$th replication, and for {two orthogonal matrices} $\Ob_1$ and $\Ob_2$ of sizes $p\times q_1$ and $p\times q_2$,

\begin{equation}\label{equ:DIS}
\cD(\Ob_1,\Ob_2)=\bigg(1-\frac{1}{\max{(q_1,q_2)}}\text{Tr}\Big(\Ob_1\Ob_1^\top\Ob_2\Ob_2^\top\Big)\bigg)^{1/2}.
\end{equation}

We can see from (\ref{equ:DIS}) that $\cD(\Ob_1,\Ob_2)$, in essence, measures the distance between the  spaces spanned by the columns of  $\Ob_1$ and $\Ob_2$, and its value is between 0 and 1. If the spaces spanned by the columns of $\Ob_1$ and $\Ob_2$  are the same, $\cD(\Ob_1,\Ob_2)=0$ and $\cD(\Ob_1,\Ob_2)=1$ if and only if the spaces are orthogonal. It is well-known that the factor loading matrix and factor score matrix are not separately identifiable, but the spaces spanned by their columns are identifiable, and thus $\cD(\cdot,\cdot)$ particularly suits to quantify  the accuracy of the estimators for factor loading/score matrices.

\begin{table}[!h]
  \caption{Simulation results for Scenario A, the values in the parentheses are the interquartile ranges for MEE-CC and standard deviations for AVE-FL and AVE-FS.}
  \label{tab:1}
  \renewcommand{\arraystretch}{1.3}
  \centering
  \selectfont
  \begin{threeparttable}
   \scalebox{0.9}{ \begin{tabular*}{16cm}{ccccccccccccccccccccccccccccc}
\toprule[2pt]
&\multirow{2}{*}{Type}&\multirow{2}{*}{Method}  &\multicolumn{3}{c}{$(p,n)=(100,100)$}&\multicolumn{3}{c}{$(p,n)=(200,200)$} \cr
\cmidrule(lr){4-6} \cmidrule(lr){7-9}
&&                 &$\text{MEE\_CC}$     &$\text{AVE\_FL}$      &$\text{AVE\_FS}$      &$\text{MEE\_CC}$      &$\text{AVE\_FL}$      &$\text{AVE\_FS}$  \\
\midrule[1pt]
&\multirow{5}{2cm}{Case \uppercase\expandafter{\romannumeral1}}
      &PCA     &0.02(0.00)&0.10(0.01)&0.10(0.01)&0.01(0.00)&0.07(0.00)&0.07(0.00)\\
&     &QFA     &0.03(1.99)&0.13(0.01)&0.13(0.01)&0.02(2.32)&0.09(0.00)&0.09(0.00)\\
&     &RTS     &0.02(0.00)&0.11(0.01)&0.10(0.01)&0.01(0.00)&0.07(0.00)&0.07(0.00)\\
&     &HPCA    &0.02(0.00)&0.10(0.01)&0.10(0.01)&0.01(0.00)&0.07(0.00)&0.07(0.00)\\
&     &IHR     &0.02(0.00)&0.10(0.01)&0.10(0.01)&0.01(0.00)&0.07(0.00)&0.07(0.00)\\
\cmidrule(lr){3-9}

&\multirow{5}{2cm}{Case \uppercase\expandafter{\romannumeral2}}
      &PCA     &0.05(0.03)&0.20(0.06)&0.12(0.04)&0.03(0.02)&0.16(0.05)&0.08(0.02)\\
&     &QFA     &0.04(1.45)&0.15(0.02)&0.13(0.02)&0.02(1.71)&0.11(0.02)&0.09(0.01)\\
&     &RTS     &0.02(0.01)&0.12(0.01)&0.10(0.01)&0.01(0.00)&0.08(0.00)&0.07(0.01)\\
&     &HPCA    &0.03(0.01)&0.14(0.03)&0.11(0.02)&0.01(0.00)&0.09(0.02)&0.07(0.01)\\
&     &IHR     &0.03(0.01)&0.15(0.04)&0.11(0.03)&0.02(0.01)&0.11(0.02)&0.08(0.01)\\
\cmidrule(lr){3-9}

&\multirow{5}{2cm}{Case \uppercase\expandafter{\romannumeral3}}
      &PCA     &0.08(0.14)&0.27(0.16)&0.27(0.16)&0.03(0.04)&0.21(0.17)&0.21(0.16)\\
&     &QFA     &0.08(1.87)&0.15(0.04)&0.22(0.06)&0.04(2.31)&0.10(0.02)&0.16(0.04)\\
&     &RTS     &0.05(0.02)&0.13(0.01)&0.18(0.04)&0.02(0.01)&0.09(0.01)&0.12(0.03)\\
&     &HPCA    &0.04(0.02)&0.13(0.03)&0.18(0.05)&0.02(0.01)&0.09(0.02)&0.12(0.03)\\
&     &IHR   &0.05(0.02)&0.18(0.13)&0.21(0.13)&0.02(0.01)&0.14(0.14)&0.17(0.13)\\
\cmidrule(lr){3-9}

&\multirow{5}{2cm}{Case \uppercase\expandafter{\romannumeral4}}
      &PCA     &0.02(0.01)&0.16(0.13)&0.16(0.13)&0.01(0.01)&0.13(0.13)&0.12(0.13)\\
&     &QFA     &0.03(0.96)&0.12(0.01)&0.12(0.01)&0.01(2.12)&0.08(0.01)&0.08(0.01)\\
&     &RTS     &0.02(0.01)&0.11(0.01)&0.12(0.04)&0.01(0.00)&0.08(0.00)&0.09(0.02)\\
&     &HPCA    &0.02(0.01)&0.11(0.02)&0.12(0.04)&0.01(0.00)&0.07(0.01)&0.09(0.02)\\
&     &IHR   &0.02(0.01)&0.13(0.12)&0.13(0.12)&0.01(0.00)&0.09(0.10)&0.09(0.10)\\
\cmidrule(lr){3-9}

&\multirow{5}{2cm}{Case \uppercase\expandafter{\romannumeral5}}
      &PCA     &0.02(0.02)&0.15(0.12)&0.15(0.12)&0.01(0.01)&0.14(0.15)&0.14(0.15)\\
&     &QFA     &0.03(1.22)&0.12(0.01)&0.12(0.01)&0.01(1.20)&0.08(0.01)&0.08(0.01)\\
&     &RTS     &0.03(0.01)&0.15(0.02)&0.12(0.04)&0.02(0.00)&0.10(0.01)&0.09(0.04)\\
&     &HPCA    &0.02(0.01)&0.11(0.04)&0.12(0.05)&0.01(0.00)&0.08(0.03)&0.09(0.05)\\
&     &IHR   &0.02(0.01)&0.12(0.10)&0.12(0.10)&0.01(0.00)&0.10(0.12)&0.10(0.12)\\
\bottomrule[2pt]
  \end{tabular*}}
  \end{threeparttable}
\end{table}

\begin{table}[!h]
  \caption{Simulation results for Scenario B, the values in the parentheses are the interquartile ranges for MEE-CC and standard deviations for AVE-FL, AVE-FS.}
  \label{tab:2}
  \renewcommand{\arraystretch}{1.3}
  \centering
  \selectfont
  \begin{threeparttable}
   \scalebox{0.9}{ \begin{tabular*}{16cm}{ccccccccccccccccccccccccccccc}
\toprule[2pt]
&\multirow{2}{*}{Type}&\multirow{2}{*}{Method}  &\multicolumn{3}{c}{$(p,n)=(100,100)$}&\multicolumn{3}{c}{$(p,n)=(200,200)$} \cr
\cmidrule(lr){4-6} \cmidrule(lr){7-9}
&&                 &$\text{MEE\_CC}$     &$\text{AVE\_FL}$      &$\text{AVE\_FS}$      &$\text{MEE\_CC}$      &$\text{AVE\_FL}$      &$\text{AVE\_FS}$  \\
\midrule[1pt]
&\multirow{5}{2cm}{$\mathcal{N}(\zero,\Ib_{p+m})$}
      &PCA     &0.02(0.01)&0.11(0.02)&0.11(0.02)&0.01(0.00)&0.07(0.01)&0.07(0.01)\\
&     &QFA     &0.04(2.09)&0.14(0.02)&0.14(0.02)&0.02(2.24)&0.09(0.01)&0.09(0.01)\\
&     &RTS     &0.02(0.01)&0.11(0.02)&0.11(0.02)&0.01(0.00)&0.08(0.01)&0.07(0.01)\\
&     &HPCA    &0.02(0.01)&0.11(0.02)&0.11(0.02)&0.01(0.00)&0.07(0.01)&0.07(0.01)\\
&     &IHR   &0.02(0.01)&0.12(0.02)&0.12(0.02)&0.01(0.00)&0.08(0.01)&0.08(0.01)\\
\cmidrule(lr){3-9}

&\multirow{5}{2cm}{$t_{3}(\zero,\Ib_{p+m})$}
      &PCA     &0.05(0.03)&0.20(0.08)&0.14(0.07)&0.03(0.02)&0.15(0.06)&0.09(0.05)\\
&     &QFA     &0.06(1.66)&0.17(0.04)&0.16(0.04)&0.03(1.91)&0.12(0.03)&0.10(0.03)\\
&     &RTS     &0.03(0.01)&0.13(0.02)&0.12(0.03)&0.01(0.00)&0.09(0.01)&0.08(0.01)\\
&     &HPCA    &0.03(0.02)&0.15(0.04)&0.12(0.04)&0.01(0.00)&0.10(0.03)&0.08(0.03)\\
&     &IHR     &0.04(0.02)&0.17(0.05)&0.13(0.05)&0.02(0.01)&0.11(0.04)&0.08(0.03)\\
\cmidrule(lr){3-9}

&\multirow{5}{2cm}{$\mathcal{N}(\zero,\Ib_{p+m})$ \& $\alpha$-stable}
      &PCA     &0.16(0.37)&0.35(0.16)&0.35(0.16)&0.05(0.08)&0.23(0.17)&0.23(0.17)\\
&     &QFA     &0.16(2.03)&0.23(0.09)&0.29(0.10)&0.05(2.31)&0.13(0.04)&0.18(0.06)\\
&     &RTS     &0.08(0.05)&0.19(0.05)&0.23(0.08)&0.03(0.01)&0.11(0.01)&0.14(0.03)\\
&     &HPCA    &0.07(0.05)&0.19(0.08)&0.23(0.09)&0.03(0.01)&0.11(0.03)&0.14(0.05)\\
&     &IHR     &0.09(0.09)&0.24(0.14)&0.28(0.13)&0.03(0.02)&0.15(0.13)&0.18(0.13)\\
\cmidrule(lr){3-9}

&\multirow{5}{2cm}{$\alpha$-stable}
      &PCA     &0.03(0.02)&0.18(0.13)&0.18(0.13)&0.02(0.01)&0.13(0.13)&0.13(0.13)\\
&     &QFA     &0.05(1.79)&0.14(0.02)&0.15(0.02)&0.02(1.49)&0.09(0.01)&0.09(0.01)\\
&     &RTS     &0.03(0.02)&0.13(0.02)&0.14(0.06)&0.01(0.00)&0.09(0.01)&0.09(0.03)\\
&     &HPCA    &0.03(0.01)&0.13(0.06)&0.14(0.07)&0.01(0.00)&0.08(0.03)&0.09(0.04)\\
&     &IHR     &0.03(0.01)&0.14(0.11)&0.14(0.11)&0.01(0.00)&0.10(0.10)&0.10(0.10)\\
\cmidrule(lr){3-9}

&\multirow{5}{2cm}{Skewed $t_{3}$\quad\& $\alpha$-stable}
      &PCA     &0.03(0.03)&0.19(0.12)&0.17(0.12)&0.02(0.01)&0.16(0.15)&0.15(0.15)\\
&     &QFA     &0.05(1.46)&0.15(0.04)&0.15(0.03)&0.02(1.68)&0.10(0.01)&0.09(0.01)\\
&     &RTS     &0.06(0.05)&0.24(0.10)&0.19(0.10)&0.02(0.01)&0.13(0.03)&0.11(0.06)\\
&     &HPCA    &0.03(0.02)&0.14(0.06)&0.14(0.06)&0.01(0.01)&0.10(0.07)&0.10(0.07)\\
&     &IHR     &0.03(0.02)&0.15(0.09)&0.14(0.09)&0.01(0.01)&0.11(0.11)&0.10(0.11)\\
\bottomrule[2pt]
  \end{tabular*}}
  \end{threeparttable}
\end{table}

The simulation results for Scenario A and Scenario B are reported in Table \ref{tab:1} and Table \ref{tab:2}, respectively. There are four main takeaways from Table \ref{tab:1} for Scenario A. {Overall, HPCA and IHR have comparable performance} and outperform other methods in heavy-tailed cases, while performing as well as PCA and RTS in Gaussian settings. Next, {the HPCA/IHR performs} slightly better than the RTS method in cases \uppercase\expandafter{\romannumeral3}-\uppercase\expandafter{\romannumeral5}, as the elliptical assumption required by RTS is violated in these cases. Besides, the QFA method performs slightly worse in cases \uppercase\expandafter{\romannumeral1}-\uppercase\expandafter{\romannumeral3} compared with HPCA/IHR, which implies that HPCA/IHR is the first choice for robust analysis as long as the second moments of the idiosyncratic errors exist. Finally, the performances of all methods grow better as $(T,N)$ gets larger. For Scenario B, from Table \ref{tab:2}, we can draw similar conclusions as for Scenario A. The results show that the HPCA procedure is also robust to the heavy tails in cases where both serial and cross-sectional correlations exist.
The QFA method is more time-consuming compared with the RTS method. {In addition, there is no significant difference between HPCA and IHR though  the IHR algorithm is computationally more demanding. For the HPCA method, we recommend the initialization value given by RTS as it can further enhance the empirical performances for non-elliptical distribution cases.}

\subsection{Estimation of the Number of Factors}
In this section,	we focus on the identification of the factor numbers. The main competitors considered here include the Eigenvalue Ratio (ER) method by \cite{ahn2013eigenvalue}, the Multivariate Kendall's tau Eigenvalue Ratio (MKER)  method for elliptical factor model by \cite{yu2019robust}  and the IQR combined with Rank Minimization (IQR-RM) method for QFA with $\tau=0.5$ by \cite{Chen2021Quantile}. { We also show the performance of the rank minimization method corresponding to HPCA and IHR and name them as HPCA-RM and IHR-RM, respectively.}  Following the same strategy in \cite{Chen2021Quantile}, we set $P=\min(T,N)^{-1/3}$. The results are shown in the form $a(b|c)$ where $a$ is the sample mean of the estimated factor numbers, and $b$ and $c$ are the underestimation and overestimation frequencies, respectively. We consider the following data-generating scenarios for comparison.

\textbf {Scenario C} Set $r=3,\theta=1,\rho=\beta=J=0$, $(N,T)=\big\{(100,100),(200,200)\big\}$, $(\bbf_t^\top,\bv_t^\top)^\top$ are generated in the following ways:

Case \uppercase\expandafter{\romannumeral1}: $(\bbf_t^\top,\bv_t^\top)^\top$ are \emph{i.i.d.} random samples from multivariate Gaussian distributions $\mathcal{N}(\zero,\Ib_{p+m})$;

Case \uppercase\expandafter{\romannumeral2}: $(\bbf_t^\top,\bv_t^\top)^\top$ are random samples from  multivariate centralized $t_5$ distributions $t_{5}(\zero,\Ib_{p+m})$;

Case \uppercase\expandafter{\romannumeral3}: $(\bbf_t^\top,\bv_t^\top)^\top$ are random samples from  multivariate centralized $t_3$ distributions $t_{3}(\zero,\Ib_{p+m})$;

\textbf {Scenario D} Set $r=3,\theta=1,\rho=0.5,\beta=0.2,J={\rm max}\{10, p/20\}$, $(N,T)=\big\{(100,100),(200,200)\big\}$ and $(\bbf_t^\top,\bv_t^\top)$ are generated in the same ways as in Scenario C.

In Scenario C, $(\bbf_t^\top,\bv_t^\top)^\top$ are from a multivariate elliptical distribution with finite second moments. Also, there are no serial and cross-sectionally correlations of the idiosyncratic errors. Thus, the setting perfectly fits all methods. Scenario D is the case where serial and cross-sectionally correlations of the idiosyncratic errors exist.

\begin{table}[!h]
\caption{Simulation results for the estimation of factor numbers in Scenario C and Scenario D.}
\label{tab:3}
\renewcommand{\arraystretch}{1.3}
\centering
\selectfont
\begin{threeparttable}
\begin{tabular}{ccccccccccccc}
\toprule[2pt]
{$(T,N)$}&{Type}&&ER&&IQR-RM&&MKER&&HPCA-RM&&IHR-RM\\
\midrule[1pt]
\multirow{3}{2cm}{Scenario C \centering $(100,100)$}&Case \uppercase\expandafter{\romannumeral1}  &&3.00(0$|$0)   &&3.00(0$|$0)  &&3.00(0$|$0) &&3.00(0$|$0) &&3.00(0$|$0)\\
&Case \uppercase\expandafter{\romannumeral2}                                  &&3.01(0$|$6)   &&2.99(3$|$1)  &&3.00(0$|$0) &&2.99(3$|$1) &&2.99(3$|$1)\\
&Case \uppercase\expandafter{\romannumeral3}                                  &&2.98(27$|$37) &&2.87(59$|$8) &&3.00(0$|$0)  &&2.87(60$|$8)&&2.87(60$|$9)\\
\cmidrule(lr){2-12}
\multirow{3}{2cm}{Scenario C \centering $(200,200)$}&Case \uppercase\expandafter{\romannumeral1}  &&3.00(0$|$0)   &&3.00(0$|$0)  &&3.00(0$|$0) &&3.00(0$|$0)  &&3.00(0$|$0)\\
&Case \uppercase\expandafter{\romannumeral2}                                  &&3.01(0$|$3)   &&3.00(0$|$0)  &&3.00(0$|$0) &&3.00(0$|$0) &&3.00(0$|$0)\\
&Case \uppercase\expandafter{\romannumeral3}                                  &&3.03(10$|$32) &&2.99(12$|$14)&&3.00(0$|$0) &&2.99(12$|$14) &&2.99(12$|$14)\\
\hline
\multirow{3}{2cm}{Scenario D \centering $(100,100)$}&Case \uppercase\expandafter{\romannumeral1}  &&3.00(0$|$0)     &&3.00(0$|$0)  &&3.00(0$|$0)&&3.00(0$|$0)&&3.00(0$|$0)\\
&Case \uppercase\expandafter{\romannumeral2}                                  &&2.99(9$|$4)  &&3.02(3$|$14) &&3.00(2$|$2)&&3.01(3$|$11)&&3.01(3$|$11)\\
&Case \uppercase\expandafter{\romannumeral3}                                  &&2.85(68$|$36)&&2.93(55$|$33)&&3.01(2$|$5)&&2.92(54$|$31)&&2.92(55$|$33)\\
\cmidrule(lr){2-12}
\multirow{3}{2cm}{Scenario D \centering $(200,200)$}&Case \uppercase\expandafter{\romannumeral1}  &&3.00(0$|$0)   &&3.00(0$|$0)  &&3.00(0$|$0) &&3.00(0$|$0)&&3.00(0$|$0)\\
&Case \uppercase\expandafter{\romannumeral2}                                  &&3.00(0$|$2)      &&3.01(0$|$3)  &&3.00(0$|$0) &&3.00(0$|$2)&&3.00(0$|$2)\\
&Case \uppercase\expandafter{\romannumeral3}                                  &&3.01(17$|$33) &&3.03(11$|$30)&&3.00(0$|$0) &&3.02(11$|$30)&&3.02(11$|$31)\\
\bottomrule[2pt]
\end{tabular}
\end{threeparttable}
\end{table}

The simulation results for Scenario C and Scenario D are reported in { Table \ref{tab:3}}. For Scenario C, we conclude that all methods perform very well in the Gaussian setting. As the tail gets heavier, MKER performs slightly better than the others as the elliptical assumption is satisfied. {The IQR-RM, HPCA-RM and IHR-RM methods have comparable performance in terms of specifying the factor numbers. Besides, HPCA-RM performs almost the same as IHR-RM}. We can draw the same conclusions for Scenario D where both the serial and cross-sectional errors exist.


\section{Real Financial Portfolio Example}
\begin{figure}[!h]
	\centerline{\includegraphics[width=17cm,height=5cm]{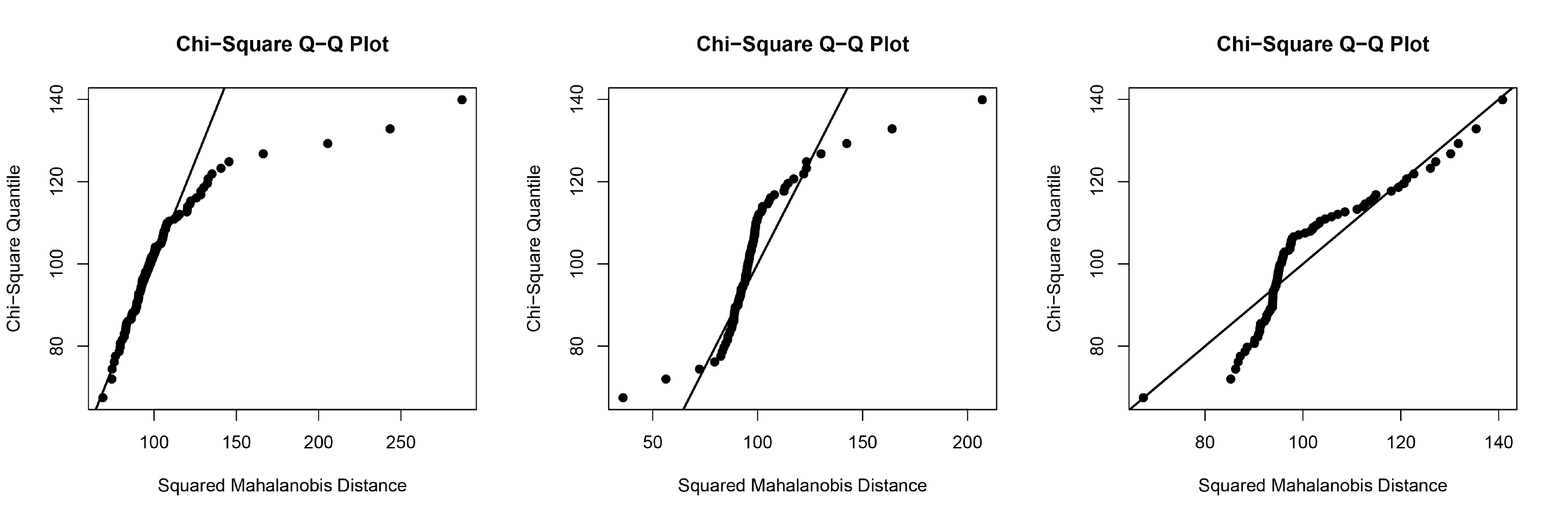}}
	\caption{QQ chart of asset pool A,B and C, from left to right.}
	\label{fig:QQ_plot}
\end{figure}

In this section, we provide empirical evidence showing that {HPCA/IHR} is a powerful tool in financial investing.  The portfolios datasets in this empirical study are freely available from the home page of Kenneth R. French\footnote{\url{http://mba.tuck.dartmouth.edu/pages/faculty/ken.french/data_library.html}.
}. Three pools of portfolios are downloaded from this website and there are non-missing values. { Pools A-C consist of 100 portfolios, each with monthly returns from June 2015 to May 2022.} The portfolios in pool A are formed on Size and Book-to-Market, the portfolios in pool B are formed by Size and Operating Profit and the portfolios in pool C are formed by Size and Investment. Accordingly, two main factors exist in each dataset and we thus set the factor numbers $r=2$ hereafter. We first conduct the Augmented Dickey-Fuller test  for each dataset under the significance level $\alpha=0.05$ by \texttt{R} function \texttt{adfTest}, which indicates that there are no significant serial correlations and all the series are stationary. Then we conduct the multivariate normality tests for the assets in pools A-C and show the QQ-chart in Figure \ref{fig:QQ_plot} by \texttt{R} package \texttt{MVN}, from which we can see that the distributions of all assets deviate far away from the normal distribution and the financial portfolios' returns are heavy-tailed.

As for the investment strategy, we employ a similar procedure as described in \cite{He2020large} and design a rolling scheme to evaluate the  returns by different methods. The  strategy is the well-known Mean-Variance (MV)
framework introduced by \cite{Markowitz1952Portofolio}, i.e.,  to determine the optimal  weights $\bomega$ by controlling the risk, that is,
$$\bomega_{opt}=\argmin_{\bm{1}^{\top}\bomega=1} \bomega^{\top}\bSigma\bomega=\frac{\bSigma^{-1}\bm{1}}{\bm{1}^{\top}\bSigma^{-1}\bm{1}}
,$$
where $\bm{1}$ is a vector with elements $1$ and $\bSigma$ is the  covariance matrix of all portfolios. One may refer to  \cite{chamberlain1982arbitrage} and \cite{owen1983class} for further details. To estimate the large matrix $\bSigma$ better, we assume the portfolio returns have a factor structure. Accordingly, at the beginning of each month $t$, we recursively use the returns during the past 72 months (a panel with $T=72, N=100$) to train the factor models by various methods and estimate $\bSigma$ by $$\hat{\bSigma}_t=\frac{1}{72}\hat{\bm{\cC}}_t^{\top}\hat{\bm{\cC}}_t+\operatorname{Hardthresh}\left (\frac{1}{72}\hat{\bE}_{t}^{\top}\hat{\bE}_{t}\right),$$
where $\hat{\bm{\cC}}_t$ and $\hat{\bE}_t$ are the estimated
common components and idiosyncratic errors and $\operatorname{Hardthresh}\left(\Ab\right)$ denotes the hard-threshold estimator proposed by \cite{10.1214/08-AOS600}. Then the $\hat{\bomega}_{t}$ can be obtained by plugging in $\hat{\bSigma}_t$ at time point $t$.

For comparison, we report the average return, Sharpe ratio and the $\tau$-th quantile of returns with $\tau=0.1,0.25,0.5,0.75,0.9$. We ignore transaction cost and report the results for  different pools of portfolios in Table \ref{tab:5}, from which we can draw the following conclusions.
\begin{table}[!h]
\caption{Large portfolio allocations for pools A-C, from left to right, under factor structures with the standard deviation in the parenthesis.}
\label{tab:5}
\renewcommand{\arraystretch}{1.3}
\centering
\selectfont
\begin{tabular}{ccccccccc}
\toprule[2pt]
\multirow{2}{2cm}{\centering Pool}&\multirow{2}{2cm}{\centering Method}&\multirow{2}{2.5cm}{\centering Average Return}&\multirow{2}{2.5cm}{\centering Sharpe Ratio}&\multicolumn{5}{c}{ $\tau$-th Quantile of Return}\\
\cmidrule(lr){5-9}
&&&&$0.1$&$0.25$&$0.5$&$0.75$&$0.9$\\
\midrule[1pt]
\multirow{4}{2cm}{\centering Pool A}
&PCA &-1.04(7.26)&-0.16(1.13)&-9.71&-6.82&-0.45&5.41&6.36\\
&QFA &\textbf{0.19(7.38)} &\textbf{0.03(1.16)} &-9.70&-4.20&0.01&6.18&8.26\\
&RTS &-0.45(6.76)&-0.07(1.09)&-9.07&-5.15&-0.19&5.32&7.62\\
&HPCA&0.08(7.14) &0.01(1.09) &-8.98&-4.43&0.19&6.15&8.29\\
&IHR &0.13(7.16)&0.02(1.09)  &-8.96&-4.49&0.25&6.56&8.28\\
\cmidrule(lr){2-9}
\multirow{4}{2cm}{\centering Pool B}
&PCA &1.28(5.32)&0.29(1.04)&-4.93&-1.80&1.94&5.08&8.04\\
&QFA &0.65(4.53)&0.17(1.00)&-4.55&-2.33&0.88&3.69&5.87\\
&RTS &1.63(5.62)&0.30(1.02)&-4.65&-2.08&2.61&5.92&8.78\\
&HPCA&1.25(5.69)&0.22(1.03)&-5.53&-2.16&0.70&5.55&8.61\\
&IHR &\textbf{1.65(5.45)}&\textbf{0.33(1.02)}&-4.70&-1.72&1.52&5.41&8.96\\
\cmidrule(lr){2-9}
\multirow{4}{2cm}{\centering Pool C}
&PCA &1.15(5.35)&0.20(0.97)&-3.92&-2.19&1.73&4.76&6.68\\
&QFA &0.98(5.36)&0.21(1.10)&-7.39&-1.60&2.24&4.59&6.42\\
&RTS &1.18(6.80)&0.36(1.02)&-4.09&-1.09&2.00&5.12&7.72\\
&HPCA&\textbf{2.25(6.34)}&\textbf{0.40(1.12)}&-3.99&-1.75&2.96&5.39&7.70\\
&IHR &1.64(5.01)&0.33(1.02)&-4.04&-1.29&2.36&5.39&6.87\\
\bottomrule[2pt]
\end{tabular}
\end{table}

{
Firstly, for the asset in pool A, the QFA performs the best in terms of the average returns and Sharpe ratio and the HPCA/IHR performs comparably from the perspective of Sharpe ratio. In addition, only the QFA, HPCA and IHR achieve a positive return though with a relatively low Sharpe ratio. In terms of the $\tau$-th quantile of returns, HPCA/IHR is also comparable with QFA. For the assets in pool B, the IHR performs the best in terms of the average returns and Sharpe ratio and the QFA performs unsatisfactorily. For the assets in pool C, the HPCA performs the best while the IHR ranks the second. For the heavy-tailed portfolio returns, the traditional PCA loses power while the robust methods QFA, RTS and HPCA/IHR perform much better. This illustrates the necessity of introducing robust factor analysis tools when conducting portfolio allocation.
We can not claim that our proposed methods are overwhelmingly superior to the others in terms of constructing portfolios. However, the empirical findings above do show the robustness and competitiveness of HPCA/IHR in maximizing the return and Sharpe ratios in high-dimensional cases.
}

\section{Conclusions and Discussions}
	In areas such as macroeconomics and finance, where factor models have been widely used, the collected data typically exhibit heavy tails. Directly removing the ``outliers" is inappropriate as the underlying distribution would be distorted. Robust factor analysis which is insensitive to the tail properties of data is of great importance and very limited literature exists on this topic. In this paper, motivated by the equivalence between PCA and least squares for factor analysis, we propose to substitute the least square loss with the Huber loss and propose two  different types of optimization problems. One is based on minimizing the $\ell_2$-norm type Huber loss, which turns out to perform PCA on the weighted sample covariance matrix and is thereby named Huber PCA. The other one is based on minimizing the element-wise type Huber loss, which can be solved by an iterative Huber regression algorithm.   We also investigate the theoretical minimizer of the element-wise type Huber loss function, and the same  convergence rates as for conventional PCA and their asymptotic distribution are derived under finite second-moment conditions on the idiosyncratic errors.
A rank minimization method is also given to determine the number of factors robustly, which is of independent interest.
Numerical results and real financial portfolio allocation results illustrate the advantage of the proposed robust methods.

The theoretical analysis of the Huber PCA method is more challenging and we leave it for future work. In the last years, the matrix factor model or even the high-order tensor factor model is growing popular, see for example \cite{han2020rank,chen2022factor,Yu2021Projected,He2023Matrix,He2021Vector}. Extending the robust Huber methods to the matrix/tensor factor model is quite  interesting but challenging, which is currently under investigation by the authors.

\section*{Acknowledgements}

This work is supported by NSF China (12171282,11801316), National
Statistical Scientific Research Key Project (2021LZ09), Project funded by
China Postdoctoral Science Foundation (2021M701997) and the Fundamental
Research Funds of Shandong University, Young Scholars Program of Shandong
University, China.

\section*{Supplementary Material}

The technical proofs of the main
results are included in the
Supplementary Material.

	\bibliographystyle{ref.bib}
	\bibliography{Ref}

\begin{thebibliography}{37}
\expandafter\ifx\csname natexlab\endcsname\relax\def\natexlab#1{#1}\fi
\providecommand{\url}[1]{\texttt{#1}}
\providecommand{\href}[2]{#2}
\providecommand{\path}[1]{#1}
\providecommand{\DOIprefix}{doi:}
\providecommand{\ArXivprefix}{arXiv:}
\providecommand{\URLprefix}{URL: }
\providecommand{\Pubmedprefix}{pmid:}
\providecommand{\doi}[1]{\href{http://dx.doi.org/#1}{\path{#1}}}
\providecommand{\Pubmed}[1]{\href{pmid:#1}{\path{#1}}}
\providecommand{\bibinfo}[2]{#2}
\ifx\xfnm\relax \def\xfnm[#1]{\unskip,\space#1}\fi
\bibitem[{Ahn and Horenstein(2013)}]{ahn2013eigenvalue}
\bibinfo{author}{Ahn, S.C.}, \bibinfo{author}{Horenstein, A.R.},
  \bibinfo{year}{2013}.
\newblock \bibinfo{title}{Eigenvalue ratio test for the number of factors}.
\newblock \bibinfo{journal}{Econometrica} \bibinfo{volume}{81},
  \bibinfo{pages}{1203--1227}.
\bibitem[{Bai(2003)}]{Bai2003Inferential}
\bibinfo{author}{Bai, J.}, \bibinfo{year}{2003}.
\newblock \bibinfo{title}{Inferential theory for factor models of large
  dimensions}.
\newblock \bibinfo{journal}{Econometrica} \bibinfo{volume}{71},
  \bibinfo{pages}{135--171}.
\bibitem[{Bai and Li(2012)}]{Bai2012Statistical}
\bibinfo{author}{Bai, J.}, \bibinfo{author}{Li, K.}, \bibinfo{year}{2012}.
\newblock \bibinfo{title}{Statistical analysis of factor models of high
  dimension}.
\newblock \bibinfo{journal}{The Annals of Statistics} \bibinfo{volume}{40},
  \bibinfo{pages}{436--465}.
\bibitem[{Bai and Li(2014)}]{Bai2014Theory}
\bibinfo{author}{Bai, J.}, \bibinfo{author}{Li, K.}, \bibinfo{year}{2014}.
\newblock \bibinfo{title}{Theory and methods of panel data models with
  interactive effects}.
\newblock \bibinfo{journal}{The Annals of Statistics} \bibinfo{volume}{42},
  \bibinfo{pages}{142--170}.
\bibitem[{Bai and Li(2016)}]{Bai2016Maximum}
\bibinfo{author}{Bai, J.}, \bibinfo{author}{Li, K.}, \bibinfo{year}{2016}.
\newblock \bibinfo{title}{Maximum likelihood estimation and inference for
  approximate factor models of high dimension}.
\newblock \bibinfo{journal}{Review of Economics and Statistics}
  \bibinfo{volume}{98}, \bibinfo{pages}{298--309}.
\bibitem[{Bai and Ng(2002)}]{Bai2002Determining}
\bibinfo{author}{Bai, J.}, \bibinfo{author}{Ng, S.}, \bibinfo{year}{2002}.
\newblock \bibinfo{title}{Determining the number of factors in approximate
  factor models}.
\newblock \bibinfo{journal}{Econometrica} \bibinfo{volume}{70},
  \bibinfo{pages}{191--221}.
\bibitem[{Barigozzi(2022)}]{barigozzi2022estimation}
\bibinfo{author}{Barigozzi, M.}, \bibinfo{year}{2022}.
\newblock \bibinfo{title}{On estimation and inference of large approximate
  dynamic factor models via the principal component analysis}.
\newblock \bibinfo{journal}{arXiv preprint arXiv:2211.01921} .
\bibitem[{Barigozzi et~al.(2022)Barigozzi, Cavaliere and
  Trapani}]{barigozzi2022inference}
\bibinfo{author}{Barigozzi, M.}, \bibinfo{author}{Cavaliere, G.},
  \bibinfo{author}{Trapani, L.}, \bibinfo{year}{2022}.
\newblock \bibinfo{title}{Inference in heavy-tailed nonstationary multivariate
  time series}.
\newblock \bibinfo{journal}{Journal of the American Statistical Association} ,
  \bibinfo{pages}{1--17}.
\bibitem[{Barigozzi and Cho(2020)}]{barigozzi2020consistent}
\bibinfo{author}{Barigozzi, M.}, \bibinfo{author}{Cho, H.},
  \bibinfo{year}{2020}.
\newblock \bibinfo{title}{Consistent estimation of high-dimensional factor
  models when the factor number is over-estimated}.
\newblock \bibinfo{journal}{Electronic Journal of Statistics}
  \bibinfo{volume}{14}, \bibinfo{pages}{2892--2921}.
\bibitem[{Barigozzi and Luciani(2019)}]{barigozzi2019quasi}
\bibinfo{author}{Barigozzi, M.}, \bibinfo{author}{Luciani, M.},
  \bibinfo{year}{2019}.
\newblock \bibinfo{title}{Quasi maximum likelihood estimation and inference of
  large approximate dynamic factor models via the em algorithm}.
\newblock \bibinfo{journal}{arXiv preprint arXiv:1910.03821} .
\bibitem[{Bickel and Levina(2008)}]{10.1214/08-AOS600}
\bibinfo{author}{Bickel, P.J.}, \bibinfo{author}{Levina, E.},
  \bibinfo{year}{2008}.
\newblock \bibinfo{title}{{Covariance regularization by thresholding}}.
\newblock \bibinfo{journal}{The Annals of Statistics} \bibinfo{volume}{36},
  \bibinfo{pages}{2577 -- 2604}.
\bibitem[{Chamberlain and Rothschild(1982)}]{chamberlain1982arbitrage}
\bibinfo{author}{Chamberlain, G.}, \bibinfo{author}{Rothschild, M.},
  \bibinfo{year}{1982}.
\newblock \bibinfo{title}{Arbitrage, factor structure, and mean-variance
  analysis on large asset markets}.
\bibitem[{Chamberlain and Rothschild(1983)}]{Chamberlain1983Arbitrage}
\bibinfo{author}{Chamberlain, G.}, \bibinfo{author}{Rothschild, M.},
  \bibinfo{year}{1983}.
\newblock \bibinfo{title}{Arbitrage, factor structure, and mean-variance
  analysis on large asset markets}.
\newblock \bibinfo{journal}{Econometrica} \bibinfo{volume}{51},
  \bibinfo{pages}{1281--1304}.
\bibitem[{Chen et~al.(2021)Chen, Dolado and Gonzalo}]{Chen2021Quantile}
\bibinfo{author}{Chen, L.}, \bibinfo{author}{Dolado, J.J.},
  \bibinfo{author}{Gonzalo, J.}, \bibinfo{year}{2021}.
\newblock \bibinfo{title}{Quantile factor models}.
\newblock \bibinfo{journal}{Econometrica} \bibinfo{volume}{89},
  \bibinfo{pages}{875--910}.
\bibitem[{Chen et~al.(2022)Chen, Yang and Zhang}]{chen2022factor}
\bibinfo{author}{Chen, R.}, \bibinfo{author}{Yang, D.}, \bibinfo{author}{Zhang,
  C.H.}, \bibinfo{year}{2022}.
\newblock \bibinfo{title}{Factor models for high-dimensional tensor time
  series}.
\newblock \bibinfo{journal}{Journal of the American Statistical Association}
  \bibinfo{volume}{117}, \bibinfo{pages}{94--116}.
\bibitem[{Cont(2001)}]{Cont2001Empirical}
\bibinfo{author}{Cont, R.}, \bibinfo{year}{2001}.
\newblock \bibinfo{title}{Empirical properties of asset returns: stylized facts
  and statistical issues}.
\newblock \bibinfo{journal}{Quantitative Finance} \bibinfo{volume}{1},
  \bibinfo{pages}{223--236}.
\bibitem[{Fama(1963)}]{Fama1963Mandelbrot}
\bibinfo{author}{Fama, E.F.}, \bibinfo{year}{1963}.
\newblock \bibinfo{title}{Mandelbrot and the stable paretian hypothesis}.
\newblock \bibinfo{journal}{Journal of Business} \bibinfo{volume}{36},
  \bibinfo{pages}{420--429}.
\bibitem[{Fan et~al.(2013)Fan, Liao and Mincheva}]{fan2013large}
\bibinfo{author}{Fan, J.}, \bibinfo{author}{Liao, Y.},
  \bibinfo{author}{Mincheva, M.}, \bibinfo{year}{2013}.
\newblock \bibinfo{title}{Large covariance estimation by thresholding principal
  orthogonal complements}.
\newblock \bibinfo{journal}{Journal of the Royal Statistical Society: Series B
  (Statistical Methodology)} \bibinfo{volume}{75}, \bibinfo{pages}{603--680}.
\bibitem[{Franklin(2012)}]{FRANKLIN2012matrix}
\bibinfo{author}{Franklin, J.N.}, \bibinfo{year}{2012}.
\newblock \bibinfo{title}{Matrix Theory}.
\newblock \bibinfo{publisher}{Courier Corporation}.
\bibitem[{Golub and Van~Loan(2013)}]{golub2013matrix}
\bibinfo{author}{Golub, G.H.}, \bibinfo{author}{Van~Loan, C.F.},
  \bibinfo{year}{2013}.
\newblock \bibinfo{title}{Matrix computations}.
\newblock \bibinfo{publisher}{JHU press}.
\bibitem[{Han et~al.(2022)Han, Chen and Zhang}]{han2020rank}
\bibinfo{author}{Han, Y.}, \bibinfo{author}{Chen, R.}, \bibinfo{author}{Zhang,
  C.H.}, \bibinfo{year}{2022}.
\newblock \bibinfo{title}{Rank determination in tensor factor model}.
\newblock \bibinfo{journal}{Electronic Journal of Statistics}
  \bibinfo{volume}{16}, \bibinfo{pages}{1726--1803}.
\bibitem[{He et~al.(2023a)He, Kong, , Yu, Zhang and Zhao}]{He2023Matrix}
\bibinfo{author}{He, Y.}, \bibinfo{author}{Kong, X.}, , \bibinfo{author}{Yu,
  L.}, \bibinfo{author}{Zhang, X.}, \bibinfo{author}{Zhao, C.},
  \bibinfo{year}{2023}a.
\newblock \bibinfo{title}{Matrix factor analysis: From least squares to
  iterative projection}.
\newblock \bibinfo{journal}{Journal of Business and Economic Statistics, in
  press} .
\bibitem[{He et~al.(2023b)He, Kong, Trapani and Yu}]{He2021Vector}
\bibinfo{author}{He, Y.}, \bibinfo{author}{Kong, X.}, \bibinfo{author}{Trapani,
  L.}, \bibinfo{author}{Yu, L.}, \bibinfo{year}{2023}b.
\newblock \bibinfo{title}{One-way or two-way factor model for matrix
  sequences?}
\newblock \bibinfo{journal}{Journal of Econometrics, in press} .
\bibitem[{He et~al.(2022)He, Kong, Yu and Zhang}]{He2020large}
\bibinfo{author}{He, Y.}, \bibinfo{author}{Kong, X.}, \bibinfo{author}{Yu, L.},
  \bibinfo{author}{Zhang, X.}, \bibinfo{year}{2022}.
\newblock \bibinfo{title}{Large-dimensional factor analysis without moment
  constraints}.
\newblock \bibinfo{journal}{Journal of Business \& Economic Statistics}
  \bibinfo{volume}{40}, \bibinfo{pages}{302--312}.
\bibitem[{He et~al.(2020)He, Kong, Yu and Zhao}]{he2020quantile}
\bibinfo{author}{He, Y.}, \bibinfo{author}{Kong, X.B.}, \bibinfo{author}{Yu,
  L.}, \bibinfo{author}{Zhao, P.}, \bibinfo{year}{2020}.
\newblock \bibinfo{title}{Quantile factor analysis for large-dimensional time
  series with statistical guarantee}.
\newblock \bibinfo{journal}{arXiv preprint arXiv:2006.08214} .
\bibitem[{Huber(1964)}]{Huber1964Robust}
\bibinfo{author}{Huber, P.J.}, \bibinfo{year}{1964}.
\newblock \bibinfo{title}{Robust estimation of a location parameter}.
\newblock \bibinfo{journal}{Annals of Mathematical Statistics}
  \bibinfo{volume}{35}, \bibinfo{pages}{73--101}.
\bibitem[{Huber(2011)}]{huber2011robust}
\bibinfo{author}{Huber, P.J.}, \bibinfo{year}{2011}.
\newblock \bibinfo{title}{Robust statistics}, in:
  \bibinfo{booktitle}{International encyclopedia of statistical science}.
  \bibinfo{publisher}{Springer}, pp. \bibinfo{pages}{1248--1251}.
\bibitem[{Markowitz(1952)}]{Markowitz1952Portofolio}
\bibinfo{author}{Markowitz, H.}, \bibinfo{year}{1952}.
\newblock \bibinfo{title}{Portfolio selection}.
\newblock \bibinfo{journal}{The Journal of finance} \bibinfo{volume}{7},
  \bibinfo{pages}{77--91}.
\bibitem[{Onatski(2009)}]{onatski2009testing}
\bibinfo{author}{Onatski, A.}, \bibinfo{year}{2009}.
\newblock \bibinfo{title}{Testing hypotheses about the number of factors in
  large factor models}.
\newblock \bibinfo{journal}{Econometrica} \bibinfo{volume}{77},
  \bibinfo{pages}{1447--1479}.
\bibitem[{Owen and Rabinovitch(1983)}]{owen1983class}
\bibinfo{author}{Owen, J.}, \bibinfo{author}{Rabinovitch, R.},
  \bibinfo{year}{1983}.
\newblock \bibinfo{title}{On the class of elliptical distributions and their
  applications to the theory of portfolio choice}.
\newblock \bibinfo{journal}{The Journal of Finance} \bibinfo{volume}{38},
  \bibinfo{pages}{745--752}.
\bibitem[{Stock and Watson(2002)}]{stock2002forecast}
\bibinfo{author}{Stock, J.H.}, \bibinfo{author}{Watson, M.W.},
  \bibinfo{year}{2002}.
\newblock \bibinfo{title}{Forecasting using principal components from a large
  number of predictors}.
\newblock \bibinfo{journal}{Journal of the American Statistical Association}
  \bibinfo{volume}{97}, \bibinfo{pages}{1167--1179}.
\bibitem[{Trapani(2018)}]{Trapani2018A}
\bibinfo{author}{Trapani, L.}, \bibinfo{year}{2018}.
\newblock \bibinfo{title}{A randomised sequential procedure to determine the
  number of factors}.
\newblock \bibinfo{journal}{Journal of the American Statistical Association}
  \bibinfo{volume}{113}, \bibinfo{pages}{1341--1349}.
\bibitem[{Van~der Vaart and Wellner(1996)}]{vanderVaart1996}
\bibinfo{author}{Van~der Vaart, A.W.}, \bibinfo{author}{Wellner, J.A.},
  \bibinfo{year}{1996}.
\newblock \bibinfo{title}{Weak Convergence and Empirical Processes}.
\newblock \bibinfo{publisher}{New York: Springer}.
\bibitem[{Venables and Ripley(2002)}]{ven2002}
\bibinfo{author}{Venables, W.N.}, \bibinfo{author}{Ripley, B.D.},
  \bibinfo{year}{2002}.
\newblock \bibinfo{title}{Modern Applied Statistics with S}.
\newblock \bibinfo{publisher}{Springer New York, NY}.
\bibitem[{Weron and Weron(2005)}]{weron2005computer}
\bibinfo{author}{Weron, A.}, \bibinfo{author}{Weron, R.}, \bibinfo{year}{2005}.
\newblock \bibinfo{title}{Computer simulation of l{\'e}vy $\alpha$-stable
  variables and processes}, in: \bibinfo{booktitle}{Chaos—The Interplay
  Between Stochastic and Deterministic Behaviour: Proceedings of the XXXIst
  Winter School of Theoretical Physics Held in Karpacz, Poland 13--24 February
  1995}. \bibinfo{publisher}{Springer}, pp. \bibinfo{pages}{379--392}.
\bibitem[{Yu et~al.(2022)Yu, He, Kong and Zhang}]{Yu2021Projected}
\bibinfo{author}{Yu, L.}, \bibinfo{author}{He, Y.}, \bibinfo{author}{Kong, X.},
  \bibinfo{author}{Zhang, X.}, \bibinfo{year}{2022}.
\newblock \bibinfo{title}{Projected estimation for large-dimensional matrix
  factor models}.
\newblock \bibinfo{journal}{Journal of Econometrics} \bibinfo{volume}{229},
  \bibinfo{pages}{201--217}.
\bibitem[{Yu et~al.(2019)Yu, He and Zhang}]{yu2019robust}
\bibinfo{author}{Yu, L.}, \bibinfo{author}{He, Y.}, \bibinfo{author}{Zhang,
  X.}, \bibinfo{year}{2019}.
\newblock \bibinfo{title}{Robust factor number specification for
  large-dimensional elliptical factor model}.
\newblock \bibinfo{journal}{Journal of Multivariate analysis}
  \bibinfo{volume}{174}, \bibinfo{pages}{104543}.

\end{thebibliography}
	
\renewcommand{\baselinestretch}{1}
\setcounter{footnote}{0}
\clearpage
\setcounter{page}{1}
\appendix
\section*{APPENDIX}
In this appendix, we provide detailed proofs of our main theoretical results.
Throughout this appendix, $C_1,C_2,\ldots$ denote some positive
constants that do not depend on $N,T$.
\section{Proof of Theorem \ref{th:1}}	
Define $d^2\l(\theta_a,\theta_b\r)=\sum_{t=1}^T \sum_{i=1}^N\left(\bl_{ai}^\top\bbf_{a t}-\bl_{bi}^\top\bbf_{b t}\right)^2/(TN)=\left\|\Lb_a\Fb_a^\top-\Lb_b\Fb_b^\top\right\|_F^2/(TN)$.	

The $\left(\Lb,\bbf_1,\ldots,\bbf_T\right)$ minimizing $L_{EH}\left(\Lb,\bbf_t\right)$ is the same as that minimizing $L_{EH}^\prime\left(\theta\right)$ where
$$
L_{EH}^\prime\left(\theta\right)=\frac{1}{TN} \sum_{t=1}^T\sum_{i=1}^N\left[H_{\tau}\left(Y_{it}-\bl_i^{\top}\bbf_t\right)-H_{\tau}\left(Y_{it}-\bl_{0i}^{\top}\bbf_{0t}\right)\right] .
$$
Let $\zeta_{it}(\bl_i,\bbf_t)=H_\tau\l(Y_{it}-\bl_{i}^\top\bbf_t\r)-H_\tau\l(Y_{it}-\bl_{0i}^\top\bbf_{0t}\r)$ and
$L_2(\theta)=\EE L_{EH}^\prime(\theta)$,
$L_1(\theta)=L_{EH}^\prime(\theta)-L_2(\theta)$. Then $L_{EH}^\prime=\sum_{t=1}^T\sum_{i=1}^N\zeta_{it}/(TN)$, $L_2(\theta)=\sum_{t=1}^T\sum_{i=1}^N\EE\zeta_{it}/(TN)$.

Note that
$$
\frac{\partial \EE H_\tau(x+\epsilon_{it})}{\partial x}=-\tau\int_{-\infty}^{-\tau-x}f_{it}(u)du+\int_{-\tau-x}^{\tau-x}(x+u)f_{it}(u)du+\tau\int_{\tau-x}^\infty f_{it}(u)du,
$$
and
$$
\frac{\partial^2 \EE H_\tau(x+\epsilon_{it})}{\partial x^2}=\int_{-\tau-x}^{\tau-x}f_{it}(u)du,
$$
In the second equation, we can see that $\tau$ and $-\tau$ are the second type of discontinuity of $H_\tau^{(2)}(\cdot)$, and we let $H^{(2)}_\tau(\tau)=H^{(2)}_\tau(-\tau)=0$ to make it meaningful over the whole real number field. In the following, we all directly use $H_\tau^{(2)}(\cdot)$ to denote the function after assignment.

Afterward, we have $\l(\partial/\partial x\r) \EE H_\tau(x+\epsilon_{it})=\EE H_\tau^{(1)}(x+\epsilon_{it})$ and $\l(\partial/\partial x\r)^2 \EE H_\tau(x+\epsilon_{it})=\EE H_\tau^{(2)}(x+\epsilon_{it})$,
where $H_{\tau}^{(1)}(z)=(\partial / \partial z) H_\tau(z)$ and $H_{\tau}^{(2)}(z)=(\partial / \partial z)^2 H_\tau(z)$.
Expanding $\EE H_{\tau}\left(Y_{it}-\bl_{i}^\top\bbf_{t}\right)$ around $\EE H_{\tau}(Y_{it}-\bl_{0i}^\top\bbf_{0t})$ gives
\beq\label{equ:expan2}
\begin{aligned}
	\EE\zeta_{it}
	=&\EE H_{\tau}^{(1)}\left(\left(Y_{it}-\bl_{0i}^\top\bbf_{0t}\right)\right)\left(\bl_{0i}^{\top}\bbf_{0t}-\bl_i^{\top}\bbf_t\right)+\frac{1}{2} \EE H_{\tau}^{(2)}\left(\xi_{it}^{(1)}\right)\left(\bl_{0i}^{\top}\bbf_{0t}-\bl_i^{\top}\bbf_t\right)^2,
\end{aligned}
\eeq
where $\xi_{it}^{(1)}$ is variable lying between $\left(Y_{it}-\bl_{i}^\top\bbf_{t}\right)$ and $(Y_{it}-\bl_{0i}^\top\bbf_{0t})$. It follows Assumption \ref{asmp:3} and the fact $\EE H_{\tau}^{(2)}\left(\xi_{it}^{(1)}\right)>0$ that $\left(\bl_{0i}^{\top}\bbf_{0t}-\bl_i^{\top}\bbf_t\right)^2 \lesssim \EE\zeta_{it}$, and then $d^2\l(\theta,\theta_0\r) \lesssim L_2(\theta)$.

To prove Theorem \ref{th:1}, first, we divide the parameter space $\Theta$ into $S_{j}=\{\theta \in \Theta$ : $\left.2^{j-1}<\sqrt{L} \cdot d\left(\theta, \theta_0\right) \leq 2^j\right\}$.
If $\sqrt{L} \cdot d\left(\widehat{\theta}, \theta_0\right)>2^V$ for a certain $V$, then $\widehat{\theta}$ is in one of the shells $S_j, j>V$, where the infimum of $L_{EH}^\prime(\theta)$ is nonpositive over this shell.
Therefore, for every $\eta>0$, we have
$$
\begin{aligned}
	\PP\left[\sqrt{L} \cdot d\left(\widehat{\theta}, \theta_0\right)>2^V\right]
	&=\PP\l[2^V < \sqrt{L}\cdot d\l(\widehat{\theta},\theta_0\r)\leq \sqrt{L}\cdot \eta\r]+\PP\l[d\l(\widehat{\theta},\theta_0\r)> \eta\r]\\
	& \leq \sum_{j>V, 2^{j-1} \leq \eta \sqrt{L}} \PP\left[\inf _{\theta \in S_{j}} L_{EH}^\prime\l(\theta\r) \leq 0\right]+\PP\left[d\left(\widehat{\theta}, \theta_0\right)>\eta\right].
\end{aligned}
$$
Based on fact $d\l(\widehat\theta,\theta_0\r)=o_p(1)$ in Lemma \ref{lem:1.1}, for arbitrarily small $\eta>0, \PP\left[d\left(\widehat{\theta}, \theta_0\right)>\eta\right]$ converges to 0 as $T, N \rightarrow \infty$.
For each $\theta$ in $S_{j}$ it holds that
$$
-d^2\left(\theta, \theta_0\right) \leq-\frac{2^{2 j-2}}{L}.
$$
Since $\inf_{\theta\in S_j}L_1\l(\theta\r)+\inf_{\theta\in S_j}L_2\l(\theta\r)\leq \inf _{\theta \in S_j} L_{EH}^\prime(\theta)\leq 0$ and $-L_2\l(\theta,\theta_0\r)\lesssim -d^2(\theta,\theta_0)$, we have that
$$
\inf _{\theta \in S_j} L_1(\theta)
\leq-\frac{2^{2 j-2}}{L}.
$$
So
$$
\PP\left[\inf _{\theta \in S_{j}} L_{EH}^\prime(\theta) \leq 0\right]
\leq \PP\left[\inf _{\theta \in S_{j}}L_1(\theta) \leq -\frac{2^{2 j-2}}{L}\right]
\leq \PP\left[\sup _{\theta \in S_{j}}\l|L_1(\theta)\r| \geq \frac{2^{2 j-2}}{L}\right]
$$
By Lemma \ref{lem:1.3} and Markov's inequality,
$$
\PP\left[\sup _{\theta \in S_{j}}|L_1(\theta)| \geq \frac{2^{2 j-2}}{L}\right]
\lesssim \frac{L}{2^{2 j}} \cdot \EE\left[\sup _{\theta \in S_j}|L_1(\theta)|\right]
\lesssim \frac{L}{2^{2 j}} \cdot \frac{2^j}{L}=2^{-j}
$$
then
$$
\sum_{j>V, 2^{j-1} \leq \eta \sqrt{L}} \PP\left[\inf _{\theta \in S_{j}} L_{EH}^\prime(\theta) \leq 0\right] \lesssim \sum_{j>V} 2^{-j}.
$$
As $V \rightarrow \infty, \sum_{j>V, 2^{j-1} \leq \eta \sqrt{L}} \PP\left[\inf _{\theta \in S_j} L_{EH}^\prime(\theta) \leq 0\right]$ converges to 0, implying $\sqrt{L} \cdot d\left(\widehat{\theta}, \theta_0\right)=O_p(1)$, e.g. $d\left(\widehat{\theta}, \theta_0\right)=O_p(1 / \sqrt{L})$.
By Lemma \ref{lem:1.2} , we have thus proved Theorem \ref{th:1} under Assumptions \ref{asmp:1}-\ref{asmp:3}.

\subsection{Complementary Lemmas for Theorem 3.1}
\begin{lemma}\label{lem:1.1}
	Under Assumptions \ref{asmp:1}-\ref{asmp:3}, $d\left(\widehat{\theta}, \theta_0\right)=o_p(1)$ as $N, T \rightarrow \infty$.
\end{lemma}	
\noindent\emph{\textbf{Proof.}}

Notice that $L_{EH}^\prime\l(\widehat\theta\r)\leq L_{EH}^\prime\l(\theta_0\r)=0$ and $L_{EH}^\prime(\theta)=L_1(\theta)+L_2(\theta)$, then we have that
$$
d^2\l(\widehat\theta,\theta_0\r)\lesssim L_2\l(\widehat\theta\r) \leq \sup_{\theta\in\Theta}|L_1(\theta)|.
$$
That is, we just need to proof $\sup_{\theta\in\Theta}|L_1(\theta)|=o_p(1)$ below.

Choose $C_1$ large enough such that $\left\|\bl_{0i}\right\|_2,\left\|\bbf_{0 t}\right\|_2,\left\|\bl_i\right\|_2,\left\|\bbf_t\right\|_2 \leq C_1$ for all $i, t$. Let $B_r(C_1)$ denote a Euclidean ball in $\mathbb{R}^r$ with radius $C_1$. For any $\epsilon>0$, let $\bl_{(1)}, \ldots, \bl_{\left(J\right)}$ be the maximal set of points in $B_{r}(C_1)$ such that $\left\|\bl_{(i)}-\bl_{(j)}\right\|_2>\epsilon / C_1$, for $\forall i \neq j$; and let $\bbf_{(1)}, \ldots, \bbf_{\left(J\right)}$ be the maximal set of points in $B_{r}(C_1)$ such that $\left\|\bbf_{(i)}-\bbf_{(j)}\right\|_2>\epsilon / C_1$, for $\forall i \neq j$. Then the packing number of $B_{r}(C_1)$ is $C_2(C_1 / \epsilon)^{r}$.

For any $\theta \in \Theta$, define $\theta^*=\left(\Lb^*,\bbf_1^*, \ldots, \bbf_T^*\right)$, where $\bl_{i}^*=\left\{\bl_{(j)}: j \leq J,\left\|\bl_{(j)}-\bl_{i}\right\|_2 \leq \epsilon / C_1\right\}$, $\bbf_t^*=\left\{\bbf_{(j)}: j \leq J,\left\|\bbf_{(j)}-\bbf_t\right\|_2 \leq \epsilon / C_1\right\}$.
Since $H_\tau^\prime(u)\leq \tau$ for all $u\in\RR$, one can see that
$$
\begin{aligned}
	|H_\tau\l(Y_{it}-\bl_i^\top\bbf_t\r)-H_\tau\l(Y_{it}-\bl_i^{*\top}\bbf_t^*\r)|
	&\leq \l|H_\tau^{(1)}\l(\xi_{it}^{(2)}\r)\r|\l|\bl_i^\top\bbf_t-\bl_i^{*\top}\bbf_t^*\r|\\
	&\leq \tau\l(\|\bl_i\|_2\|\bbf_t-\bbf_t^*\|_2+\|\bbf_t^*\|_2\|\bl_i-\bl_i^*\|_2\r)\leq 2\tau\epsilon
\end{aligned}
$$
with $\xi_{it}^{(2)}$ lying between $\l(Y_{it}-\bl_i^\top\bbf_t\r)$ and $\l(Y_{it}-\bl_i^{*\top}\bbf_t^*\r)$. Thus,
\beq\label{equ:I-I}
\sup _{\theta \in \Theta}\left|L_1(\theta)-L_1\left(\theta^*\right)\right| \leq 2\tau \epsilon .
\eeq
Similarly, $\l|\zeta_{it}\l(\bl_i^*,\bbf_t^*\r)\r| \leq \tau \l|\bl_i^{*\top}\bbf_t^*-\bl_{0i}^\top\bbf_{0t}\r|$. By Hoeffding's inequality, we have that
$$
\PP(|L_1(\theta^*)|\geq \delta)\leq 2\exp\l\{-\dfrac{2TN\delta^2}{\tau^2d^2(\theta^*,\theta_0)}\r\}.
$$

For any random variable $X$, the Orlicz norm is defined as
$$
\|X\|_\psi=\inf \{Z>0: \EE \psi(|X| / Z) \leq 1\} .
$$
Especially, when $\psi(x)=e^{x^2}-1$, the norm is denoted as $\|X\|_{\psi_2}$.

By Lemma 2.2.1 of \cite{vanderVaart1996}, we have that
\begin{equation}\label{equ:psi2}
	\left\|L_1\left(\theta^*\right)\right\|_{\psi_2} \lesssim \frac{1}{\sqrt{TN}} d\left(\theta^*, \theta_0\right).
\end{equation}
Since $\theta^*$ can take at most $J^{T+N} \lesssim(C_1 / \epsilon)^{Nr+T r}$ different values, and $d\left(\theta^*, \theta_0\right) \lesssim C_1$, we have
$$
\EE\left[\sup _{\theta \in \Theta}\left|L_1\left(\theta^*\right)\right|\right] \leq\left\|\sup _{\theta \in \Theta}\left|L_1\left(\theta^*\right)\right|\right\|_{\psi_2} \lesssim \sqrt{Tr+Nr} \sqrt{\log (C_1/\epsilon)} / \sqrt{TN}
\lesssim \sqrt{\log (C_1 / \epsilon)} / \sqrt{L}
$$
from Lemma 2.2.2 of \cite{vanderVaart1996}.

Finally, by Markov's inequality and (\ref{equ:I-I}), for any $\delta>0$,
$$
\begin{aligned}
	\PP\left[\sup _{\theta \in \Theta}|L_1(\theta)|>\delta\right]
	& \leq \PP\left[\sup _{\theta \in \Theta}\left|L_1\left(\theta^*\right)\right|>\delta / 2\right]+\PP\left[\sup _{\theta \in \Theta}\left|I(\theta)-L_1\left(\theta^*\right)\right|>\delta / 2\right] \\
	& \leq 2 / \delta \cdot \EE\left[\sup _{\theta \in \Theta}\left|L_1\left(\theta^*\right)\right|\right]+\PP\left[2\tau\epsilon>\delta / 2\right] .
\end{aligned}
$$
Thus,
$$
\sup _{\theta \in \Theta}|L_1(\theta)|=o_p(1)
$$
holds as $\epsilon$ is arbitrarily small.

\begin{lemma}\label{lem:1.2}
	Under Assumptions \ref{asmp:1}-\ref{asmp:3} and for sufficiently small $\delta>0$, for any $\theta \in \Theta(\delta)=\left\{\theta \in \Theta: d\left(\theta, \theta_0\right) \leq \delta\right\}$, it holds that
	$$
	\frac{1}{N}\left\|\Lb-\Lb_0 \mathbf{S}\right\|_F^2+\frac{1}{T}\left\|\Fb-\Fb_{0 }\mathbf{S}\right\|_F^2 \lesssim \delta^2,
	$$
	where $\mathbf{S}=\operatorname{sgn}\left(\Fb^\top\Fb_0/T\right)$.
\end{lemma}
\noindent\emph{\textbf{Proof.}}
First, let $\mathbf{U} \in \mathbb{R}^{r \times r}$ be a diagonal matrix whose diagonal elements are either 1 or -1, respectively. By the fact that $\Lb^{\top}\Lb/N=\Lb_0^{\top}\Lb_0/N=\Ib_r$ and Assumption \ref{asmp:1}, one can see that $\left\|\Fb_0\right\|_F^2/T \leq C_{3}$.
Thus, we have
$$
\begin{aligned}
	\left\|\Fb-\Fb_0 \mathbf{U}\right\|_F^2 / T
	& =\left\|\left(\Fb-\Fb_0 \mathbf{U}\right) \Lb^{\top}\right\|_F^2 / (TN)=\left\|\Fb \Lb^{\top}-\Fb_0 \Lb_0^{\top}+\Fb_0 \Lb_0^{\top}-\Fb_0 \mathbf{U} \Lb^{\top}\right\|_F^2 / (TN) \\
	& \leq2(\left\|\Fb \Lb^{\top}-\Fb_0 \Lb_0^{\top}\right\|_F^2 / (TN)+\left\|\Fb_0\right\|_F^2 / T \cdot\left\|\Lb-\Lb_0 \mathbf{U}\right\|_F^2 / N) \\
	& \leq 2(d^2\left(\theta, \theta_0\right)+C_{3}\left\|\Lb-\Lb_0 \mathbf{U}\right\|_F^2 / N) .
\end{aligned}
$$
Thus, for $\theta \in \Theta(\delta)$,
$$
\left\|\Fb-\Fb_0 \mathbf{U}\right\|_F^2 / T+\left\|\Lb-\Lb_0 \mathbf{U}\right\|_F^2 / N \leq 2\delta^2+\left(1+2C_{3}\right)\left\|\Lb-\Lb_0 \mathbf{U}\right\|_F^2 / N. $$
Second,
$$
\begin{aligned}
	\left\|\Lb-\Lb_0 \mathbf{U}\right\|_F^2 / N & =\left\|\Lb_0 \mathbf{U}-\Lb\left(\Lb^{\top} \Lb_0 \mathbf{U} / N\right)+\Lb\left(\Lb^{\top} \Lb_0 \mathbf{U} / N\right)-\Lb\right\|_F^2 / N \\
	& \leq2(\left\|\Lb_0 \mathbf{U}-\Lb\left(\Lb^{\top} \Lb_0 \mathbf{U} / N\right)\right\|_F^2 / N+\left\|\Lb\left(\Lb^{\top} \Lb_0 \mathbf{U} / N\right)-\Lb\right\|_F^2 / N) \\
	& =2(\left\|\mathbf{M}_{\Lb} \Lb_0\right\|_F^2 / N+\left\|\Lb^{\top} \Lb_0/N-\mathbf{U}\right\|_F^2)\\
	& =:\cI+\mathcal{II}
\end{aligned}
$$
where $\Pb_{\Ab}=\Ab\l(\Ab^{\top}\Ab\r)^{-1}\Ab^{\top}$, $\Mb_{\Ab}=\Ib-\Pb_{\Ab}$.

For $\cI$, we have
\beq\label{equ:M1}
\begin{aligned}
	& \frac{1}{T N} \left\|\mathbf{M}_{\Lb}\left(\Lb\Fb^{\top}-\Lb_0\Fb_0^{\top}\right)\right\|_F^2 \\
	\leq & \frac{1}{T N}  \operatorname{rank}\left[\mathbf{M}_{\Lb}\left(\Lb \Fb^{\top} -\Lb_0 \mathbf{F}^{\top}_{0} \right)\right]\left\|\mathbf{M}_{\Lb}\right\|^2\left\|\Lb\Fb^{\top}_t- \Lb_0\Fb_{0 t}^{\top}\right\|_2^2 \\
	\lesssim & \frac{1}{T N} \left\|\Lb\Fb_t^{\top}- \Lb_0\Fb_{0 t}^{\top} \right\|_F^2=d^2\left(\theta, \theta_0\right),
\end{aligned}
\eeq
and
\beq\label{equ:M2}
\begin{aligned}
	& \frac{1}{T N} \left\|\mathbf{M}_{\Lb}\left(\Lb\Fb^{\top}-\Lb_0\Fb_0^{\top}\right)\right\|_F^2 \\
	=& \frac{1}{T N} \left\|\mathbf{M}_{\Lb} \Lb_0 \mathbf{F}_{0}^{\top} \right\|_F^2=\frac{1}{T N}  \operatorname{Tr}\left(\mathbf{F}_{0}^{\top} \mathbf{F}_{0} \Lb_0^{\top} \mathbf{M}_{\Lb} \Lb_0\right) \\
	\geq & \frac{1}{T N} \lambda_{\min }\left(\mathbf{F}_{0}^{\top} \mathbf{F}_{0}\right) \operatorname{Tr}\left(\Lb_0^{\top} \mathbf{M}_{\Lb} \Lb_0\right)
	\geq \frac{1}{TN}\lambda_{\min }\left(\mathbf{F}_{0}^{\top} \mathbf{F}_{0}\right)\left\|\mathbf{M}_{\Lb} \Lb_0\right\|_F^2 .
\end{aligned}
\eeq
Thus we have that
\beq\label{equ:ML}
\frac{1}{N}\left\|\mathbf{M}_{\Lb} \Lb_0\right\|_F^2
\lesssim \frac{1}{\lambda_{\min }\left(\mathbf{F}_{0}^{\top} \mathbf{F}_{0}\right)/T}d^2\left(\theta, \theta_0\right)
\lesssim d^2\left(\theta, \theta_0\right),
\eeq
from Assumption \ref{asmp:1}.

For $\mathcal{II}$, let $\Zb=\Lb^{\top}\Lb_0/N$, $\Vb=\diag \l(\l(\bz_{1}\bz_{1}^{\top}\r)^{-1/2},\ldots,\l(\bz_{r}\bz_{r}^{\top}\r)^{-1/2}\r)$, where $\bz_{j}$ is the $j$th row of $\Zb$, and $\Ub=\Sbb$. Then we have
$$
\l\|\frac{1}{N}\Lb^\top\Lb_0-\Sbb\r\|_F^2
=\l\|\Zb^{\top}-\Sbb\r\|_F^2
\lesssim \l\|\Zb^{\top}\Vb-\Sbb\r\|_F^2+\l\|\Zb^{\top}\Vb-\Zb^\top\r\|^2_F
\leq \l\|\Zb^{\top}\Vb-\Sbb\r\|_F^2+\l\|\Zb\r\|^2_F\l\|\Vb-\Ib_{r}\r\|^2_F.
$$
For the first term,
by Assumption \ref{asmp:1} and the perturbation theory for eigenvectors (see Section 6.12 of \cite{FRANKLIN2012matrix}),
$$
\l\|\Zb^{\top}\Vb-\Sbb\r\|_F^2=\l\|\Zb^\top\Vb\Sbb-\Ib_{r}\r\|^2_F \lesssim d^2\l(\theta,\theta_0\r).
$$
For the second term, we have that
$$
\begin{aligned}
	\l\|\Vb-\Ib_{r}\r\|^2_F
	&\lesssim \l\|\Zb\Zb^\top-\Ib_{r}\r\|_F^2
	=\l\|\frac{1}{N}\Lb^\top\Pb_{\Lb_0}\Lb-\frac{1}{N}\Lb^\top\Lb\r\|_F^2\\
	&=\l\|\frac{1}{N}\Lb^\top\Mb_{\Lb_0}\Lb\r\|_F^2
	\leq \frac{1}{N^2}\l\|\Lb\r\|_F^2\l\|\Mb_{\Lb_0}\Lb\r\|_F^2
	=\frac{1}{N}\l\|\Mb_{\Lb_0}\Lb\r\|_F^2
\end{aligned}
$$
Similar to the (\ref{equ:M1}) and (\ref{equ:M2}), we have
$$
\frac{1}{N}\left\|\mathbf{M}_{\Lb_0} \Lb\right\|_F^2 \lesssim \frac{1}{  \lambda_{\min }\left(\mathbf{F}^{\top} \mathbf{F}\right)/T} d^2\left(\theta, \theta_0\right) .
$$
In the following we prove that $\lambda_{\min }\left(\mathbf{F}^{\top} \mathbf{F}\right)/T$ is bounded.

Notice that
$$
\begin{aligned}
	\frac{1}{T N} \left\|\mathbf{P}_{\Lb}\left(\Lb\Fb^{\top} -\Lb_0 \mathbf{F}_{0}^{\top}\right)\right\|_F^2
	\leq & \frac{1}{T N} \left\|\mathbf{P}_{\Lb}\right\|_F^2\left\|\Lb\Fb^{\top}-\Lb_0\mathbf{F}_{0}^{\top} \right\|_F^2
	=r d^2\left(\theta, \theta_0\right) .
\end{aligned}
$$
Thus
\beq \label{equ:F1}
\begin{aligned}
	\frac{1}{T} \left\|\mathbf{F}^{\top}-\frac{1}{N} \Lb^{\top} \Lb_0 \mathbf{F}_{0}^{\top} \right\|_F^2
	&= \frac{1}{T N} \left\|\Lb\Fb^{\top} -\frac{1}{N} \Lb\Lb^{\top} \Lb_0 \mathbf{F}_{0}^{\top} \right\|_F^2\\
	&=\frac{1}{T N} \left\|\Lb\Fb^{\top} -\mathbf{P}_{\Lb} \Lb_0 \mathbf{F}_{0}^{\top} \right\|_F^2\\
	&=\frac{1}{T N} \left\|\mathbf{P}_{\Lb}\left(\Lb \mathbf{F}^{\top} -\Lb_0\mathbf{F}_{0}^{\top} \right)\right\|_F^2
	\lesssim d^2\left(\theta, \theta_0\right) .
\end{aligned}
\eeq
Similarly,
\beq \label{equ:F2}
\frac{1}{T} \left\|\mathbf{F}_{0}^{\top} -\frac{1}{N} \Lb_0^{\top} \Lb\Fb^{\top} \right\|_F^2 \lesssim d^2\left(\theta, \theta_0\right) .
\eeq

Note that $\Lb \Zb= \Lb\Lb^{\top} \Lb_0/N=\mathbf{P}_{\Lb} \Lb_0$, which implies that
\beq\label{equ:I1}
\begin{aligned}
	\mathbf{I}_{r} &=\frac{1}{N} \Lb_0^{\top} \Lb_0
	=\Zb^{\top}\left(\frac{1}{N} \Lb^{\top} \Lb\right) \Zb+\frac{1}{N} \Lb_0^{\top} \Lb_0-\Zb^{\top}\left(\frac{1}{N} \Lb^{\top} \Lb\right) \Zb \\
	&=\Zb^{\top} \Zb+\frac{1}{N} \Lb_0^{\top} \Lb_0-\frac{1}{N} \Lb_0^{\top} \Lb \Zb
	=\Zb^{\top} \Zb+\Lb_0^{\top} \frac{1}{N}\left(\Lb_0-\Lb \Zb\right)\\
	&=\Zb^{\top} \Zb+\frac{1}{N} \Lb_0^{\top} \mathbf{M}_{\Lb} \Lb_0.
\end{aligned}
\eeq
Similarly,
$$
\mathbf{I}_{r}=\Zb \Zb^{\top}+\frac{1}{N}\Lb^{\top} \left(\Lb-\Lb_0 \Zb^{\top}\right)=\Zb \Zb^{\top}+\frac{1}{N} \Lb^{\top} \mathbf{M}_{\Lb_0} \Lb .
$$
In addition,
$$
\begin{aligned}
	\frac{1}{T} \mathbf{F}_{0} ^{\top}  \mathbf{F}_{0}
	=& \Zb^{\top}\left(\frac{1}{T}  \mathbf{F}^{\top} \mathbf{F} \right) \Zb
	+\frac{1}{T }  \mathbf{F}_{0}^{\top} \mathbf{F}_{0} -\Zb^{\top}\left(\frac{1}{T} \mathbf{F}^{\top} \mathbf{F}\right) \Zb \\
	=& \Zb^{\top}\left(\frac{1}{T}  \mathbf{F}^{\top} \mathbf{F} \right)\l(\Zb^\top\r)^{-1}\Zb^\top \Zb
	+\frac{1}{T}  \mathbf{F}_{0}^{\top}\l(\mathbf{F}_{0} -\Fb\Zb\r) +\frac{1}{T}\l(\mathbf{F}_{0} -\mathbf{F}\Zb\r)^\top\mathbf{F}\Zb\\
	=& \Zb^{\top}\left(\frac{1}{T} \mathbf{F}^{\top} \mathbf{F} \right) \l(\Zb^\top\r)^{-1}
	+\Zb^{\top}\left(\frac{1}{T}  \mathbf{F}^{\top} \mathbf{F} \right) \l(\Zb^\top\r)^{-1}\l(\Zb^\top\Zb-\Ib_{r}\r)\\
	&+\frac{1}{T}\mathbf{F}_{0} ^{\top} \l(\mathbf{F}_{0} -\Fb\Zb\r)
	+\frac{1}{T}\l(\mathbf{F}_{0} -\mathbf{F}\Zb\r)^\top\mathbf{F}\Zb.
\end{aligned}
$$
Then it follows from the above equation and (\ref{equ:I1}) that
$$
\left(\frac{1}{T}  \mathbf{F}_{0}^{\top} \mathbf{F}_{0} +\mathbf{D}\right) \Zb^{\top}=\Zb^{\top}\left(\frac{1}{T} \mathbf{F}^{\top} \mathbf{F}\right),
$$
where
$$
\begin{aligned}
	\mathbf{D}=& \Zb^{\top}\left(\frac{1}{T} \mathbf{F}^{\top} \mathbf{F} \right)\left(\Zb^{\top}\right)^{-1}\left(\mathbf{I}_{r}-\Zb^{\top} \Zb\right)
	-\frac{1}{T} \mathbf{F}_{0}^{\top}\left(\mathbf{F}_{0}- \mathbf{F}\Zb\right)
	-\frac{1}{T} \left(\mathbf{F}_{0} -\mathbf{F}\Zb \right)^{\top} \mathbf{F}\Zb  \\
	=& \Zb^{\top}\left(\frac{1}{T}\mathbf{F}^{\top} \mathbf{F} \right)\left(\Zb^{\top}\right)^{-1}\left(\frac{1}{N} \Lb_0^{\top} \mathbf{M}_{\Lb} \Lb_0\right)
	-\frac{1}{T}\mathbf{F}_{0}^{\top} \left(\mathbf{F}_{0} - \mathbf{F}\Zb\right)
	-\frac{1}{T}\left(\mathbf{F}_{0} -\mathbf{F}\Zb \right)^{\top} \mathbf{F} \Zb .
\end{aligned}
$$
From (\ref{equ:ML}), (\ref{equ:F1}) and (\ref{equ:F2}), we have that
$$
\begin{aligned}
	\left\|\mathbf{D}\right\|_F \lesssim & \left\|\Zb^{\top}\left(\frac{1}{T} \mathbf{F}^{\top} \mathbf{F} \right)\left(\Zb^{\top}\right)^{-1}\left(\frac{1}{N} \Lb_0^{\top} \mathbf{M}_{\Lb} \Lb_0\right)\right\|_F +\left\|\frac{1}{T}\mathbf{F}_{0}^{\top} \left(\mathbf{F}_{0} - \mathbf{F}\Zb\right) \right\|_F
	+\left\|\frac{1}{T}\left(\mathbf{F}_{0} -\mathbf{F}\Zb \right)^{\top} \mathbf{F} \Zb\right\|_F\\
	\lesssim& \l\|\Zb\r\|_F\frac{1}{T}\l\|\Fb\r\|_F^2\l(\l\|\Zb\r\|_F\r)^{-1}\frac{1}{N}\l\|\Mb_{\Lb}\Lb_0\r\|_F^2
	+\frac{1}{T}\l\|\Fb_0\r\|_F\l\|\Fb_0-\Fb\Zb\r\|_F
	+\frac{1}{T}\l\|\Fb_0-\Fb\Zb\r\|_F\l\|\Fb\r\|_F\l\|\Zb\r\|_F\\
	\lesssim& d\left(\theta, \theta_0\right) .
\end{aligned}
$$
By the Bauer-Fike theorem in \cite{golub2013matrix}, there is a eignvector of $\mathbf{F}_{0}^{\top} \mathbf{F}_{0}/T$ , $\mu\left(\mathbf{F}_{0}^{\top} \mathbf{F}_{0}/T\right)$, such that
$$
\begin{aligned}
	&\left|\lambda_{\min }\left(\frac{1}{T} \mathbf{F}^{\top} \mathbf{F}\right)-\mu\left(\frac{1}{T}  \mathbf{F}_{0}^{\top} \mathbf{F}_{0}\right)\right|
	\leq\left\|\mathbf{D}\right\|_2 \leq\left\|\mathbf{D}\right\|_F \lesssim d\left(\theta, \theta_0\right) .
\end{aligned}
$$
Therefore, we have $\lambda_{\min }\left(\mathbf{F}^{\top} \mathbf{F}\right) / T$ is bounded below by a positive constant.
Thus we have that
$$
\frac{1}{N}\left\|\mathbf{M}_{\Lb_0} \Lb\right\|_F^2
\lesssim d^2\left(\theta, \theta_0\right).
$$
This concludes the proof.

\begin{lemma}\label{lem:1.3}
	Under Assumptions \ref{asmp:1}-\ref{asmp:3}, we have that
	$$
	\EE\left[\sup _{\theta \in \Theta(\delta)}|L_1(\theta)|\right] \lesssim \frac{\delta}{\sqrt{L}},
	$$
	where $L=\min \left\{N, T\right\}$
\end{lemma}
\noindent\emph{\textbf{Proof.}}
According to (\ref{equ:psi2}) in Lemma \ref{lem:1.1}, it is easy to see that for any $\theta_a,~\theta_b\in\Theta$,
$$
\left\|\sqrt{T N}\left|L_1\left(\theta_a\right)-L_1\left(\theta_b\right)\right|\right\|_{\psi_2} \lesssim d\left(\theta_a, \theta_b\right) .
$$
Since the process $L_1(\theta)$ is separable, it follows from Theorem 2.2.4 of \cite{vanderVaart1996} that
$$
\EE\left[\sup _{\theta \in \Theta(\delta)} \sqrt{T N}|L_1(\theta)|\right] \lesssim\left\|\sup _{\theta \in \Theta(\delta)} \sqrt{T N}|L_1(\theta)|\right\|_{\psi_2} \lesssim \int_0^\delta \sqrt{\log D(\epsilon, d, \Theta(\delta))} d \epsilon,
$$
where $D(\cdot, g, \mathcal{G})$ is the packing number of space $\mathcal{G}$ with semimetric $g$.
Then we need to show that
$$\int_0^\delta \sqrt{\log D(\epsilon, d, \Theta(\delta))} d \epsilon /\sqrt{TN}=O_p\l(\delta/\sqrt{L}\r).$$

From Lemma \ref{lem:1.2}, it is clear that
$$
\Theta(\delta) \subset \bigcup_{\mathbf{U} \in \mathcal{S}} \Theta\left(\delta ; \mathbf{U}\right),
$$
where $\mathcal{S}=\left\{\mathbf{U} \in \mathbb{R}^{r \times r}: \mathbf{U}=\operatorname{diag}\left(u_1, \ldots, u_r\right), u_j \in\{-1,1\} \text{ for } j=1, \ldots, r\right\}$  and \\ $\Theta\left(\delta, \mathbf{U}\right)=\left\{\theta \in \Theta: \left\|\Lb-\Lb_0 \mathbf{U}\right\|_F/\sqrt{N}+\left\|\left(\Fb-\Fb_0\mathbf{U}\right)\right\|_F/\sqrt{T} \leq C_{4} \delta\right\}$.
Because there are $2^{r}$ elements in $\mathcal{S}$, we only to need  study the property of $\int_0^\delta \sqrt{\log D\left(\epsilon, d, \Theta(\delta) ; \mathbf{U}\right)} d \epsilon$ for each $\mathbf{U} \in \mathcal{S}$. Without loss of generality, we set $\mathbf{U}=\mathbf{I}_{r}$.
For any $\theta_a, \theta_b \in \Theta$, we have
$$
\begin{aligned}
	d\left(\theta_a, \theta_b\right) & =\frac{1}{\sqrt{TN}}\left\|\Lb_a \Fb_a^{\top}-\Lb_b \Fb_b^{\top}\right\|_F=\frac{1}{\sqrt{TN}}\left\|\Lb_a \Fb_a^{\top}-\Lb_b \Fb_a^{\top}+\Lb_b \Fb_a^{\top}-\Lb_b \Fb_b^{\top}\right\|_F \\
	& \leq \frac{1}{\sqrt{N}}\left\|\Lb_a-\Lb_b\right\|_F+\frac{\left\|\Lb_b\right\|_F}{\sqrt{N}} \cdot \frac{\left\|\Fb_a-\Fb_b\right\|_F}{\sqrt{T}} \leq C_{5}\left(\frac{\left\|\Lb_a-\Lb_b\right\|_F}{\sqrt{N}}+\frac{\left\|\Fb_a-\Fb_b\right\|_F}{\sqrt{T}}\right),
\end{aligned}
$$
where $C_{5} \geq 1$. Define
$$
d^*\left(\theta_a, \theta_b\right)=2 C_{5} \sqrt{\frac{\left\|\Lb_a-\Lb_b\right\|_F^2}{N}+\frac{\left\|\Fb_a-\Fb_b\right\|_F^2}{T}}.
$$
It is clearly that $d\l(\theta_a,\theta_b\r) \leq d^*\l(\theta_a,\theta_b\r)$. And $\Theta\l(\delta;\Ib_{r}\r) \subset \Theta^*(\delta)= \l\{\theta \in \Theta: d^*\left(\theta, \theta_0\right) \leq C_{6} \delta\r\}$, where $C_{6}=2C_{4} C_{5}$.
Then
\beq \label{equ:D}
D\left(\epsilon, d, \Theta\left(\delta ; \mathbf{I}_{r}\right)\right) \leq D\left(\epsilon, d^*, \Theta\left(\delta ; \mathbf{I}_{r}\right)\right) \leq D\left(\epsilon / 2, d^*, \Theta^*(\delta)\right) \leq C\left(\epsilon / 4, d^*, \Theta^*(\delta)\right) \text {, }
\eeq
where $C(\cdot, g, \mathcal{G})$ is the covering numberof space $\mathcal{G}$ with semimetric $g$.
Next we find an upper bound for $C\left(\epsilon / 4, d^*, \Theta^*(\delta)\right)$.

Let $\eta=\epsilon / 4$, $\theta_1^*, \ldots, \theta_J^*$ be the maximal set in $\Theta^*(\delta)$ such that $d^*\left(\theta_j^*, \theta_l^*\right)>\eta$, for all $j \neq l$.
Set $B(\theta, c)=\left\{\gamma \in \Theta: d^*(\gamma, \theta) \leq c\right\}$. Then $B\left(\theta_1^*, \eta\right), \ldots, B\left(\theta_J^*, \eta\right)$ cover $\Theta^*(\delta)$ and $C\left(\epsilon / 4, d^*, \Theta^*(\delta)\right) \leq J$. Moreover, $B\left(\theta_1^*, \eta / 4\right), \ldots, B\left(\theta_J^*, \eta / 4\right)$ are disjoint and
$$
\bigcup_{j=1}^J B\left(\theta_j^*, \eta / 4\right) \subset \Theta^*(\delta+\eta / 4).
$$
The volume of a ball defined by $d^*$ with radius $c$ is equal to $h_M c^M$, where $h_M$ is a constant. So
$$
J \cdot h_M\left(\frac{\eta}{4}\right)^M \leq h_M\left(C_{6}\left(\delta+\frac{\eta}{4}\right)\right)^M .
$$
Then
\beq \label{equ:J}
J \leq\left(\frac{C_{6}(4 \delta+\eta)}{\eta}\right)^M=\left(\frac{C_{6}(16 \delta+\epsilon)}{\epsilon}\right)^M \leq\left(\frac{C_{7} \delta}{\epsilon}\right)^M
\eeq
for $\epsilon \leq \delta$.
From (\ref{equ:D}) and (\ref{equ:J}), we have
$$
\begin{aligned}
	\int_0^\delta \sqrt{\log D\left(\epsilon, d, \Theta\left(\delta ; \mathbf{I}_{r}\right)\right)} d \epsilon & \leq \int_0^\delta \sqrt{\log C\left(\epsilon / 4, d^*, \Theta^*(\delta)\right)} d \epsilon \\
	& \leq \sqrt{rN+rT} \int_0^\delta \sqrt{\log (C_{7} \delta / \epsilon)} d \epsilon
\end{aligned}
$$
Thus
$$
\EE\left[\sup _{\theta \in \Theta(\delta)}|L_1(\theta)|\right] \lesssim \int_0^\delta \sqrt{\log D(\epsilon, d, \Theta(\delta))} d \epsilon / \sqrt{TN} \lesssim \frac{\delta}{\sqrt{L}},
$$
since $\int_0^\delta \sqrt{\log (C_{7} \delta / \epsilon)} d \epsilon=O\l(\delta\r)$.
This completes the proof.

\section{Proof of Theorem \ref{th:3}}	
\noindent\emph{\textbf{Definitions and Notations.}}
Let
$H_{\tau}^{(j)}(z)=(\partial / \partial z)^j H_\tau(z)$ for $j=1,2$.
For fixed $\bl_i, \bbf_t$, define
$$
\begin{aligned}
	& \bar{H}_{\tau}^{(j)}\left(Y_{i t}-\bl_i^{\top} \bbf_t\right)=\mathbb{E}\left[H_{\tau}^{(j)}\left(Y_{i t}-\bl_i^{\top} \bbf_t\right)\right] \text { and } \\
	& \tilde{H}_{\tau}^{(j)}\left(Y_{i t}-\bl_i^{\top} \bbf_t\right)=H_{\tau}^{(j)}\left(Y_{i t}-\bl_i^{\top} \bbf_t\right)-\bar{H}_{\tau}^{(j)}\left(Y_{i t}-\bl_i^{\top} \bbf_t\right) \quad \text { for } j=1,2.
\end{aligned}
$$
From Lemma \ref{lem:1.1}, we have that
$$
\bar{H}_{\tau}^{(1)}(x+\epsilon_{it})=\partial\EE H_{\tau}(x+\epsilon_{it})/\partial x=-\tau\int_{-\infty}^{-\tau-x}f_{it}(u)du+\int_{-\tau-x}^{\tau-x}(x+u)f_{it}(u)du+\tau\int_{\tau-x}^\infty f_{it}(u)du,
$$
and
$$
\bar{H}_{\tau}^{(2)}(x+\epsilon_{it})=\l(\partial/\partial x\r)^2\EE H_{\tau}(x+\epsilon_{it})=\int_{-\tau-x}^{\tau-x}f_{it}(u)du.
$$
To further simplify the notations, we define
$H_{it}=H_\tau(Y_{it}-\bl_{0i}^{\top} \bbf_{0t})$, $\bar H_{it}=\bar H_\tau(Y_{it}-\bl_{0i}^{\top} \bbf_{0t})$, $\widetilde H_{it}=\widetilde H_\tau(Y_{it}-\bl_{0i}^{\top} \bbf_{0t})$, and $H^{(j)}_{it}=H_\tau^{(j)}(Y_{it}-\bl_{0i}^{\top} \bbf_{0t})$, $\bar H_{it}^{(j)}=\bar H^{(j)}_\tau(Y_{it}-\bl_{0i}^{\top} \bbf_{0t})$, $\widetilde H^{(j)}_{it}=\widetilde H^{(j)}_\tau(Y_{it}-\bl_{0i}^{\top} \bbf_{0t})$ for $j=1,2$.

In addition, we have
$$
\l(\partial/\partial x\r)^3\EE H_{\tau}(x+\epsilon_{it})=-f_{it}(\tau-x)+f_{it}(-\tau-x)
$$
which is obviously meaningful on $\RR$, and we denote $\bar H_\tau^{(3)}(x+\epsilon_{it})=\l(\partial/\partial x\r)^3\EE H_{\tau}(x+\epsilon_{it}).$

To obtain the stochastic expansion of $\widehat{\bbf}_t$, define
$$
\mathbb{P}(\theta)=b\left[\frac{1}{2 N} \sum_{p=1}^r \sum_{q>p}^r\left(\sum_{i=1}^N \bl_{i p} \bl_{i q}\right)^2+\frac{1}{8 N} \sum_{k=1}^r\left(\sum_{i=1}^N \bl_{i k}^2-N\right)^2+\frac{1}{2 T} \sum_{p=1}^r \sum_{q>p}^r\left(\sum_{t=1}^T \bbf_{t p} \bbf_{t q}\right)^2\right]
$$
for some $b>0$. Define
$$
\bar{\mathcal{S}}^*(\theta)=\big[\underbrace{\ldots,-\frac{1}{\sqrt{TN}} \sum_{t=1}^T \bar{H}_\tau^{(1)}\left(Y_{i t}-\bl_i^{\top} \bbf_t\right) \bbf_t^{\top}, \ldots}_{1 \times N r} ,  \underbrace{\ldots,-\frac{1}{\sqrt{TN}} \sum_{i=1}^N \bar{H}_\tau^{(1)}\left(Y_{i t}-\bl_i^{\top} \bbf_t\right) \bl_i^{\top}, \ldots}_{1 \times T r}\big]^{\top},
$$
$\mathcal{S}(\theta)=\bar{\mathcal{S}}^*(\theta)+\partial \mathbb{P}(\theta) / \partial \theta, \quad \mathcal{H}(\theta)=\partial \bar{\mathcal{S}}^*(\theta) / \partial \theta^{\top}+\partial^2 \mathbb{P}(\theta) / \l(\partial \theta \partial \theta^{\top}\r)$,
and let $\mathcal{H}=\mathcal{H}\left(\theta_0\right)$.
Expanding $\mathcal{S}(\widehat{\theta})$ around $\mathcal{S}\left(\theta_0\right)$ gives
\beq\label{equ:S=S+H+R}
\mathcal{S}(\widehat{\theta})=\mathcal{S}\left(\theta_0\right)+\mathcal{H}\left(\widehat{\theta}-\theta_0\right)+\frac{1}{2} \mathcal{R}(\widehat{\theta}),
\eeq
where
$$
\mathcal{R}(\widehat{\theta})=\left(\sum_{j=1}^{(N+T)r} \partial \mathcal{H}\left(\theta^*\right) / \partial \theta_j \cdot\left(\widehat{\theta}_j-\theta_{0 j}\right)\right)\left(\widehat{\theta}-\theta_0\right)
$$
and $\theta^*$ lies between $\widehat{\theta}$ and $\theta_0$.
Further, define
$$
\begin{aligned}
	\mathcal{H}_d & =\left(\begin{array}{cc}
		\mathcal{H}_d^{L} & 0 \\
		0 & \mathcal{H}_d^F
	\end{array}\right), \quad \mathcal{H}_d^{L}=\frac{\sqrt{T}}{\sqrt{N}} \operatorname{diag}\left[\bPhi_{T, 1}, \ldots, \bPhi_{T, i}, \ldots, \bPhi_{T, N}\right], \\
	\mathcal{H}_d^F & =\frac{\sqrt{N}}{\sqrt{T}} \operatorname{diag}\left[\bPsi_{N, 1}, \ldots, \bPsi_{N, t}, \ldots, \bPsi_{N, T}\right],
\end{aligned}
$$
where
$$
\bPhi_{T, i}=\frac{1}{T} \sum_{t=1}^T \bar{H}_{i t}^{(2)} \bbf_{0 t} \bbf_{0 t}^{\top}, \quad \bPsi_{N, t}=\frac{1}{N} \sum_{i=1}^N \bar{H}_{i t}^{(2)} \bl_{0 i} \bl_{0 i}^{\top} .
$$

In this section, $\bl^*$, $\bbf^*$, and $\bl_i^*$, $\bbf^*_t$ just mean some median vectors, and may not be the same in different equalities.

\noindent\emph{\textbf{Proof.}}
We only prove the conclusion of $\widehat\bl_i$, because the proof of $\widehat\bbf_t$ is symmetric. Without loss of generality, we assume that $\widehat{\Sbb}=\operatorname{sgn}\left(\widehat{\Fb}^{\top} \Fb_0 / T\right)=\Ib_r$.

From the expansion in the proof of Lemma \ref{lem:3.2}, we have
$$
\begin{aligned}
	\bPhi_{T, i}\left(\widehat{\bl}_i-\bl_{0 i}\right)
	= & -\frac{1}{T} \sum_{t=1}^T \bar{H}_\tau^{(1)}\left(Y_{i t}-\widehat{\bl}_i^{\top} \widehat{\bbf}_t\right) \widehat{\bbf}_t+\frac{1}{T} \sum_{t=1}^T \bar{H}_{i t}^{(1)} \bbf_{0 t}+\frac{1}{T} \sum_{t=1}^T \bar{H}_{i t}^{(1)}\left(\widehat{\bbf}_t-\bbf_{0 t}\right) \\
	& -\frac{1}{T} \sum_{t=1}^T \bar{H}_{i t}^{(2)} \bbf_{0 t}\left(\widehat{\bbf}_t-\bbf_{0 t}\right)^{\top} \bl_{0 i}+O_p\left(\l\|\widehat\Fb-\Fb_0\r\|_F^2/T\right)\\
	&+O_p\left(\l\|\widehat\Fb-\Fb_0\r\|_F\l\|\widehat\bl_i-\bl_{0i}\r\|_2/\sqrt{T}\right)+O_p\left(\l\|\widehat\bl_i-\bl_{0i}\r\|_2^2\right)\\
	=& -\frac{1}{T} \sum_{t=1}^T \bar{H}_\tau^{(1)}\left(Y_{i t}-\widehat{\bl}_i^{\top} \widehat{\bbf}_t\right) \widehat{\bbf}_t +O_p(1/L)
\end{aligned}
$$
by Assumption \ref{asmp:3}, Lemma \ref{lem:3.2} and Lemma \ref{lem:3.4}.
Note that
$$
\begin{aligned}
	-\frac{1}{T} \sum_{t=1}^T \bar{H}_\tau^{(1)}\left(Y_{i t}-\widehat{\bl}_i^{\top} \widehat{\bbf}_t\right) \widehat{\bbf}_t
	=&\frac{1}{T} \sum_{t=1}^T \tilde{H}_\tau^{(1)}\left(Y_{i t}-\widehat{\bl}_i^{\top} \widehat{\bbf}_t\right) \widehat{\bbf}_t \\
	\quad=& \frac{1}{T} \sum_{t=1}^T \tilde{H}_{i t}^{(1)} \widehat{\bbf}_t-\frac{1}{T} \sum_{t=1}^T \tilde{H}_{i t}^{(2)} \left(\widehat{\bl}_i^{\top} \widehat{\bbf}_t-\bl_{0 i}^{\top} \bbf_{0 t}\right) \widehat{\bbf}_t+1/T\sum_{t=1}^To_p\l(\widehat\bl_i^\top\widehat\bbf_t-\bl_{0i}^\top\bbf_{0t}\r).
\end{aligned}
$$

First, following Lemma \ref{lem:3.5}, we have
$$
\frac{1}{T} \sum_{t=1}^T \tilde{H}_{i t}^{(1)} \widehat{\bbf}_t=\frac{1}{T} \sum_{t=1}^T \tilde{H}_{i t}^{(1)} \bbf_{0 t}+\frac{1}{T} \sum_{t=1}^T \tilde{H}_{i t}^{(1)}\left(\widehat{\bbf}_t-\bbf_{0 t}\right)=\frac{1}{T} \sum_{t=1}^T \tilde{H}_{i t}^{(1)} \bbf_{0 t}+O_p\left(1/L\right) .
$$
Then, we have that
$$
\begin{aligned}
	& \frac{1}{T} \sum_{t=1}^T \tilde{H}_{i t}^{(2)} \left(\widehat{\bl}_i^{\top} \widehat{\bbf}_t-\bl_{0 i}^{\top} \bbf_{0 t}\right) \widehat{\bbf}_t \\
	=& \frac{1}{T} \sum_{t=1}^T \tilde{H}_{i t}^{(2)} \widehat{\bbf}_t \left(\widehat{\bbf}_t-\bbf_{0 t}\right)^{\top} \widehat{\bl}_i+\frac{1}{T} \sum_{t=1}^T \tilde{H}_{i t}^{(2)} \widehat{\bbf}_t \bbf_{0 t}^{\top} \cdot\left(\widehat{\bl}_i-\bl_{0 i}\right) \\
	=& \frac{1}{T} \sum_{t=1}^T \tilde{H}_{i t}^{(2)} \bbf_{0 t} \left(\widehat{\bbf}_t-\bbf_{0 t}\right)^{\top} \widehat{\bl}_i+\frac{1}{T} \sum_{t=1}^T \tilde{H}_{i t}^{(2)}\left(\widehat{\bbf}_t-\bbf_{0 t}\right) \left(\widehat{\bbf}_t-\bbf_{0 t}\right)^{\top} \widehat{\bl}_i \\
	& +\frac{1}{T} \sum_{t=1}^T \tilde{H}_{i t}^{(2)} \bbf_{0 t} \bbf_{0 t}^{\top} \left(\widehat{\bl}_i-\bl_{0 i}\right)  +\frac{1}{T} \sum_{t=1}^T \tilde{H}_{i t}^{(2)}\left(\widehat{\bbf}_t-\bbf_{0 t}\right) \bbf_{0 t}^{\top} \left(\widehat{\bl}_i-\bl_{0 i}\right) .
\end{aligned}
$$
It is easy to see that the first term is $O_p\left(1/L\right)$ by Lemma \ref{lem:3.5},
the second term is $O_p\left(\l\|\widehat\Fb-\Fb_0\r\|_F^2/T\right)=O_p(1/L)$ by Theorem \ref{th:1},
the third term is $O_p(1/\sqrt{T})O_p\left(\left\|\widehat{\bl}_i-\bl_{0 i}\right\|_2\right)=O_p(1/L)$
and the forth term is $O_p\l(\l\|\widehat\Fb-\Fb_0\r\|_F/\sqrt{T}\r)O_p\left(\left\|\widehat{\bl}_i-\bl_{0 i}\right\|_2\right)=O_p(1/L)$. Thus, we have
$$
\frac{1}{T} \sum_{t=1}^T \tilde{H}_{i t}^{(2)} \left(\widehat{\bl}_i^{\top} \widehat{\bbf}_t-\bl_{0 i}^{\top} \bbf_{0 t}\right) \widehat{\bbf}_t=O_p(1/L).
$$
Next, following Lemma \ref{lem:3.2}
$$
\frac{1}{T} \sum_{t=1}^T\left(\widehat{\bl}_i^{\top} \widehat{\bbf}_t-\bl_{0 i}^{\top} \bbf_{0 t}\right) \lesssim\left\|\widehat{\bl}_i-\bl_{0 i}\right\|_2 +\frac{1}{T} \sum_{t=1}^T \left\|\widehat{\bbf}_t-\bbf_{0 t}\right\|_2=O_p(1/\sqrt{L}).
$$
Combining all the above results, we get
$$
\Phi_{T, i}\left(\widehat{\bl}_i-\bl_{0 i}\right)=\frac{1}{T} \sum_{t=1}^T \tilde{H}_{i t}^{(1)} \bbf_{0 t}+o_p\left(\l\|\widehat{\bl}_i-\bl_{0 i}\r\|_2\right),
$$
and it is easy to show that
$$
\frac{1}{T} \sum_{t=1}^T \bar{H}_{i t}^{(2)} \bbf_{0 t} \bbf_{0 t}^{\top} \rightarrow \Phi_i>0 \quad \text { and } \quad \frac{1}{\sqrt{T}} \sum_{t=1}^T \tilde{H}_{i t}^{(1)} \bbf_{0 t} \stackrel{d}{\rightarrow} \mathcal{N}\left(0, \bSigma_{L,i}\right),
$$
where $\bSigma_{L,i}=\lim_{T\rightarrow\infty}\sum_{t=1}^T\l(\int_{-\infty}^{\infty}\min\{u^2,\tau^2\}f_{it}(u)du\r) \bbf_{0t}\bbf_{0t}^\top/T$.
This concludes the proof.

\subsection{Complementary Lemmas}
\begin{lemma}\label{lem:3.1}
	Under Assumptions \ref{asmp:1}-\ref{asmp:4}, we have $\left\|\widehat{\bl}_i-\bl_{0 i}\right\|_2=o_p(1)$ for each $i$.
\end{lemma}	
\noindent\emph{\textbf{Proof.}}
Define:
$$
\begin{aligned}
	& L^\prime(\bl, \Fb)_{i,T}=\frac{1}{T} \sum_{t=1}^T\left[H_\tau\left(Y_{it}-\bl^{\top} \bbf_t\right)-H_\tau\left(Y_{i t}-\bl_{0i}^{\top} \bbf_{0t}\right)\right] \text {, } \\
	& \bar L^\prime(\bl, \Fb)_{i,T}=\frac{1}{T} \sum_{t=1}^T \mathbb{E}\left[H_\tau\left(Y_{it}-\bl^{\top} \bbf_t\right)-H_\tau\left(Y_{i t}-\bl_{0i}^{\top} \bbf_{0t}\right)\right].
\end{aligned}
$$

Note that
\beq\label{equ:l=argminL}
\widehat{\bl}_i=\underset{\bl \in \cL_r}{\operatorname{argmin}} L^\prime_{i,T}(\bl, \widehat\Fb).
\eeq
First we proof that $\sup_{\bl\in\cL_r}|L^\prime_{i,T}(\bl, \widehat\Fb)-\bar L^\prime_{i,T}(\bl, \Fb_0)|=o_p(1)$.
Since
$$
\begin{aligned}
	|L^\prime_{i,T}(\bl, \widehat\Fb)- L^\prime_{i,T}(\bl, \Fb_0)|
	&=|\frac{1}{T}\sum_{t=1}^T\left[H_\tau\left(Y_{it}-\bl^{\top} \widehat\bbf_t\right)-H_\tau\left(Y_{i t}-\bl^{\top} \bbf_{0t}\right)\right]|\\
	&\leq \frac{1}{T}\sum_{t=1}^T|H_\tau\left(Y_{it}-\bl^{\top} \widehat\bbf_t\right)-H_\tau\left(Y_{i t}-\bl^{\top} \bbf_{0t}\right)|\\
	&\leq \frac{1}{T}\sum_{t=1}^T\tau|\bl^\top\widehat\bbf_t-\bl^\top\bbf_{0t}|\leq \frac{1}{T}\sum_{t=1}^T\tau\|\bl\|_2\|\widehat\bbf_t-\bbf_{0t}\|_2,
\end{aligned}
$$
then
$$
\sup_{\bl\in\cL_r}|L^\prime_{i,T}(\bl, \widehat\Fb)- L^\prime_{i,T}(\bl, \Fb_0)|\lesssim \sup_{\bl\in\cL_r}\|\bl\|_2 \frac{1}{T}\sum_{t=1}^T\|\widehat\bbf_t-\bbf_{0t}\|_2 \lesssim \|\widehat\Fb-\Fb_0\|_F/\sqrt{T}=O_p(1/\sqrt{L}) .
$$
In addation, $\sup_{\bl\in\cL_r}|L^\prime_{i,T}(\bl, \Fb_0)-\bar L^\prime_{i,T}(\bl, \Fb_0)|=o_p(1)$.
Thus,
\beq\label{equ:L-Lbar}
\sup_{\bl\in\cL_r}|L^\prime_{i,T}(\bl, \widehat\Fb)-\bar L^\prime_{i,T}(\bl, \Fb_0)|\leq \sup_{\bl\in\cL_r}|L^\prime_{i,T}(\bl, \widehat\Fb)- L^\prime_{i,T}(\bl, \Fb_0)|+\sup_{\bl\in\cL_r}|L^\prime_{i,T}(\bl, \Fb_0)-\bar L^\prime_{i,T}(\bl, \Fb_0)|=o_p(1).
\eeq
Second, one can see that
$$
\begin{aligned}
	\bar L^\prime_{i,T}(\bl, \Fb_0)
	&=\frac{1}{T} \sum_{t=1}^T \mathbb{E}\left[H_\tau\left(Y_{it}-\bl^{\top} \bbf_{0t}\right)-H_\tau\left(Y_{i t}-\bl_{0i}^{\top} \bbf_{0t}\right)\right]\\
	&=\frac{1}{T} \sum_{t=1}^T\left\{\bar H^{(1)}_{it}\l(\bl-\bl_{0i}\r)^\top\bbf_{0t} +  \bar H_\tau^{(2)}\left(Y_{it}-\bl^{*\top} \bbf_{0t}\right)\l[\l(\bl-\bl_{0i}\r)^\top\bbf_{0t}\r]^2\right\}\\
	&> 0 =\bar L^\prime_{i,T}(\bl_{0i}, \Fb_0),
\end{aligned}
$$
where $\bl^*$ is lying between $\bl$ and $\bl_{0i}$.
Then for any $\eta>0$,
$$
\bar L^\prime_{i,T}\l(\widehat \bl_i, \Fb_0\r)
< L^\prime_{i,T}\l(\widehat \bl_i, \widehat \Fb\r)+\eta/2
< L^\prime_{i,T}\l(\bl_{0i}, \widehat \Fb\r)+\eta/2
< \bar L^\prime_{i,T}\l(\bl_{0i}, \Fb_0\r)+\eta.
$$
Let $B_i(\epsilon)=\{\bl\in \cL_r:~\|\bl-\bl_{0i}\|_2\leq\epsilon\}$, and $\bar L^\prime_{i,T}(\tilde\bl, \Fb_0)=\inf_{\bl\in B^c_i(\epsilon)}\bar L^\prime_{i,T}(\bl, \Fb_0)>\bar L^\prime_{i,T}\l(\bl_{0i}, \Fb_0\r)$ for any $\epsilon>0$. Then let $\eta=\bar L^\prime_{i,T}(\tilde\bl, \Fb_0)-\bar L^\prime_{i,T}\l(\bl_{0i}, \Fb_0\r)$, we have that $\bar L^\prime_{i,T}\l(\widehat \bl_i, \Fb_0\r) < \bar L^\prime_{i,T}(\tilde\bl, \Fb_0)$, which means that $\widehat \bl_i\in B_i(\epsilon)$. This concludes the proof.

\begin{lemma}\label{lem:3.2}
	Under Assumptions \ref{asmp:1}-\ref{asmp:4}, we have $\left\|\widehat{\bl}_i-\bl_{0 i}\right\|_2=O_p(1/\sqrt{L})$ for each $i$.
\end{lemma}	
\noindent\emph{\textbf{Proof.}}
It can be obtained by expanding $\bar H_\tau^{(1)}\left(Y_{i t}-\bl_i^{\top} \bbf_t\right) \bbf_t$ at $\bl_i=\bl_{0i}$ and $\bbf_t=\bbf_{0t}$ respectively that
$$
\begin{aligned}
	\bar H_\tau^{(1)}\left(Y_{i t}-\bl_i^{\top} \bbf_t\right) \bbf_t
	= &\bar H_\tau^{(1)}\left(Y_{i t}-\bl_{0i}^{\top} \bbf_t\right) \bbf_t-\bar H_\tau^{(2)}\left(Y_{i t}-\bl_{0 i}^{\top} \bbf_t\right) \bbf_t \bbf_t^{\top} \left(\bl_i-\bl_{0i}\right) \\
	& +\frac{1}{2} \bar H_{\tau}(Y_{it}-\bl_i^{*\top} \bbf_t) \bbf_t\left[\left(\bl_i-\bl_{0i}\right)^{\top} \bbf_t\right]^2 \\
	= &\bar H_{i t}^{(1)} \bbf_{0 t}+\bar H_\tau^{(1)}\left(Y_{i t}-\bl_{0 i}^{\top} \bbf_t^*\right)\left(\bbf_t-\bbf_{0 t}\right)-\bar H_\tau^{(2)}\left(Y_{i t}-\bl_{0 i}^{\top} \bbf_t^*\right) \bbf_t^* \bl_{0 i}^{\top}\left(\bbf_t-\bbf_{0 t}\right) \\
	& -\bar H_{i t}^{(2)} \bbf_t \bbf_t^{\top} \cdot\left(\bl_i-\bl_{0 i}\right) \\
	& +\bar H_{\tau}^{(3)}(Y_{it}-\bl_{0i}^\top\bbf_t^*)\bbf_t \bbf_t^{\top} \cdot\left(\bl_i-\bl_{0 i}\right) \bl_{0 i}^{\top}\left(\bbf_t-\bbf_{0 t}\right) \\
	& +\frac{1}{2} \bar H_{\tau}(Y_{it}-\bl_i^{*\top} \bbf_t) \bbf_t\left[\left(\bl_i-\bl_{0 i}\right)^{\top} \bbf_t\right]^2, \\
\end{aligned}
$$
where $\bl_i^*$ lies between $\bl_i$ and $\bl_{0 i}$ and $\bbf_t^*$ lies between $\bbf_t$ and $\bbf_{0 t}$.
Setting $\bl_i=\widehat\bl_i$ and $\bbf_t=\widehat\bbf_t$ and taking the average of both sides of the above equation, we have that
$$
\begin{aligned}
	\frac{1}{T} \sum_{t=1}^T \bar{H}_\tau^{(1)}\left(Y_{i t}-\widehat{\bl}_i^{\top} \widehat{\bbf}_t\right) \widehat{\bbf}_t
	=& \frac{1}{T} \sum_{t=1}^T \bar{H}_{i t}^{(1)} \bbf_{0 t}-\left(\frac{1}{T} \sum_{t=1}^T \bar{H}_{i t}^{(2)} \widehat{\bbf}_t \widehat{\bbf}_t^{\top}\right)\left(\widehat{\bl}_i-\bl_{0 i}\right) +O_p\left(\left\|\widehat{\Fb}-\Fb_0\right\|_F/\sqrt{T}\right)\\
	& +O_p\left(\left\|\widehat{\bl}_i-\bl_{0 i}\right\|_2\right) \cdot O_p\left(\left\|\widehat{\Fb}-\Fb_0\right\|_F/\sqrt{T}\right)+O_p\left(\left\|\widehat{\bl}_i-\bl_{0 i}\right\|_2^2\right) \text {. }
\end{aligned}
$$
Since
$$
\frac{1}{T} \sum_{t=1}^T \bar{H}_{i t}^{(2)} \widehat{\bbf}_t \widehat{\bbf}_t^{\top}=\frac{1}{T} \sum_{t=1}^T \bar{H}_{i t}^{(2)} \bbf_{0 t} \bbf_{0 t}^{\top}+o_p(1)=\bPhi_i+o_p(1) ,
$$
we get
\beq\label{equ:phi(l-l)}
\bPhi_i\left(\widehat{\bl}_i-\bl_{0 i}\right)+o_p\left(\left\|\widehat{\bl}_i-\bl_{0 i}\right\|_2\right)=O_p\left(1 / \sqrt{L}\right)-\frac{1}{T} \sum_{t=1}^T \bar{H}_\tau^{(1)}\left(Y_{i t}-\widehat{\bl}_i^{\top} \widehat{\bbf}_t\right) \widehat{\bbf}_t
\eeq
by Lemma \ref{lem:3.1} and Theorem \ref{th:1}.

Note that
$$
\begin{aligned}
	\frac{1}{T} \sum_{t=1}^T \bar{H}_\tau^{(1)}\left(Y_{i t}-\widehat{\bl}_i^{\top} \widehat{\bbf}_t\right) \widehat{\bbf}_t
	=&-\frac{1}{T} \sum_{t=1}^T \widetilde{H}_\tau^{(1)}\left(Y_{i t}-\widehat{\bl}_i^{\top} \widehat{\bbf}_t\right) \widehat{\bbf}_t\\
	=&-\frac{1}{T} \sum_{t=1}^T\l[\widetilde{H}_\tau^{(1)}\left(Y_{i t}-\widehat{\bl}_i^{\top} \widehat{\bbf}_t\right) \widehat{\bbf}_t-\widetilde{H}_\tau^{(1)}\left(Y_{i t}-\widehat{\bl}_i^{\top} \bbf_{0t}\right) \bbf_{0t}\r]\\
	&-\frac{1}{T} \sum_{t=1}^T\l[\widetilde{H}_\tau^{(1)}\left(Y_{i t}-\widehat{\bl}_i^{\top} \bbf_{0t}\right)-\widetilde{H}_{it}^{(1)}\r]\bbf_{0t}
	-\frac{1}{T} \sum_{t=1}^T \widetilde{H}_{it}^{(1)}\bbf_{0t}\\
	=&:\mathcal{III}+\mathcal{IV}+\mathcal{V}.
\end{aligned}
$$
For the first term,
$$
\begin{aligned}
	-\mathcal{III}=&\frac{1}{T} \sum_{t=1}^T\l\{\widetilde{H}_\tau^{(1)}\left(Y_{i t}-\widehat{\bl}_i^{\top} \widehat{\bbf}_t\right) \l(\widehat\bbf_t-\bbf_{0t}\r)
	+\l[\widetilde{H}_\tau^{(1)}\left(Y_{i t}-\widehat{\bl}_i^{\top} \widehat\bbf_t\right)-\widetilde{H}_\tau^{(1)}\left(Y_{i t}-\widehat{\bl}_i^{\top} \bbf_{0t}\right)\r] \bbf_{0t}\r\}\\
	=& O_p(\|\widehat\Fb-\Fb_0\|_F/\sqrt{T})+O_p\l(\frac{1}{T} \sum_{t=1}^T\l[\widetilde H_\tau^{(2)}\l(Y_{it}-\widehat\bl^\top\bbf_t^*\r)\widehat\bl_i^\top\l(\widehat\bbf_t-\bbf_{0t}\r)\r]\r)\\
	=&O_p( 1/\sqrt{L}),
\end{aligned}
$$
where $\bbf_t^*$ lies between $\widehat{\bbf}_t$ and $\bbf_{0t}$.

For the second term,
$$
\begin{aligned}
	-\mathcal{IV}=&\frac{1}{T} \sum_{t=1}^T\l[\widetilde{H}_\tau^{(2)}\left(Y_{i t}-\bl_i^{*\top} \bbf_{0t}\right)\l(\widehat\bl_i-\bl_{0i}\r)^\top\bbf_{0t}\r]\bbf_{0t}\\
	=&o_p(\|\widehat\bl_i-\bl_{0i}\|_2),
\end{aligned}
$$
where $\bl_i^*$ lies between $\widehat{\bl}_i$ and $\bl_{0i}$.

For the last term, following the fact that
$\|H^{(1)}_{it}\bbf_{0t}\|_2$ is bounded,
we have $\mathcal{V}=O_p(1/\sqrt{T})$ by Lyapunov’s CLT.

Combining the above results yields
$$
\frac{1}{T} \sum_{t=1}^T \bar{H}_\tau^{(1)}\left(Y_{i t}-\widehat{\bl}_i^{\top} \widehat{\bbf}_t\right) \widehat{\bbf}_t=O_p\left(\frac{1}{\sqrt{L}}\right)+o_p\left(\left\|\widehat{\bl}_i-\bl_{0 i}\right\|_2\right)
$$
and the desired result follows from (\ref{equ:phi(l-l)}).

\begin{lemma}\label{lem:3.3}
	The matrix $\mathcal{H}$ is invertible and $\left\|\mathcal{H}^{-1}-\mathcal{H}_d^{-1}\right\|_{\max }=O(1 / L)$.
\end{lemma}	
\noindent\emph{\textbf{Proof.}}
Without loss of generality, we assume that  $r=2$, i.e. $\bl_{0 i}=\left(\bl_{0 i, 1}, \bl_{0 i, 2}\right)^{\top}$ and $\bbf_{0 t}=\left(\bbf_{0 t, 1}, \bbf_{0 t, 2}\right)^{\top}$, and $\mathbb{P}(\theta)$ simplifies to
$$
\begin{aligned}
	\mathbb{P}(\theta)= & b\left[\frac{1}{2 N}\left(\sum_{i=1}^N \bl_{i 1} \bl_{i 2}\right)^2+\frac{1}{8 N}\left(\sum_{i=1}^N \bl_{i 1}^2-N\right)^2+\frac{1}{8 N}\left(\sum_{i=1}^N \bl_{i 2}^2-N\right)^2+\frac{1}{2 T}\left(\sum_{t=1}^T \bbf_{t 1} \bbf_{t 2}\right)^2\right].
\end{aligned}
$$
Denote $\partial^2 \mathbb{P}\left(\theta_0\right) / \l(\partial \theta \partial \theta^{\top}\r)=b\left(\sum_{k=1}^4 \bgamma_k \bgamma_k^{\top}\right)$, where
$$
\begin{aligned}
	& \bgamma_1^{\top}=\left[\left(\bl_{01,1}, 0\right), \ldots,\left(\bl_{0 i, 1}, 0\right), \ldots,\left(\bl_{0 N, 1}, 0\right), \mathbf{0}_{1 \times 2 T}\right] / \sqrt{N}, \\
	& \bgamma_2^{\top}=\left[\left(0, \bl_{01,2}\right), \ldots,\left(0, \bl_{0 i, 2}\right), \ldots,\left(0, \bl_{0 N, 2}\right), \mathbf{0}_{1 \times 2 T}\right] / \sqrt{N}, \\
	& \bgamma_3^{\top}=\left[\left(\bl_{01,2}, \bl_{01,1}\right), \ldots,\left(\bl_{0 i, 2}, \bl_{0 i, 1}\right), \ldots,\left(\bl_{0 N, 2}, \bl_{0 N, 1}\right), \mathbf{0}_{1 \times 2 T}\right] / \sqrt{N}, \\
	& \bgamma_4^{\top}=\left[\mathbf{0}_{1 \times 2 N},\left(\bbf_{01,2}, \bbf_{01,1}\right), \ldots,\left(\bbf_{0 t, 2}, \bbf_{0 t, 1}\right), \ldots,\left(\bbf_{0 T, 2}, \bbf_{0 T, 1}\right)\right] / \sqrt{T},
\end{aligned}
$$
and define
$$
\begin{aligned}
	& \bomega_1^{\top}=[\underbrace{\left(\bl_{01,1}, 0\right) / \sqrt{N}, \ldots,\left(\bl_{0 N, 1}, 0\right) / \sqrt{N}}_{\bomega_{1 L}^{\top}}, \underbrace{\left(-\bbf_{01,1}, 0\right) / \sqrt{T}, \ldots,\left(-\bbf_{0 T, 1}, 0\right) / \sqrt{T}}_{\bomega_{1 F}^{\top}}], \\
	& \bomega_2^{\top}=[\underbrace{\left(0, \bl_{01,2}\right) / \sqrt{N}, \ldots,\left(0, \bl_{0 N, 2}\right) / \sqrt{N}}_{\bomega_{2 L}^{\top}}, \underbrace{\left(0,-\bbf_{01,2}\right) / \sqrt{T}, \ldots,\left(0,-\bbf_{0 T, 2}\right) / \sqrt{T}}_{\bomega_{2 F}^{\top}}], \\
	& \bomega_3^{\top}=[\underbrace{\left(\bl_{01,2}, 0\right) / \sqrt{N}, \ldots,\left(\bl_{0 N, 2}, 0\right) / \sqrt{N}}_{\bomega_{3 L}^{\top}}, \underbrace{\left(0,-\bbf_{01,1}\right) / \sqrt{T}, \ldots,\left(0,-\bbf_{0 T, 1}\right) / \sqrt{T}}_{\bomega_{3 F}^{\top}}],\\
	&\bomega_4^{\top}=[\underbrace{\left(0, \bl_{01,1}\right) / \sqrt{N}, \ldots,\left(0, \bl_{0 N, 1}\right) / \sqrt{N}}_{\bomega_{4 L}^{\top}}, \underbrace{\left(-\bbf_{01,2}, 0\right) / \sqrt{T}, \ldots,\left(-\bbf_{0 T, 2}, 0\right) / \sqrt{T}}_{\bomega_{4 F}^{\top}}],
\end{aligned}
$$
and $\bomega=\left[\bomega_1, \bomega_2, \bomega_3, \bomega_4\right]$. We have
\beq\label{equ:ww}
\bomega \bomega^{\top}=\sum_{k=1}^4 \bomega_k \bomega_k^{\top}=\left(\begin{array}{cc}
	\sum_{k=1}^4 \bomega_{k L} \bomega_{k L}^{\top} & -(1/\sqrt{TN})\left\{\bbf_{0 t} \bl_{0 i}^{\top}\right\}_{i \leq N, t \leq T} \\
	-(1/\sqrt{TN})\left\{\bl_{0 i} \bbf_{0 t}^{\top}\right\}_{t \leq T, i \leq N,} & \sum_{k=1}^4 \bomega_{k F} \bomega_{k F}^{\top}
\end{array}\right),
\eeq
where $\left\{\bbf_{0 t} \bl_{0 i}^{\top}\right\}_{i \leq N, t \leq T}$ denotes a $2 N \times 2 T$ matrix whose $\{i, t\}$ th block is $\bbf_{0 t} \bl_{0 i}^{\top}$. Further, it is easy to see that
$$
\bomega^{\top} \bomega=\left(\begin{array}{cccc}
	\sigma_{T 1}+1 & 0 & 0 & 0 \\
	0 & \sigma_{T 2}+1 & 0 & 0 \\
	0 & 0 & \sigma_{T1}+1 & 0 \\
	0 & 0 & 0 & \sigma_{T2}+1
\end{array}\right) .
$$
with our identifiability conditions (\ref{equ:identify}).

Next, we project $\bgamma_k$ onto $\bomega$, and denote $\bgamma_k=\bomega \bbeta_k+\bzeta_k$ for $k=1, \ldots, 4$, where $\bbeta_k=$ $\left(\bomega^{\top} \bomega\right)^{-1} \bomega^{\top} \bgamma_k$. It can be obtained by simple calculation that
$$
\begin{aligned}
	& \bbeta_1=\left(\begin{array}{c}
		-\frac{1}{\sigma_{T 1}+1} \\
		0 \\
		0 \\
		0
	\end{array}\right), \quad \bbeta_2=\left(\begin{array}{c}
		0 \\
		-\frac{1}{\sigma_{T 2}+1} \\
		0 \\
		0
	\end{array}\right), \\
	& \bbeta_3=\left(\begin{array}{c}
		0 \\
		-\frac{1}{\sigma_{T1}+1} \\
		-\frac{1}{\sigma_{T2}+1}
	\end{array}\right), \quad \bbeta_4=\left(\begin{array}{c}
		0 \\
		\frac{\sigma_{T1}}{\sigma_{T1}+1} \\
		\frac{\sigma_{T2}}{\sigma_{T2}+1}
	\end{array}\right) .
\end{aligned}
$$
Define $\Bb_T=\sum_{k=1}^4 \bbeta_k \bbeta_k^{\top}$. Following our Assumption \ref{asmp:1} that $\sigma_{T 1} \rightarrow \sigma_{01}, \sigma_{T 2} \rightarrow \sigma_{02}$, and $\sigma_{01}>\sigma_{02}$, it can be seen that $\sigma_{T 1}-\sigma_{T 2}$ is bounded by a positive constant for all large $T$, and then there exists $\underline{\rho}>0$ such that $\rho_{\min }\left(\Bb_T\right)>\underline{\rho}$ for all large $T$.
Then we have that
\beq\label{equ:partial^2P}
\begin{aligned}
	\partial^2 \mathbb{P}\left(\theta\right) / \l(\partial \theta \partial \theta^{\top}\r)\big|_{\theta=\theta_0} & =b\left(\sum_{k=1}^4 \bgamma_k \bgamma_k^{\top}\right)=b \cdot \bomega\left(\sum_{k=1}^4 \bbeta_k \bbeta_k^{\top}\right) \bomega^{\top}+b\left(\sum_{k=1}^4 \bzeta_k \bzeta_k^{\top}\right) \\
	& =b \underline{\rho} \cdot \bomega \bomega^{\top}+b \cdot \bomega\left(\Bb_T-\underline{\rho} \cdot \Ib_4\right) \bomega^{\top}+b\left(\sum_{k=1}^4 \bzeta_k \bzeta_k^{\top}\right) .
\end{aligned}
\eeq
Let $\underline{b}=\min \{2\tau\underline{f}, b \underline{\rho}\}$. Then it follows from (\ref{equ:partial^2P}) that
$$
\begin{aligned}
	\mathcal{H}
	&=\partial \bar{\mathcal{S}}^*\left(\theta\right) / \partial \theta^{\top}\big|_{\theta=\theta_0}+\partial^2 \mathbb{P}\left(\theta\right) / \l(\partial \theta \partial \theta^{\top}\r)\big|_{\theta=\theta_0}\\
	& =\partial \bar{\mathcal{S}}^*\left(\theta\right) / \partial \theta^{\top}\big|_{\theta=\theta_0}+\underline{b} \cdot \bomega \bomega^{\top}+\underbrace{(b \underline{\rho}-\underline{b}) \cdot \bomega \bomega^{\top}}_{\geq 0}+\underbrace{b \cdot \bomega\left(\Bb_T-\underline{\rho} \Ib_4\right) \bomega^{\top}}_{\geq 0}+\underbrace{\left(\sum_{k=1}^4 \bzeta_k \bzeta_k^{\top}\right)}_{\geq 0} \\
	& \geq \partial \bar{\mathcal{S}}^*\left(\theta\right) / \partial \theta^{\top}\big|_{\theta=\theta_0}+\underline{b} \cdot \bomega \bomega^{\top} .
\end{aligned}
$$
The first term can be denoted as
$$
\begin{aligned}
	\partial \bar{\mathcal{S}}^*\left(\theta\right) / \partial \theta^{\top}\big|_{\theta=\theta_0}
	=& \left(\begin{array}{cc}
		\frac{1}{\sqrt{TN}} \operatorname{diag}\left[\left\{\sum_{t=1}^T \bar{H}_{i t}^{(2)} \bbf_{0 t} \bbf_{0 t}^{\top}\right\}_{i \leq N}\right] & \frac{1}{\sqrt{TN}}\left\{\bar{H}_{i t}^{(2)} \bbf_{0 t} \bl_{0 i}^{\top}\right\}_{i \leq N, t \leq T} \\
		\frac{1}{\sqrt{TN}}\left\{\bar{H}_{i t}^{(2)} \bl_{0 i} \bbf_{0 t}^{\top}\right\}_{t \leq T, i \leq N} & \frac{1}{\sqrt{TN}} \operatorname{diag}\left[\left\{\sum_{i=1}^N \bar{H}_{i t}^{(2)} \bl_{0 i} \bl_{0 i}^{\top}\right\}_{t \leq N}\right]
	\end{array}\right)\\
	=&\underline b \left(\begin{array}{cc}
		\operatorname{diag}\left[\left\{\frac{1}{\sqrt{TN}} \sum_{t=1}^T \bbf_{0 t} \bbf_{0 t}^{\top}\right\}_{i \leq N}\right] & \mathbf{0}_{2N\times 2T} \\
		\mathbf{0}_{2T\times 2N} & \operatorname{diag}\left[\left\{\frac{1}{\sqrt{TN}} \sum_{i=1}^N \bl_{0 i} \bl_{0 i}^{\top}\right\}_{t \leq N}\right]
	\end{array}\right)\\
	&+\underline b\left(\begin{array}{cc}
		\mathbf{0}_{2N\times 2N} & \frac{1}{\sqrt{TN}}\left\{\bbf_{0 t} \bl_{0 i}^{\top}\right\}_{i \leq N, t \leq T} \\
		\frac{1}{\sqrt{TN}}\left\{\bl_{0 i} \bbf_{0 t}^{\top}\right\}_{t \leq T, i \leq N} & \mathbf{0}_{2T\times 2T}
	\end{array}\right)\\
	&+\left(\begin{array}{cc}
		\frac{1}{\sqrt{TN}} \operatorname{diag}\left[\left\{\sum_{t=1}^T (\bar{H}_{i t}^{(2)}-\underline b)  \bbf_{0 t} \bbf_{0 t}^{\top}\right\}_{i \leq N}\right] & \frac{1}{\sqrt{TN}}\left\{(\bar{H}_{i t}^{(2)}-\underline b)  \bbf_{0 t} \bl_{0 i}^{\top}\right\}_{i \leq N, t \leq T} \\
		\frac{1}{\sqrt{TN}}\left\{(\bar{H}_{i t}^{(2)}-\underline b)  \bl_{0 i} \bbf_{0 t}^{\top}\right\}_{t \leq T, i \leq N} & \frac{1}{\sqrt{TN}} \operatorname{diag}\left[\left\{\sum_{i=1}^N (\bar{H}_{i t}^{(2)}-\underline b)  \bl_{0 i} \bl_{0 i}^{\top}\right\}_{t \leq N}\right]
	\end{array}\right)\\
	=&:\mathcal{VI}+\mathcal{VII}+\mathcal{VIII}.
\end{aligned}
$$

Following our Assumptions \ref{asmp:1} and \ref{asmp:4}, there exists a constant $\underline{c}>0$ such that
$$
\mathcal{VI}=\underline{b}\left(\begin{array}{cc}
	\sqrt{T / N} \cdot \Ib_{N}\otimes \operatorname{diag}\left(\sigma_{T 1}, \sigma_{T 2}\r) & \mathbf{0}_{2 N \times 2 T} \\
	\mathbf{0}_{2 T \times 2 N} & \sqrt{N / T} \cdot \Ib_{2T}
\end{array}\right) \geq \underline{c} \cdot \Ib_{2(N+T)} .
$$
From (\ref{equ:ww}), we have
$$
\mathcal{VII}+\underline{b} \cdot \bomega \bomega^{\top}=\underline{b} \cdot\left(\begin{array}{cc}
	\sum_{k=1}^4 \bomega_{k L} \bomega_{k L}^{\top} & \mathbf{0}_{2 N \times 2 T} \\
	\mathbf{0}_{2 T \times 2 N} & \sum_{k=1}^4 \bomega_{k F} \bomega_{k F}^{\top}
\end{array}\right) \geq 0 .
$$
For the last term, we have that
$$
\mathcal{VIII}=\frac{1}{\sqrt{TN}} \sum_{i=1}^N \sum_{t=1}^T\left(\bar{H}_{i t}^{(2)}-\underline{b}\right) \mu_{i t} \mu_{i t}^{\top} \geq 0,
$$
for $N, T$ large enough, where $\mu_{i t}=[\underbrace{\mathbf0_{1 \times 2}, \ldots, \bbf_{0 t}^{\top}, \ldots, \mathbf{0}_{1 \times 2}}_{1 \times 2 N}, \underbrace{\mathbf{0}_{1 \times 2}, \ldots, \bl_{0 i}^{\top}, \ldots, \mathbf{0}_{1 \times 2}}_{1 \times 2 T}]^{\top}$, following the fact that $\bar{H}_{i t}^{(2)}>2\tau\underline{f}$. Combining the above results yields
$$
\mathcal{H} \geq \partial \bar{\mathcal{S}}^*\left(\theta\right) / \partial \theta^{\top}\big|_{\theta=\theta_0}+\underline{b} \cdot \bomega \bomega^{\top}=\mathcal{VI}+\mathcal{VII}+\mathcal{VIII}+\underline{b} \cdot \bomega \bomega^{\top} \geq \underline{c} \cdot \Ib_{2(N+T)},
$$
and thus
\beq\label{equ:H-1}
\mathcal{H}^{-1} \leq \underline{c}^{-1} \cdot \Ib_{2(N+T)}
\eeq
Finally, denote $\mathcal{H}=\mathcal{H}_d+\mathcal{C}$, where
$$
\mathcal{C}=\left(\begin{array}{cc}
	\mathbf{0}_{2 N \times 2 N} & (1/\sqrt{TN})\left\{\bar{H}_{i t}^{(2)} \bbf_{0 t} \bl_{0 i}^{\top}\right\}_{i \leq N, t \leq T} \\
	(1/\sqrt{TN})\left\{\bar{H}_{i t}^{(2)} \bl_{0 i} \bbf_{0 t}^{\top}\right\}_{t \leq T, i \leq N} & \mathbf{0}_{2 T \times 2 T}
\end{array}\right)+b\left(\sum_{k=1}^4 \bgamma_k \bgamma_k^{\top}\right) .
$$
Following Sherman-Morrison formula, we have that $\mathcal{H}^{-1}=\mathcal{H}_d^{-1}-\mathcal{H}_d^{-1} \mathcal{C H}_d^{-1}+\mathcal{H}_d^{-1} \mathcal{C H}^{-1} \mathcal{C H}_d^{-1}$,
and thus $\left\|\mathcal{H}^{-1}-\mathcal{H}_d^{-1}\right\|_{\max } \leq$ $\left\|\mathcal{H}_d^{-1} \mathcal{C H}_d^{-1}\right\|_{\max }+\left\|\mathcal{H}_d^{-1} \mathcal{C H}^{-1} \mathcal{C H}_d^{-1}\right\|_{\max }$.
It is easy to see from inequality (\ref{equ:H-1}) that $\mathcal{H}_d^{-1} \mathcal{C H}^{-1} \mathcal{C H}_d^{-1} \leq$ $\underline{c}^{-1} \mathcal{H}_d^{-1} \mathcal{C}^2 \mathcal{H}_d^{-1}$, and thus the $j$ th diagonal element of $\mathcal{H}_d^{-1} \mathcal{CH}^{-1} \mathcal{CH}_d^{-1}$ is smaller than the $j$ th diagonal element of $\underline{c}^{-1} \mathcal{H}_d^{-1} \mathcal{C}^2 \mathcal{H}_d^{-1}$. Then we have
$\left\|\mathcal{H}_d^{-1} \mathcal{C H}^{-1} \mathcal{C H}_d^{-1}\right\|_{\max } \leq$ $\underline{c}^{-1}\left\|\mathcal{H}_d^{-1} \mathcal{C}^2 \mathcal{H}_d^{-1}\right\|_{\max }$
from the fact that the elements with the largest absolute value of the positive semidefinite matrix are all on the diagonal.
Therefore,
$$
\left\|\mathcal{H}^{-1}-\mathcal{H}_d^{-1}\right\|_{\max } \leq\left\|\mathcal{H}_d^{-1} \mathcal{C} \mathcal{H}_d^{-1}\right\|_{\max }+\underline{c}^{-1}\left\|\mathcal{H}_d^{-1} \mathcal{C}^2 \mathcal{H}_d^{-1}\right\|_{\max }.
$$
Since the elements of $\mathcal{H}_d^{-1}$ are all $O(1)$ by Assumption \ref{asmp:4}, and both $\|\mathcal{C}\|_{\max }$ and $\left\|\mathcal{C}^2\right\|_{\max }$ are $O(1 / L)$, the desired result follows.

\begin{lemma}\label{lem:3.4}
	Let $c_1, \ldots, c_T$ be a sequence of uniformly bounded constants. Then, under Assumptions \ref{asmp:1}-\ref{asmp:4}, we have
	$$
	\frac{1}{T} \sum_{t=1}^T c_t\left(\widehat{\bbf}_t-\bbf_{0 t}\right)=O_p\left(\frac{1}{L}\right)
	$$
\end{lemma}	
\noindent\emph{\textbf{Proof.}}
Following our identifiability condition (\ref{equ:identify}), we have that $\partial \mathbb{P}(\theta) / \partial \theta\big|_{\theta=\widehat\theta}=\partial \mathbb{P}\left(\theta\right) / \partial \theta\big|_{\theta=\theta_0}=0$. Then it can be see from (\ref{equ:S=S+H+R}) that
\beq\label{equ:theta-theta}
\widehat{\theta}-\theta_0=\mathcal{H}^{-1} \bar{\mathcal{S}}^*(\widehat{\theta})-\mathcal{H}^{-1} \bar{\mathcal{S}}^*\left(\theta_0\right)-\frac{1}{2} \mathcal{H}^{-1} \mathcal{R}(\widehat{\theta})
\eeq
Define
$$
\mathcal{S}^*(\theta)
=\big[\underbrace{\ldots,-\frac{1}{\sqrt{TN}} \sum_{t=1}^T H_\tau^{(1)}\left(Y_{i t}-\bl_i^{\top} \bbf_t\right) \bbf_t^{\top}, \ldots}_{1 \times N r},  \underbrace{\ldots,-\frac{1}{\sqrt{TN}} \sum_{i=1}^N H_\tau^{(1)}\left(Y_{i t}-\bl_i^{\top} \bbf_t\right) \bl_i^{\top}, \ldots}_{1 \times T r}\big]^\top,
$$
$\widetilde{\mathcal{S}}^*(\theta)=\mathcal{S}^*(\theta)-\bar{\mathcal{S}}^*(\theta)$, and $\mathcal{D}=\mathcal{H}^{-1}-\mathcal{H}_d^{-1}$. Note that by (\ref{equ:l=argminL}), $\mathcal{S}^*(\widehat{\theta})=0$. Then,
\beq\label{equ:H^-1S}
\begin{aligned}
	\mathcal{H}^{-1} \bar{\mathcal{S}}^*(\widehat{\theta})
	=&\mathcal{H}_d^{-1} \bar{\mathcal{S}}^*(\widehat{\theta})+\mathcal{D} \bar{\mathcal{S}}^*(\widehat{\theta})=-\mathcal{H}_d^{-1} \widetilde{\mathcal{S}}^*(\widehat{\theta})+\mathcal{D} \bar{\mathcal{S}}^*(\widehat{\theta})\\
	= & -\mathcal{H}_d^{-1} \widetilde{\mathcal{S}}^*\left(\theta_0\right)-\mathcal{H}_d^{-1}\left(\widetilde{\mathcal{S}}^*(\widehat{\theta})-\widetilde{\mathcal{S}}^*\left(\theta_0\right)\right)+\mathcal{D} \bar{\mathcal{S}}^*(\widehat{\theta}) \\
	= & -\mathcal{H}_d^{-1} \widetilde{\mathcal{S}}^*\left(\theta_0\right)-\mathcal{H}_d^{-1}\left(\widetilde{\mathcal{S}}^*(\widehat{\theta})-\widetilde{\mathcal{S}}^*\left(\theta_0\right)\right) \\
	& -\mathcal{D} \widetilde{\mathcal{S}}^*\left(\theta_0\right)-\mathcal{D}\left(\widetilde{\mathcal{S}}^*(\widehat{\theta})-\widetilde{\mathcal{S}}^*\left(\theta_0\right)\right) .
\end{aligned}
\eeq
Let $\mathcal{R}(\widehat{\theta})_j$ denote the vector containing the $(j-1) r+1$ th to the $j r$ th elements of $\mathcal{R}(\widehat{\theta})$ for $j=1, \ldots, N+T$, and let $\bar{O}_p(\cdot)$ denote a stochastic order that is uniformly in $i$ and $t$. Then, it can be shown that
\beq\label{equ:Ri}
\mathcal{R}(\widehat{\theta})_i=\bar{O}_p(1 / L) \text{, for $i=1, \ldots, N$ }
\eeq
and
\beq\label{equ:Rt}
\mathcal{R}(\widehat{\theta})_{N+t}=\bar{O}_p(1 / L) \text{, for $t=1, \ldots, T$ }
\eeq
by Theorem \ref{th:1} and Lemma \ref{lem:3.2}.
Write $\mathcal{D}_{j, s}$ as the $r \times r$ matrix containing the $(j-1) r+1$ to $j r$ rows and $(s-1) r+1$ to $s r$ columns of $\mathcal{D}$. Note that $\left\|\mathcal{H}^{-1} \bar{\mathcal{S}}^*\left(\theta_0\right)\right\|_{\max }=0$. Then, we can obtain the following equation from (\ref{equ:theta-theta}) and (\ref{equ:H^-1S}) that
\beq\label{equ:f-f}
\begin{aligned}
	\widehat{\bbf}_t-\bbf_{0 t}
	= & \left(\bPsi_{N, t}\right)^{-1} \frac{1}{N} \sum_{j=1}^N \tilde{H}_{j t}^{(1)} \bl_{0 j} +\left(\bPsi_{N, t}\right)^{-1} \frac{1}{N} \sum_{j=1}^N\left\{\tilde{H}_\tau^{(1)}\left(Y_{j t}-\widehat{\bl}_j^{\top} \widehat{\bbf}_t\right) \widehat{\bl}_j-\tilde{H}_{j t}^{(1)} \bl_{0 j}\right\} \\
	& -\frac{1}{\sqrt{TN}} \sum_{j=1}^N \sum_{s=1}^T \mathcal{D}_{N+t, j}\left\{\bar{H}_\tau^{(1)}\left(Y_{j s}-\widehat{\bl}_j^{\top} \widehat{\bbf}_s\right) \widehat{\bbf}_s\right\}\\
	& -\frac{1}{\sqrt{TN}} \sum_{j=1}^N \sum_{s=1}^T \mathcal{D}_{N+t, N+s}\left\{\bar{H}_\tau^{(1)}\left(Y_{j s}-\widehat{\bl}_j^{\top} \widehat{\bbf}_s\right) \widehat{\bl}_j\right\} \\
	& -\frac{1}{2}\left(\bPsi_{N, t}\right)^{-1} \mathcal{R}(\widehat{\theta})_{N+t}-\frac{1}{2} \sum_{j=1}^N \mathcal{D}_{N+t, j} \mathcal{R}(\widehat{\theta})_j -\frac{1}{2} \sum_{s=1}^T \mathcal{D}_{N+t, N+s} \mathcal{R}(\widehat{\theta})_{N+s} \\
	= & \left(\bPsi_{N, t}\right)^{-1} \frac{1}{N} \sum_{j=1}^N \tilde{H}_{j t}^{(1)} \bl_{0 j}+\frac{1}{\sqrt{TN}} \sum_{j=1}^N \sum_{s=1}^T \mathcal{D}_{N+t, j} \cdot \tilde{H}_{j s}^{(1)} \cdot \bbf_{0 s}\\
	&  +\frac{1}{\sqrt{TN}} \sum_{j=1}^N \sum_{s=1}^T \mathcal{D}_{N+t, N+s} \cdot \tilde{H}_{j s}^{(1)} \cdot \bl_{0 j} \\
	& +\left(\bPsi_{N, t}\right)^{-1} \frac{1}{N} \sum_{j=1}^N\left\{\tilde{H}_\tau^{(1)}\left(Y_{j t}-\widehat{\bl}_j^{\top} \widehat{\bbf}_t\right) \widehat{\bl}_j-\tilde{H}_{j t}^{(1)} \bl_{0 j}\right\} \\
	& +\frac{1}{\sqrt{TN}} \sum_{j=1}^N \sum_{s=1}^T \mathcal{D}_{N+t, j}\left\{\tilde{H}_\tau^{(1)}\left(Y_{j s}-\widehat{\bl}_j^{\top} \widehat{\bbf}_s\right) \widehat{\bbf}_s-\tilde{H}_{j s}^{(1)} \bbf_{0 s}\right\} \\
	& +\frac{1}{\sqrt{TN}} \sum_{j=1}^N \sum_{s=1}^T \mathcal{D}_{N+t, N+s}\left\{\tilde{H}_\tau^{(1)}\left(Y_{j s}-\widehat{\bl}_j^{\top} \widehat{\bbf}_s\right) \widehat{\bl}_j-\tilde{H}_{j s}^{(1)} \bl_{0 j}\right\} \\
	& -\frac{1}{2}\left(\bPsi_{N, t}\right)^{-1} \mathcal{R}(\widehat{\theta})_{N+t}-\frac{1}{2} \sum_{j=1}^N \mathcal{D}_{N+t, j} \mathcal{R}(\widehat{\theta})_j  -\frac{1}{2} \sum_{s=1}^T \mathcal{D}_{N+t, N+s} \mathcal{R}(\widehat{\theta})_{N+s} .
\end{aligned}
\eeq

Define $d_j=\sqrt{TN} \cdot T^{-1} \sum_{t=1}^T c_t \mathcal{D}_{N+t, j}$ for $j=1, \ldots, N+T$, $\max _{1 \leq j \leq N+T}\left\|d_j\right\|$ is bounded by Lemma \ref{lem:3.3}.
From (\ref{equ:f-f}), we have
$$
\begin{aligned}
	& \frac{1}{T} \sum_{t=1}^T c_t\left(\widehat{\bbf}_t-\bbf_{0 t}\right) \\
	&= \frac{1}{TN} \sum_{j=1}^N \sum_{t=1}^T c_t\left(\bPsi_{N, t}\right)^{-1} \tilde{H}_{j t}^{(1)} \bl_{0 j}+\frac{1}{TN} \sum_{j=1}^N \sum_{s=1}^T d_j \tilde{H}_{j s}^{(1)} \bbf_{0 s}+\frac{1}{TN} \sum_{j=1}^N \sum_{s=1}^T d_{N+s} \tilde{H}_{j s}^{(1)} \bl_{0 j} \\
	&+\frac{1}{TN} \sum_{j=1}^N \sum_{t=1}^T c_t\left(\bPsi_{N, t}\right)^{-1}\left\{\tilde{H}_\tau^{(1)}\left(Y_{j t}-\widehat{\bl}_j^{\top} \widehat{\bbf}_t\right) \widehat{\bl}_j-\tilde{H}_{j t}^{(1)} \bl_{0 j}\right\} \\
	&+\frac{1}{TN} \sum_{j=1}^N \sum_{s=1}^T d_j\left\{\tilde{H}_\tau^{(1)}\left(Y_{j s}-\widehat{\bl}_j^{\top} \widehat{\bbf}_s\right) \widehat{\bbf}_s-\tilde{H}_{j s}^{(1)} \bbf_{0 s}\right\} \\
	&+\frac{1}{TN} \sum_{j=1}^N \sum_{s=1}^T d_{N+s}\left\{\tilde{H}_\tau^{(1)}\left(Y_{j s}-\widehat{\bl}_j^{\top} \widehat{\bbf}_s\right) \widehat{\bl}_j-\tilde{H}_{j s}^{(1)} \bl_{0 j}\right\}\\
	&-\frac{1}{2} \frac{1}{T} \sum_{t=1}^T c_t\left(\bPsi_{N, t}\right)^{-1} \mathcal{R}(\widehat{\theta})_{N+t} -\frac{1}{2} \frac{1}{\sqrt{TN}} \sum_{j=1}^N d_j \mathcal{R}(\widehat{\theta})_j-\frac{1}{2} \frac{1}{\sqrt{TN}} \sum_{s=1}^T d_{N+s} \mathcal{R}(\widehat{\theta})_{N+s}.
\end{aligned}
$$
First, since $\|c_t\left(\bPsi_{N, t}\right)^{-1} \tilde{H}_{j t}^{(1)} \bl_{0 j}\|_2$, $\|d_j \tilde{H}_{j s}^{(1)} \bbf_{0 s}\|_2$ and $\|d_{N+s} \tilde{H}_{j s}^{(1)} \bl_{0 j}\|_2$ are bounded, it is easy to see that the first three terms are all $O_p(1 / \sqrt{TN})$ by Lyapunov’s CLT.
Next, the last four terms are all $O_p\left(1 / L\right)$  following from (\ref{equ:Ri}) and (\ref{equ:Rt}).
Finally, we will show that the remaining three terms are all $O_p(1 /L)$.

Define
$$
\mathbb{V}_{TN}(\theta)=\frac{1}{TN} \sum_{j=1}^N \sum_{s=1}^T d_j\left\{\tilde{H}_\tau^{(1)}\left(Y_{j s}-\bl_j^{\top} \bbf_s\right) \bbf_s-\tilde{H}_{j s}^{(1)} \bbf_{0 s}\right\},
$$
and $\Delta_{TN}\left(\theta_a, \theta_b\right)=\sqrt{TN} \left[\mathbb{V}_{TN}\left(\theta_a\right)-\mathbb{V}_{TN}\left(\theta_b\right)\right]$. Note that
$$
\begin{aligned}
	\Delta_{TN}\left(\theta_a, \theta_b\right)
	=&\frac{1}{\sqrt{TN}} \sum_{j=1}^N \sum_{s=1}^T d_j \cdot \tilde{H}_\tau^{(1)}\left(Y_{j s}-\bl_{a j}^{\top} \bbf_{a s}\right) \cdot\left(\bbf_{a s}-\bbf_{b s}\right)\\
	&+\frac{1}{\sqrt{TN}} \sum_{j=1}^N \sum_{s=1}^T d_j \cdot\left[\tilde{H}_\tau^{(1)}\left(Y_{j s}-\bl_{a j}^{\top} \bbf_{a s}\right)-\tilde{H}_\tau^{(1)}\left(Y_{j s}-\bl_{b j}^{\top} \bbf_{b s}\right)\right] \cdot \bbf_{b s}\\
	=&:\Delta_{1,TN}(\theta_a,\theta_b) +\Delta_{2,TN}(\theta_a,\theta_b).
\end{aligned}
$$
It is easy to see that
$$
\begin{aligned}
	& \left|d_j \cdot \tilde{H}_\tau^{(1)}\left(Y_{j s}-\bl_{a j}^{\top} \bbf_{a s}\right) \cdot\left(\bbf_{a s}-\bbf_{b s}\right)\right| \lesssim\left\|\bbf_{a s}-\bbf_{b s}\right\|_2, \\
	& \left|H_\tau^{(1)}\left(Y_{j s}-\bl_{a j}^{\top} \bbf_{a s}\right)-H_\tau^{(1)}\left(Y_{j s}-\bl_{b j}^{\top} \bbf_{b s}\right)\right| \lesssim\left|\bl_{a j}^{\top} \bbf_{a s}-\bl_{b j}^{\top} \bbf_{b s}\right| .
\end{aligned}
$$
By Hoeffding's inequality and Lemma 2.2.1 of \cite{vanderVaart1996}, we can show that for $d_F(\Fb_a,\Fb_b)=:\sqrt{\sum_{t=1}^T\left\|\bbf_{a s}-\bbf_{b s}\right\|_2^2/T}$ and $d\left(\theta_a, \theta_b\right)$ sufficiently small,
$$
\left\|\Delta_{1, TN}\left(\theta_a, \theta_b\right)\right\|_{\psi_2} \lesssim d_F(\Fb_a,\Fb_b) \quad \text { and } \quad\left\|\Delta_{2, TN}\left(\theta_a, \theta_b\right)\right\|_{\psi_2} \lesssim d\left(\theta_a, \theta_b\right) .
$$
Similar to the proof of Lemma \ref{lem:1.3},
$$
\mathbb{E}\left[\sup _{\theta \in \Theta(\delta)}\left|\Delta_{1, TN}\left(\theta,\theta_0\right)\right|\right]\lesssim O_p(\delta/\sqrt{N})
\text{ and }\mathbb{E}\left[\sup _{\theta \in \Theta(\delta)}\left|\Delta_{2,TN}(\theta,\theta_0)\right|\right] \lesssim O_p(\delta/\sqrt{L}).
$$
Thus, $\mathbb{V}_{TN}(\widehat{\theta})=O_p(1 /L)$ following the fact that $d(\widehat\theta,\theta_0)=O_p(1/\sqrt{L})$ which indicates that the fifth term is $O_p(1 /L)$. Similarly, we can get the same result for the fourth and sixth terms, which concludes the result.

\begin{lemma}\label{lem:3.5}
	Under Assumptions \ref{asmp:1}-\ref{asmp:4}, for each $i$ we have
	$$
	\frac{1}{T} \sum_{t=1}^T \tilde{H}_{i t}^{(1)}\left(\widehat{\bbf}_t-\bbf_{0 t}\right)=O_p\left(\frac{1}{L}\right)
	\text{ and }
	\frac{1}{T} \sum_{t=1}^T \tilde{H}_{i t}^{(2)} \bbf_{0 t} \left(\widehat{\bbf}_t-\bbf_{0 t}\right)^{\top}=O_p\left(\frac{1}{L}\right).
	$$
\end{lemma}	
\noindent\emph{\textbf{Proof.}}
To save space, we only prove the first result here, and the proof of the second is similar.
Using (\ref{equ:f-f}), we have that
$$
\begin{aligned}
	& \frac{1}{T} \sum_{t=1}^T \tilde{H}_{i t}^{(1)} \left(\widehat{\bbf}_t-\bbf_{0 t}\right) \\
	=& \frac{1}{TN} \sum_{t=1}^T \sum_{j=1}^N \tilde{H}_{i t}^{(1)} \tilde{H}_{j t}^{(1)} \left(\bPsi_{N, t}\right)^{-1}\bl_{0 j} \\
	&+ \frac{1}{TN} \sum_{t=1}^T \sum_{j=1}^N \tilde{H}_{i t}^{(1)}\left(\bPsi_{N, t}\right)^{-1}  \left\{\tilde{H}_\tau^{(1)}\left(Y_{j t}-\widehat{\bl}_j^{\top} \widehat{\bbf}_t\right) \widehat{\bl}_j-\tilde{H}_{j t}^{(1)} \bl_{0 j}\right\} \\
	&- \frac{1}{\sqrt{TN}} \sum_{j=1}^N \sum_{s=1}^T\left(\frac{1}{T} \sum_{t=1}^T \tilde{H}_{i t}^{(1)}  \mathcal{D}_{N+t, j}\right) \bar{H}_\tau^{(1)}\left(Y_{j s}-\widehat{\bl}_j^{\top} \widehat{\bbf}_s\right) \widehat{\bbf}_s \\
	&- \frac{1}{\sqrt{TN}} \sum_{j=1}^N \sum_{s=1}^T\left(\frac{1}{T} \sum_{t=1}^T \tilde{H}_{i t}^{(1)} \mathcal{D}_{N+t, N+s}\right) \bar{H}_\tau^{(1)}\left(Y_{j s}-\widehat{\bl}_j^{\top} \widehat{\bbf}_s\right)\widehat{\bl}_j \\
	&- \frac{1}{2 T} \sum_{t=1}^T \tilde{H}_{i t}^{(1)}\left(\bPsi_{N, t}\right)^{-1} \mathcal{R}(\widehat{\theta})_{N+t} -\frac{1}{2 T} \sum_{j=1}^N \sum_{t=1}^T \tilde{H}_{i t}^{(1)} \mathcal{D}_{N+t, j} \mathcal{R}(\widehat{\theta})_j \\
	&- \frac{1}{2 T} \sum_{s=1}^T \sum_{t=1}^T \tilde{H}_{i t}^{(1)}\mathcal{D}_{N+t, N+s} \mathcal{R}(\widehat{\theta})_{N+s} \\
	=&: (i)+(ii)+(iii)+(iv)+(v)+(vi)+(vii) .
\end{aligned}
$$
First, we can write
$$
(i)=\frac{1}{TN} \sum_{t=1}^T \tilde{H}_{i t}^{(1)} \tilde{H}_{i t}^{(1)} \left(\bPsi_{N, t}\right)^{-1}\bl_{0 i}+\frac{1}{TN} \sum_{t=1}^T \sum_{j=1, j \neq i}^N \tilde{H}_{i t}^{(1)} \tilde{H}_{j t}^{(1)} \left(\bPsi_{N, t}\right)^{-1}\bl_{0 j} .
$$
Since $H_{i t}^{(1)}(\cdot)$ is uniformly bounded and $\max _{t \leq T}\left\|\left(\bPsi_{N, t}\right)^{-1}\right\|_F=O(1)$ for large $N$ by Assumption \ref{asmp:4}, the first term is $O_p\left(1/N\right)$. Using Lyapunov's CLT and the fact that $\tilde{H}_{i t}^{(1)} $ is independent with $\tilde{H}_{j t}^{(1)} $, the second term can be shown to be $O_p\left(1/\sqrt{TN}\right)$. Thus, $(i)=O_p\left(1/L\right)$.

Second, similar to the proof of Lemma \ref{lem:3.4}, one can see that $(ii)=O_p\left(1/L\right).$

Next, the $p$th element of $(iii)$ can be denoted as
$$
\frac{1}{TN} \sum_{j=1}^N \sum_{s=1}^T \chi_{i, j} \cdot \bar{H}_\tau^{(1)}\left(Y_{j s}-\widehat{\bl}_j^{\top} \widehat{\bbf}_s\right) \widehat{\bbf}_s
$$
where $\chi_{i, j}=1/T \sum_{t=1}^T\left(\sqrt{TN} \mathcal{D}_{N+t, j, p}\right) \tilde{H}_{i t}^{(1)}$, and $\mathcal{D}_{N+t, j, p}$ is the $p$th row of $\mathcal{D}_{N+t, j}$. Then we have
$$
\left\|(iii)\right\|_2
\leq \sqrt{\frac{1}{N} \sum_{j=1}^N\left\|\chi_{i, j}\right\|_2^2} \sqrt{\frac{1}{TN} \sum_{j=1}^N \sum_{s=1}^T\left[\bar{H}_\tau^{(1)}\left(Y_{j s}-\widehat{\bl}_j^{\top} \widehat{\bbf}_s\right)\right]^2\left\|\widehat{\bbf}_s\right\|_2^2 .}
$$
Following the fact that $\left\|\sqrt{TN} \mathcal{D}_{N+t, j, p}\right\|_2$ is uniformly bounded by Lemma \ref{lem:3.3}, one can see that $\mathbb{E}\left\|\chi_{i, j}\right\|_2^2=O\left(1/T\right)$. Moreover, expanding $\bar{H}_\tau^{(1)}\left(X_{j s}-\widehat{\bl}_j^{\top} \widehat{\bbf}_s\right)$ at $\bar{H}_{i t}^{(1)}$ yields that
$$
\left[\bar{H}_\tau^{(1)}\left(X_{j s}-\widehat{\bl}_j^{\top} \widehat{\bbf}_s\right)\right]^2=\left[\bar{H}_{i t}^{(1)}+\bar{H}_\tau^{(2)}\left(X_{j s}-\bl_j^* \bbf_s^*\right) \cdot\left(\bl_{0 i}^{\top} \bbf_{0 t}-\widehat{\bl}_j^{\top} \widehat{\bbf}_s\right)\right]^2 \lesssim \left(\bl_{0 i}^{\top} \bbf_{0 t}-\widehat{\bl}_j^{\top} \widehat{\bbf}_s\right)^2,
$$
where $\bl_{j}^*$ lies between $\bl_{0j}$ and $\widehat\bl_{j}$, $\bbf_{s}^*$ lies between $\bbf_{0s}$ and $\widehat\bbf_{s}$.
Therefore, by Lemma \ref{lem:1.3}
$$
\sqrt{\frac{1}{TN} \sum_{j=1}^N \sum_{s=1}^T\left[\bar{H}_\tau^{(1)}\left(X_{j s}-\widehat{\bl}_j^{\top} \widehat{\bbf}_s\right)\right]^2\left\|\widehat{\bbf}_s\right\|_2^2} =O_p\l(d\left(\widehat{\theta}, \theta_0\right)\r)=O_p\left(1 / \sqrt{L}\right).
$$
So we get $(iii)$ is $O_p\left(1/L\right)$, while $(iv)$ can be shown to be $O_p\left(1/L\right)$ in the same way.

Finally, by (\ref{equ:Rt}), $(v)$ is $O_p\left(1/L\right)$. The $p$th element of $(vi)$ can be denoted as $1/(2 \sqrt{TN}) \sum_{j=1}^N \chi_{i, j} \mathcal{R}(\widehat{\theta})_j$, which is bounded by
$$
\frac{\sqrt{N}}{2 \sqrt{T}} \sqrt{\frac{1}{N} \sum_{j=1}^N\left\|\chi_{i, j}\right\|_2^2} \sqrt{\frac{1}{N} \sum_{j=1}^N\left\|\mathcal{R}(\widehat{\theta})_j\right\|_2^2}=O_p\left(1/\sqrt{T}\right) O_p\left(1/L\right)=o_p\left(1/L\right).
$$
It is easy to see that $(vii)$ is also $o_p(1/L)$ by the same method.
Combining the results for $(i)-(vii)$ above, we obtain that
$$
\frac{1}{T} \sum_{t=1}^T \tilde{H}_{i t}^{(1)} \left(\widehat{\bbf}_t-\bbf_{0 t}\right)=O_p\left(1/L\right),
$$
so that the proof concludes.

\section{Proof of Theorem \ref{th:2}}	
For $\theta_a\in \Theta^k$, $\theta_b\in \Theta^k$, let $$
d_k\left(\theta_a, \theta_b\right)=\left\|\Lambda_a F_a^{\top}-\Lambda_b F_b^{\top}\right\| / \sqrt{TN}=\sqrt{\frac{1}{T N} \sum_{t=1}^T \sum_{i=1}^N\left(\bl_{ai}^\top\bbf_{a t}-\bl_{bi}^\top\bbf_{b t}\right)^2}.
$$
Denote $\bl^k_{0i}=\left(\bl_{0i}^\top~\mathbf{0}_{1\times (k-r)}^\top\right)^\top$,
$\bbf^k_{0t}=\left(\bbf_{0t}^\top~\mathbf{0}_{1\times (k-r)}^\top\right)^\top$,
then $\Lb^k_0=\l(\bl_{01}^k,\ldots,\bl_{0N}^k\r)^\top=\l(\Lb_0~\mathbf{0}_{N\times(k-r)}\r)$,
$\Fb^k_0=\l(\bbf_{01}^k,\ldots,\bbf_{0T}^k\r)^\top=\l(\Fb_0~\mathbf{0}_{T\times(k-r)}\r)$, and $\theta^k_0=\l(\bl_{01}^{k\top}, \ldots, \bl_{0N}^{k\top},\bbf_{01}^{k\top}, \ldots, \bbf_{0T}^{k\top}\r)^{\top}$.
Similar to the proof of Theorem \ref{th:1}, it can be shown that
\beq\label{equ:d_k}
d_k(\widehat\theta^k,\theta_0^k)=O_p(1/\sqrt{L}).
\eeq

Let $\Lb^{k,r}$ and $\Fb^{k,r}$ be the first $r$ columns of $\Lb^k$ and $\Fb^k$ respectively, and $\Lb^{k,-r}$ and $\Fb^{k,-r}$ be the remaining $(k-r)$ columns of $\Lb^k$ and $\Fb^k$.
	Define $\Theta^k(\delta)=\{\theta^k\in\Theta^k:d_k(\theta^k,\theta_0^k)\leq\delta\}$.
	Similar to the proof of Lemma \ref{lem:1.2}, for any $\theta^k\in \Theta^k(\delta)$, it holds that
	$$
	\frac{1}{N}\|\Lb^{k}-\Lb_0^k\Sbb^k\|_F^2+\frac{1}{T}\|\Fb^{k}-\Fb_0^k \Sbb^k\|_F^2\lesssim\delta^2
	$$
	where $\Sbb^k=\operatorname{sgn}\left((\Fb^{k})^\top\Fb_0^k/T\right)$.
	And we have
	\beq\label{equ:LFF}
	\frac{1}{N}\|\Lb^{k,r}-\Lb_0\Sbb^{k,r}\|_F^2\lesssim \delta^2,~\frac{1}{T}\|\Fb^{k,r}-\Fb_0 \Sbb^{k,r}\|_F^2\lesssim\delta^2,~\frac{1}{T}\|\Fb^{k,-r}\|_F^2\lesssim\delta^2
	\eeq
	where $\Sbb^{k,r}=\operatorname{sgn}\left((\Fb^{k,r})^\top\Fb_0/T\right)$.
	Since $\|\Fb^{k,r}-\Fb_0 \Sbb^{k,r}\|_2^2/T\leq \|\Fb^{k,r}-\Fb_0 \Sbb^{k,r}\|_F^2/T\lesssim\delta^2$ for any $\delta>0$,
	\beq\label{equ:sigma_j<r}
	|\widehat\sigma_{Tj}^k-\sigma_{0j}|=o_p(1) \text{ for } j=1,2,\ldots,r.
	\eeq
	
	By (\ref{equ:d_k}) and (\ref{equ:LFF}),
	\beq\label{equ:sigma_j>r}
	\sum_{j=r+1}^k \hat{\sigma}_{T j}^k=\left\|\widehat{\Fb}^{k,-r}\right\|_F^2 / T = O_p\left(1/L\right) .
	\eeq
	Finally, by (\ref{equ:sigma_j<r}) and (\ref{equ:sigma_j>r}), we have
	$$
	\begin{aligned}
		\PP\left(\widehat{r} \neq r\right) & =\PP\left(\widehat{r}<r\right)+\PP\left(\widehat{r}>r\right) \\
		& \leq\PP\left(\widehat{\sigma}_{Tr}^k \leq P\right)+\PP\left(\widehat{\sigma}_{Tr+1}^k>P\right)=o(1)
	\end{aligned}
	$$
	with the assumption $P\rightarrow 0$ and $PL\rightarrow \infty$ as $T,N\rightarrow \infty$.
	It then follows that $\PP\left(\widehat{r}=r\right) \rightarrow 1$, which completes the proof.

\end{document}